\documentclass[12pt]{article}

\usepackage{newtxtext,newtxmath}

\usepackage{graphicx}

\usepackage[letterpaper,margin=1in]{geometry}

\renewenvironment{abstract}
	{\quotation}
	{\endquotation}

\date{}

\makeatletter
\renewcommand{\fnum@figure}{\textbf{Figure \thefigure}}
\renewcommand{\fnum@table}{\textbf{Table \thetable}}
\makeatother

\usepackage{url}

\usepackage{booktabs} 
\usepackage{hyperref}
\usepackage{nameref}
\usepackage{makecell}
\usepackage{multirow}
\usepackage[labelfont=bf]{subcaption}
\def\mytitle{
	Control of a commercially available vehicle by a tetraplegic human using a brain-computer interface
}
\title{\bfseries \boldmath \mytitle}

\author
{
Xinyun~Zou$^{1,2\ast}$, Jorge~Gamez$^{1,2\ast}$, Meghna~Menon$^{3}$, Phillip~Ring$^{3}$,\and
Chadwick~Boulay$^{4}$, Likhith~Chitneni$^{4}$, Jackson~Brennecke$^{4}$, Shana~R.~Melby$^{4}$,\and
Gracy~Kureel$^{4}$, Kelsie~Pejsa$^{1,2}$, Emily~R.~Rosario$^{5}$, Ausaf~A.~Bari$^{6}$,\and
Aniruddh~Ravindran$^{3}$, Tyson~Aflalo$^{1,2}$, Spencer~S.~Kellis$^{4,7,8}$,\and
Dimitar~Filev$^{3}$, Florian~Solzbacher$^{4,9,10,11}$, Richard~A.~Andersen$^{1,2}$\and
\small$^{1}$Division of Biology and Biological Engineering, California Institute of Technology,\and
\small Pasadena, CA 91125, USA.\and
\small$^{2}$T$\&$C Chen Brain-Machine Interface Center, California Institute of Technology,\and
\small  \hspace*{2cm} Pasadena, CA 91125, USA. \hspace*{2cm} \and
\small$^{3}$Ford Motor Company, Dearborn, MI 48126, USA.\and
\small$^{4}$Blackrock Neurotech, Salt Lake City, UT 84108, USA.\and
\small$^{5}$Casa Colina Hospital and Centers for Healthcare, Pomona, CA 91767, USA.\and
\small$^{6}$Department of Neurological Surgery, University of California Los Angeles, \and
\small \hspace*{2cm} Los Angeles, CA 90095, USA. \hspace*{2cm} \and
\small$^{7}$Department of Neurological Surgery, Keck School of Medicine, \and
\small University of Southern California, Los Angeles, CA 90033, USA.\and
\small$^{8}$Neurorestoration Center, University of Southern California, Los Angeles, CA 90033, USA.\and
\small$^{9}$Department of Electrical and Computer Engineering, The University of Utah, \and
\small Salt Lake City, UT 84112, USA.\and
\small$^{10}$Department of Biomedical Engineering, The University of Utah, Salt Lake City, UT 84112, USA.\and
\small$^{11}$Department of Materials Science and Engineering, The University of Utah, \and
\small Salt Lake City, UT 84112, USA.\and
\small$^\ast$Corresponding authors. E-mails: xzou@caltech.edu, jgamez@caltech.edu
}

\begin{document}

\maketitle

\begin{abstract} \bfseries \boldmath
Brain-computer interfaces (BCIs) read neural signals directly from the brain to infer motor planning and execution. However, the implementation of this technology has been largely limited to laboratory settings, with few real-world applications. We developed a BCI system to drive a vehicle in both simulated and real-world environments. We demonstrate that an individual with tetraplegia, implanted with intracortical BCI electrodes in the posterior parietal cortex (PPC) and the hand knob region of the motor cortex (MC), reacts at least as fast and precisely as motor intact participants. This BCI participant, living in California, could also remotely drive a Ford Mustang Mach-E vehicle in Michigan. Our teledriving tasks relied on cursor movement control for speed and steering in a closed urban test facility and through a predefined obstacle course. These two tasks serve as a proof-of-concept that takes into account the safety and feasibility of BCI-controlled driving. The final BCI system added click control for full-stop braking and thus enabled bimanual cursor-and-click control for simulated town driving with the same proficiency level as the motor intact control group through a virtual town with traffic. This first-of-its-kind implantable BCI application not only highlights the versatility and innovative potentials of BCIs but also illuminates the promising future for the development of life-changing solutions to improve independent mobility for those who suffer catastrophic neurological injury. 
\end{abstract}

\section*{Summary}

This study develops an intracortical BCI system that allows a tetraplegic person to volitionally drive a vehicle.

\section*{INTRODUCTION}
Brain computer interfaces (BCIs) are changing the ways in which humans interact with their environment. BCIs translate neural activity from the brain and can help individuals overcome motor disabilities, enabling their control of robotic limbs, wheelchairs, exoskeletons, and functional electrical stimulation of the body \cite{hochberg_reach_2012,salazar-gomez_correcting_2017,tonin_learning_2022,guan_decoding_2023,kim_noir_2024}. BCIs also facilitate communication by decoding speech articulation, interpreting motor imagery for handwriting, and enabling interactions with keyboards \cite{willett_high-performance_2023,wandelt_representation_2024}. In addition, they support the use of computers, tablets, and smartphones by controlling cursor movement, clicks, and keyboards \cite{aflalo_decoding_2015,nuyujukian_cortical_2018,deo_brain_2024}. For those with severe paralysis, BCI technology increases their opportunities to improve independent mobility, return to work, and enhance social interactions \cite{van_erp_brain-computer_2012}.

In 2019, approximately 20.6 million people worldwide lived with spinal cord injury (SCI) \cite{safdarian_global_2023}, roughly 60\% of whom were estimated to have tetraplegia, meaning that all four of their limbs were affected by paralysis. Tetraplegia has devastating effects on an individual's independence, functionality, mental health, and overall quality of life. Beyond activities of daily living, individuals with SCI are often dependent on caregivers, partners, and other family members for transportation needs. Driving is frequently regarded as a key component of individual autonomy, particularly in societies such as the United States, where the ownership of personal vehicles and the necessity of driving are widespread. These capabilities facilitate independent mobility and access to social, economic, and healthcare resources. The capacity of personally controlling a vehicle not only encompasses transportation but also preserves a sense of agency and independence, and its loss constitutes a decline in functional independence and autonomy.

\begin{figure}
     \centering
    \includegraphics[width=\textwidth]{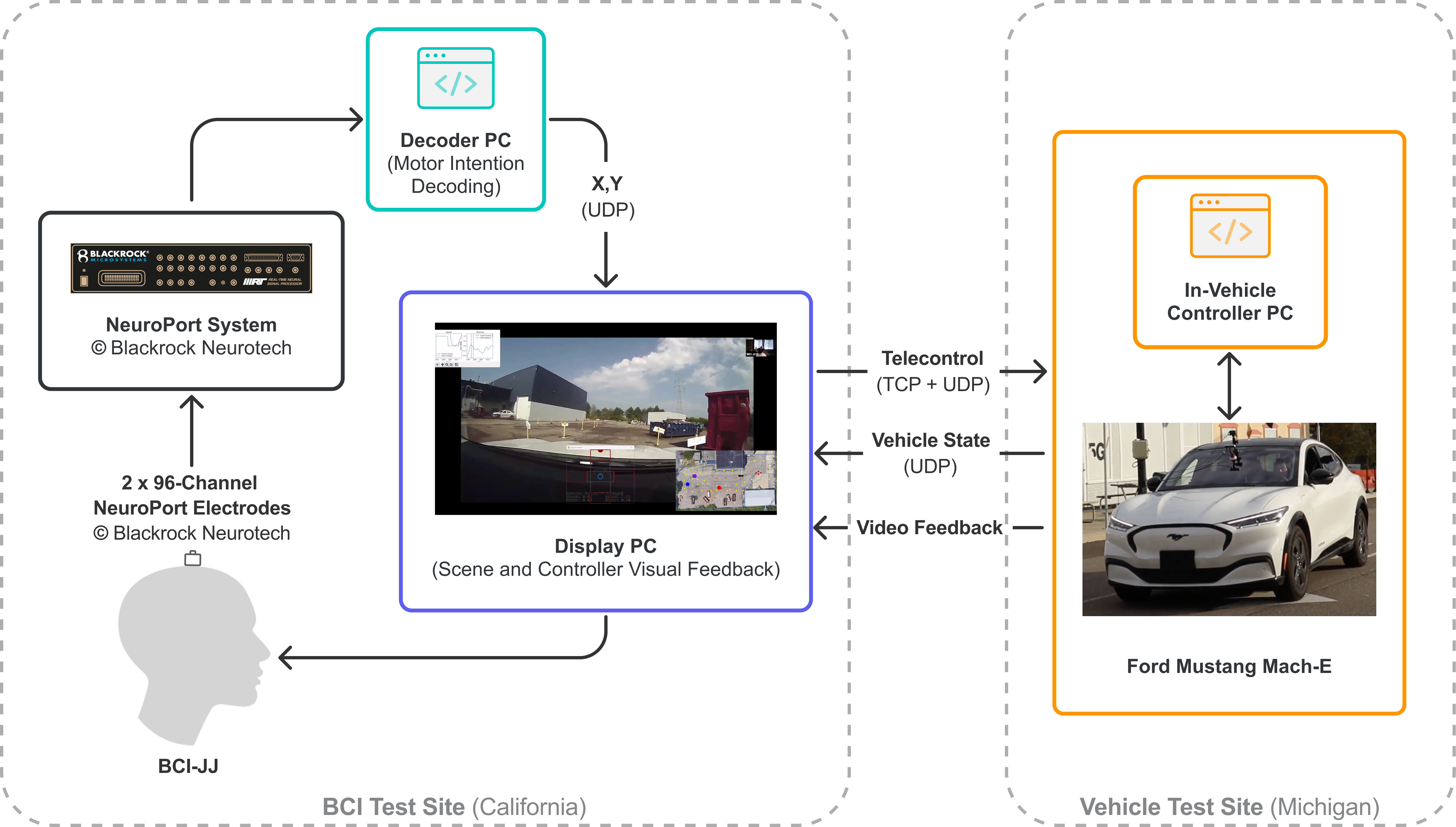}
    \caption{\textbf{The BCI control system diagram for teledriving a Ford Mustang Mach-E vehicle.} Our setup included the decoder and display computers at the BCI test site in California and the in-vehicle controller computer at the vehicle test site in Michigan. Brain signals were recorded from BCI-JJ's posterior parietal cortex (PPC) and the hand knob region of the motor cortex (MC) via the intracortical NeuroPort Electrodes \copyright~Blackrock Neurotech and processed by the NeuroPort system \copyright~Blackrock Neurotech. The decoder computer extracted the neural features from the acquired signals, decoded the motor intention, and generated a corresponding motor command. This motor command was transmitted to the display computer and then sent to the in-vehicle controller via TCP and UDP. At the vehicle test site, the in-vehicle controller computer executed the motor command for the vehicle. This computer also recorded a live-stream video from a camera mounted on the vehicle. The video feedback and the vehicle state were transferred to the display computer at the BCI test site for BCI-JJ to watch and respond in real-time. }
    \label{fig:bcisys}
\end{figure}

The non-invasive electroencephalogram (EEG) technology has previously been used to develop BCIs for vehicle driving that assess cognitive states \cite{zhang_eeg-based_2015,vecchiato_electroencephalographic_2019,cao_multi-channel_2019,rito_lima_neurobehavioural_2020,liu_human-machine_2023}, predict intentions \cite{haufe_electrophysiology-based_2014,yang_driving_2018,lu_modeling_2024}, and perform basic vehicle control \cite{di_liberto_robust_2021,zhou_novel_2021}. However, EEG faces challenges such as susceptibility to noise, lack of individual generalizability, and the need for extensive training \cite{lu_modeling_2024,lebedev_brainmachine_2006,waldert_invasive_2016,yu_toward_2016,stawicki_driving_2016,zhuang_motion_2017,li_human-vehicle_2018,abiri_comprehensive_2019,zhang_motor_2020,zhang_brain-controlled_2023,sharma_recent_2023}. In contrast, intracortical BCIs, which involve surgical implantation of recording electrodes, offer greater signal fidelity and stability, allowing for more precise decoding and possibilities of broader applications \cite{hochberg_reach_2012,willett_high-performance_2023,aflalo_decoding_2015,andersen_preserved_2022}. 

In this study, we present the first-of-its-kind intracortical BCI system that allows remote driving control of a commercially available vehicle by a tetraplegic human participant (see Figure~\ref{fig:bcisys}). A multidisciplinary group spanning academia and industry built a BCI driving system that allowed a tetraplegic BCI participant, BCI-JJ, to remotely operate a Mustang Mach-E (Ford Motor Company, Dearborn, MI) in closed test environments. Two real-world driving tasks were conducted in a closed urban test facility and on an obstacle course without other traffic or pedestrians, with controlled speed limits, predefined safety boundaries, and emergency procedures in place, to ensure appropriate safety control. As part of an FDA-approved clinical trial (ClinicalTrials.gov number NCT01958086), we implanted BCI-JJ intracortical NeuroPort Electrodes (Blackrock Neurotech, Salt Lake City, UT) in both the posterior parietal cortex (PPC) and the hand knob region of the motor cortex (MC) (see figure~\ref{Sfig:brainimplant}). For the teledriving tasks with the Mach-E vehicle, BCI-JJ employed mental imagery of the right thumb for cursor-movement control of speed and steering. By employing additional mental imagery of the left index finger for click control of full-stop braking, we further upgraded our BCI driving system with a more intuitive separation of driving functions across the left and right limbs. Using this bimanual control scheme, BCI-JJ could volitionally drive a virtual vehicle with the same proficiency level as a motor intact control group through a designated route in a simulated town driving environment with heavy traffic and other complex traffic elements. In experimentally constrained laboratory tests, our BCI system enabled BCI-JJ to react at least as fast and precisely as the motor intact control group performing the same reaction-time tasks with a standard computer mouse.

\begin{figure}
    \centering
     \begin{subfigure}[b]{0.78\textwidth}
         \centering
         \caption{}
         \includegraphics[width=\textwidth]{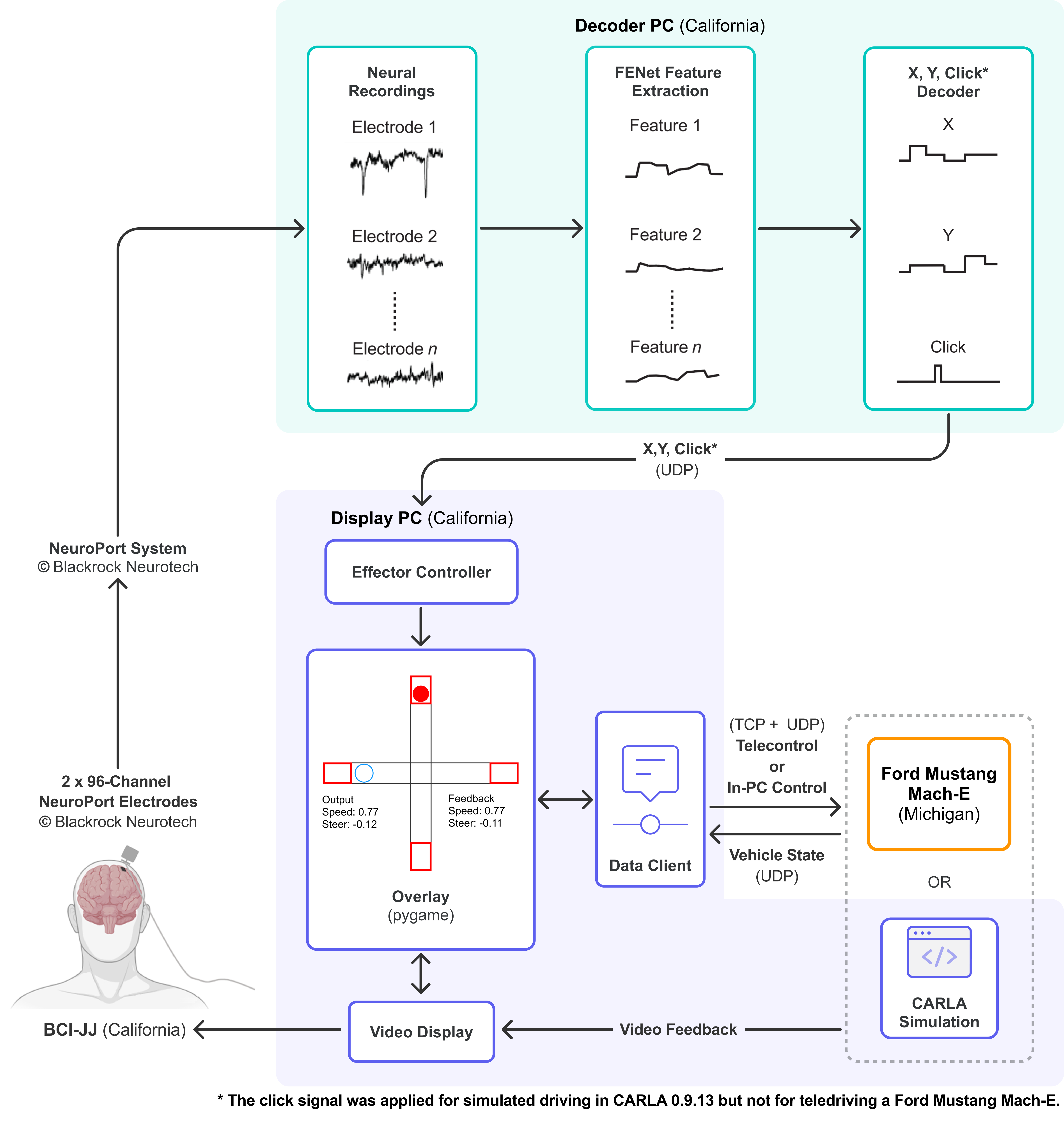}
         \label{fig:bcimechanism}
     \end{subfigure}
     
     \begin{subfigure}[b]{0.78\textwidth}
         \centering
         \caption{}
         \includegraphics[width=0.68\textwidth]{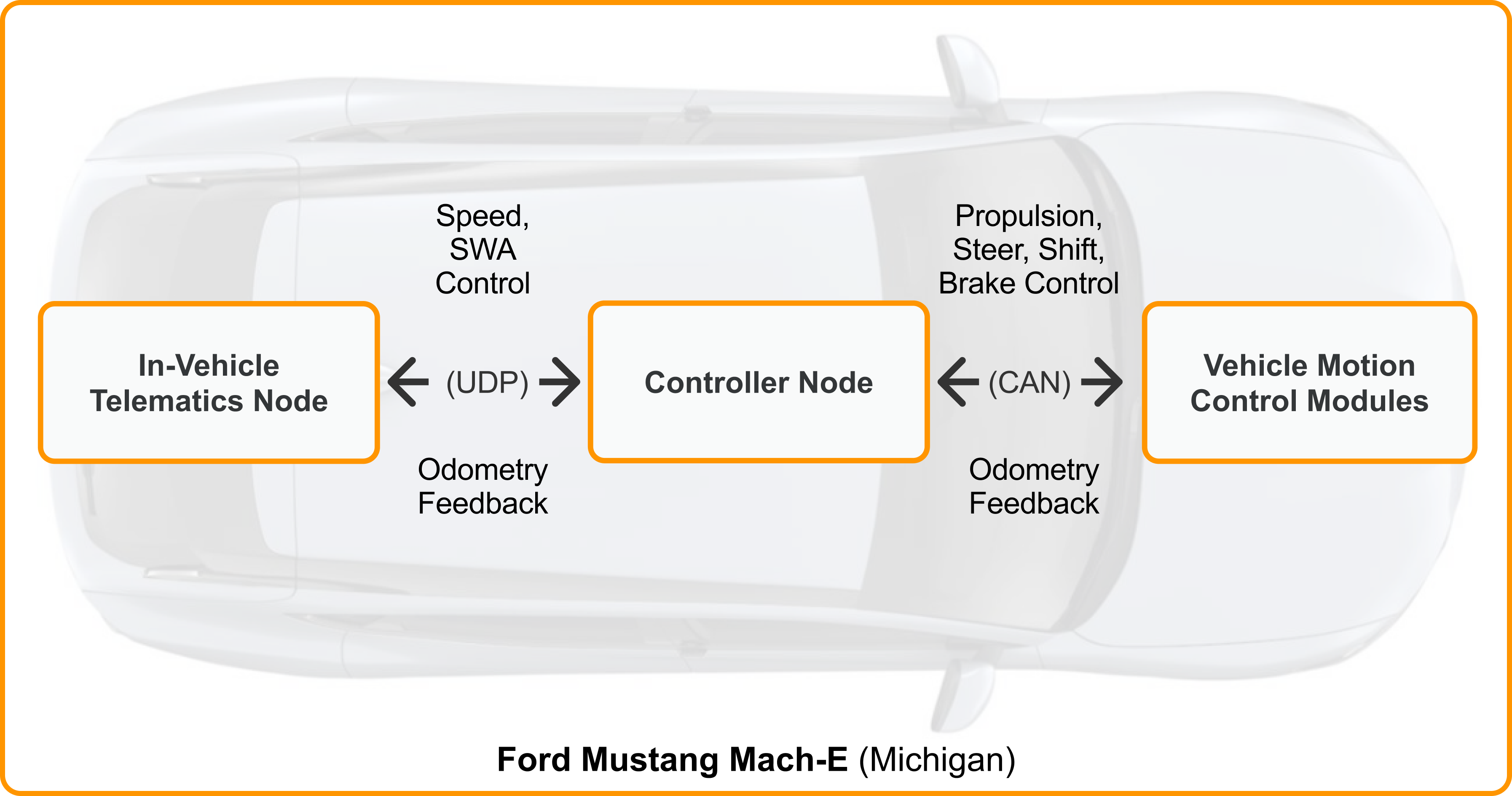}
         \label{fig:teledriving_car}
     \end{subfigure}
    \caption{\textbf{Detailed flowcharts of the BCI-enabled driving system.} (\textbf{A}) The detailed mechanism of BCI neural signal decoding and driving control of either a Ford Mustang Mach-E vehicle remotely or a CARLA simulated vehicle. (\textbf{B}) The vehicle control architecture of a Ford Mach-E vehicle.}
    \label{fig:bciflowcharts}
\end{figure}

This work represents both a BCI-enabled driving system with the most advanced capability to date and a model for meaningful academic-industry collaboration in the rapidly growing field of BCIs. Building this system required neuroscientific development to provide reliable cursor movement decoding and click classification enabled by implants across the cortical motor system, as well as significant technology development to allow moment-by-moment BCI control of a full-sized commercially available vehicle remotely as a proof-of-concept and of a virtual vehicle in a simulated town with the same proficiency level as motor intact behavior.

\section*{RESULTS}
\subsection*{System and tasks overview}
We used FENet \cite{haghi_enhanced_2024}, a multi-layer, one-dimensional convolutional neural network, to process neural signals collected from the NeuroPort Electrodes via the NeuroPort System (Blackrock Neurotech, Salt Lake City, UT) and generate neural features to be used as input by a linear decoder to estimate motor intention (see the ``\nameref{sec:decoder}'' section) and enable the control of three key variables of a vehicle: steering, speed, and braking (see Figure~\ref{fig:bcisys}). The ``\nameref{sec:driving-tests}'' section will explain each module of our BCI-enabled driving system at the BCI test site and the vehicle test site in details, as shown in Figures~\ref{fig:bcimechanism} and~\ref{fig:teledriving_car} respectively.

\begin{table}
  \caption{\textbf{BCI reaction and driving task summary.} ``MI'' means comparisons with motor intact participants. ``CARLA Lb 2.0'' stands for the CARLA Autonomous Driving Leaderboard 2.0. Green circles represent effectors tested for click movement, including left and right index fingers, ring fingers and power grips. Blue circles represent effectors tested for cursor movement, including the right thumb. }
  \label{tab:tasksummary}
  \centering
  \begin{tabular}{l|l|l|l|l}
    \toprule
    \bf{Task Name} & \bf{Platform} & \bf{Tested Components} & \bf{Tested BCI Effectors} & \bf{MI?}\\
    \midrule
    \makecell[l]{Simple \\ Reaction \\ Time} & Unity 2018.4.23f1 & \makecell[l]{Simple reaction time $\&$ \\ performance measures} & \makecell{\includegraphics[height=0.6in]{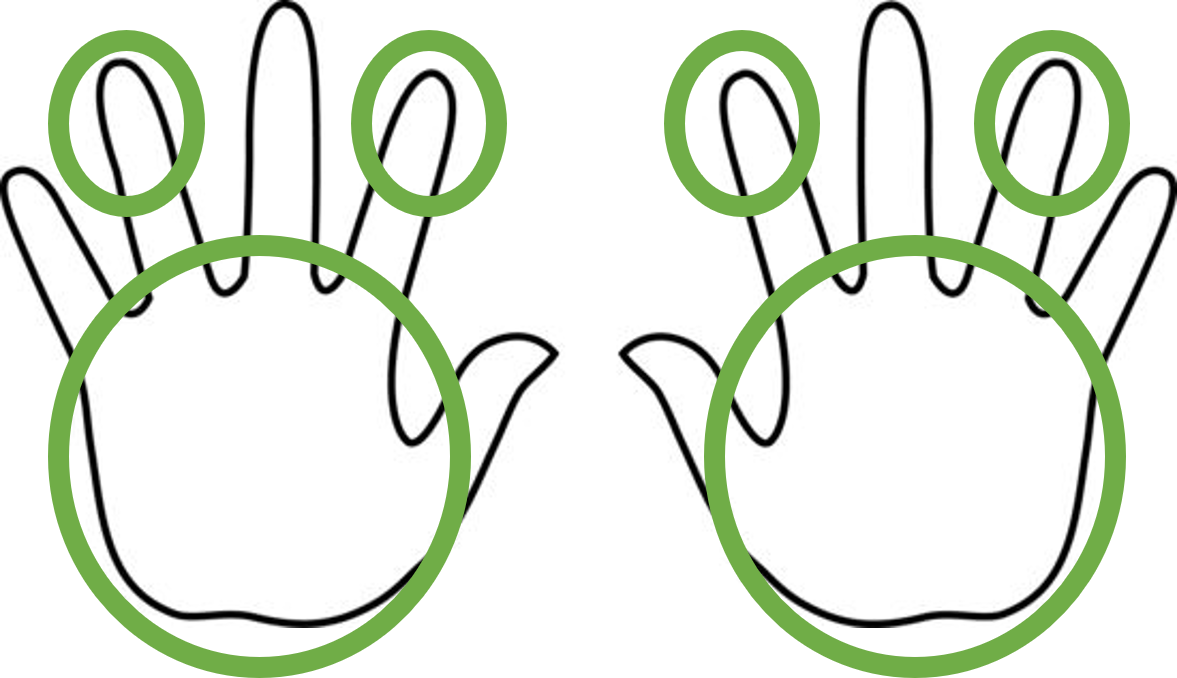}} & Yes \\
    \midrule
    \makecell[l]{Braking \\ Reaction \\ Time} & \makecell[l]{CARLA 0.9.13 $+$ \\ braking overlay} & \makecell[l]{Braking reaction time $\&$ \\ performance measures} & \makecell{\includegraphics[height=0.6in]{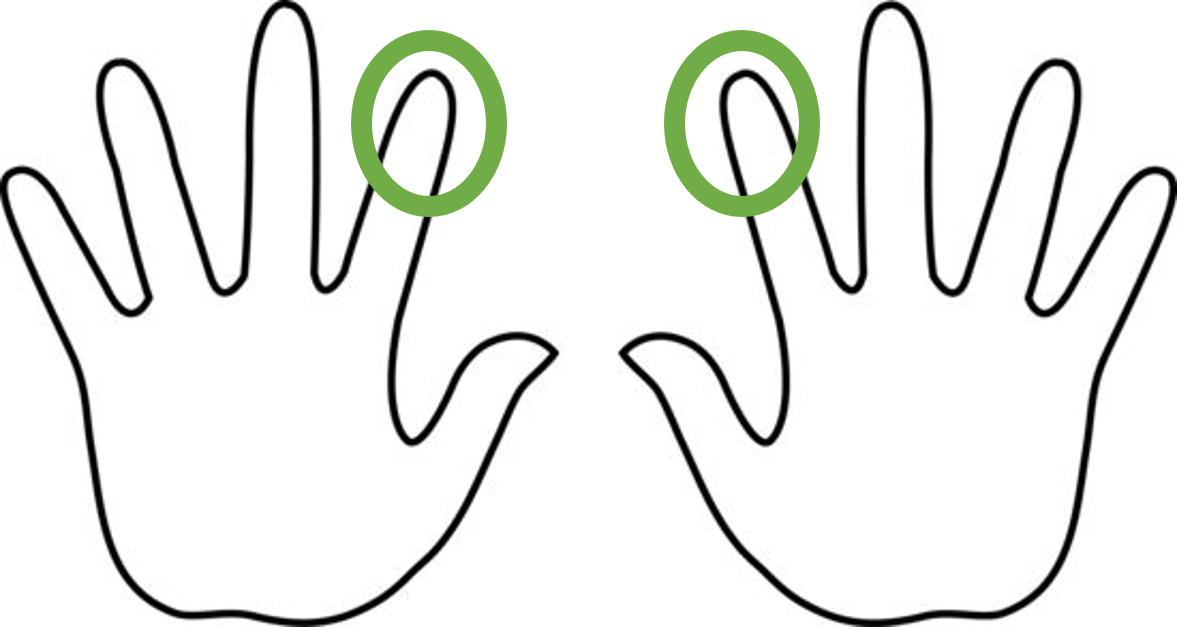}} & Yes \\
    \midrule
    \makecell[l]{Mcity \\ Teledriving} & \makecell[l]{Ford Mach-E $+$ \\ speed and steering \\ overlay} & \makecell[l]{4 random routes (with \\ turnings, stops, lane \\ switches, roundabouts)} & \makecell{\includegraphics[height=0.6in]{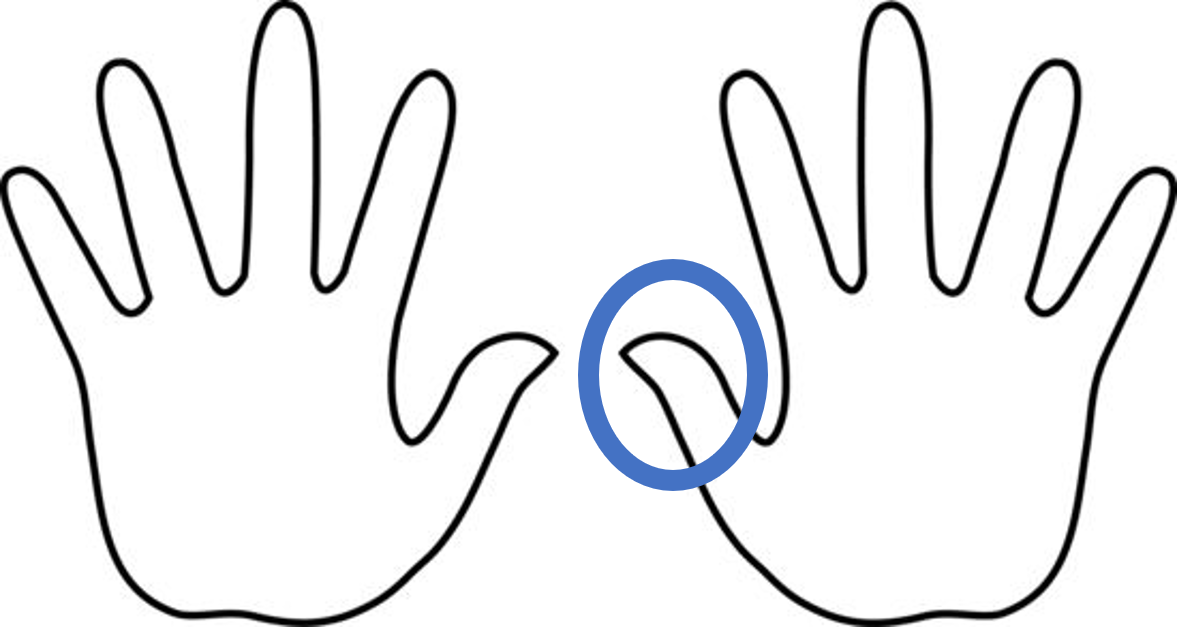}} & No\\
    \midrule
    \makecell[l]{Obstacle \\ -course \\ Teledriving} & \makecell[l]{Ford Mach-E $+$ \\ speed and steering \\ overlay} & \makecell[l]{4 fixed routes (with \\ stops, turnings, lane \\ switches, roundabouts)} & \makecell{\includegraphics[height=0.6in]{table_1_picC.png}} & No\\
    \midrule
    \makecell[l]{Simulated \\ Town \\ Driving} & \makecell[l]{CARLA Lb 2.0 $+$ \\ speed, steering and \\ braking overlay} & \makecell[l]{A fixed route (with \\ traffic lights, turnings, \\ lane switches)} & \makecell{\includegraphics[height=0.6in]{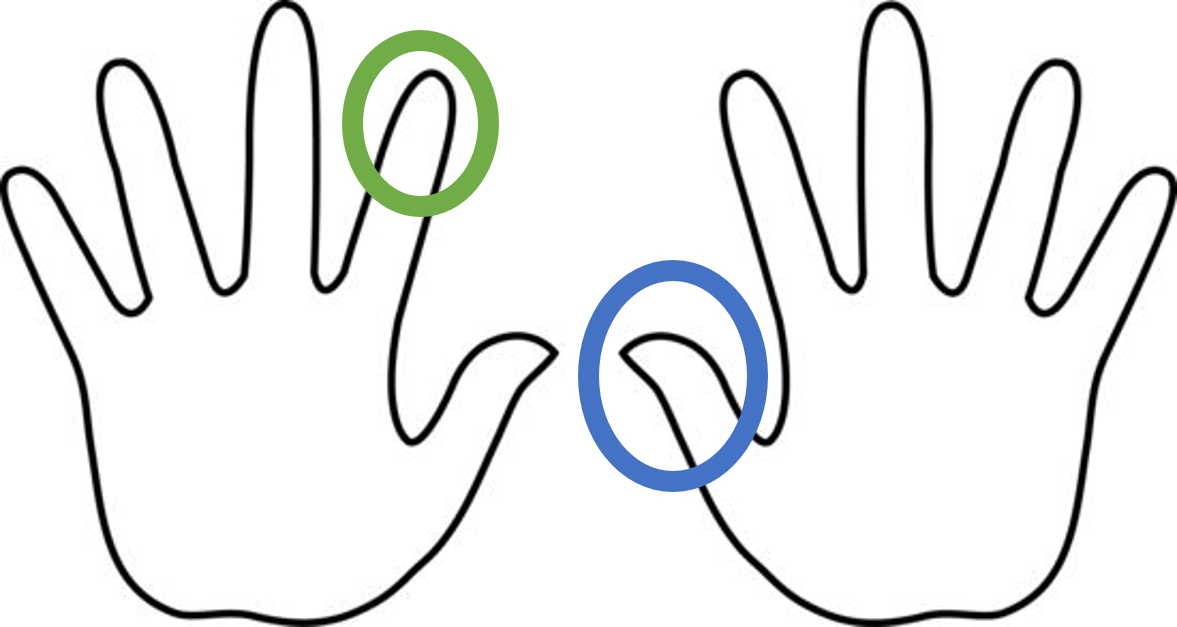}} & Yes \\
  \bottomrule
\end{tabular}
\end{table}

\begin{figure}
     \centering
     \begin{subfigure}[b]{0.47\textwidth}
         \centering
         \caption{}
         \includegraphics[width=0.8\textwidth]{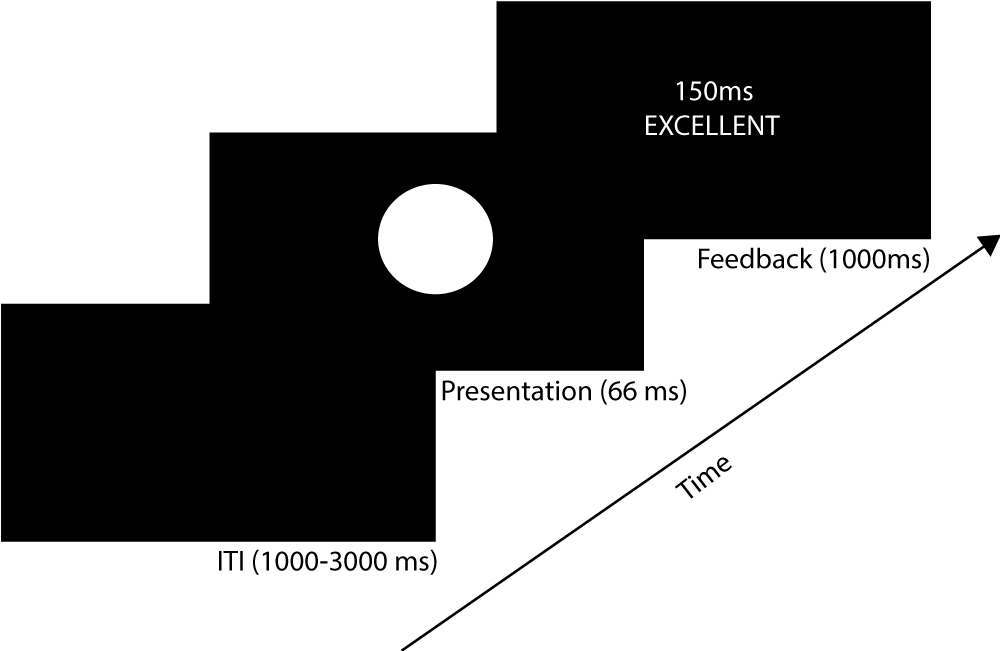}
         \label{fig:simpletask}
     \end{subfigure}
     \hfill
     \begin{subfigure}[b]{0.5\textwidth}
         \centering
         \caption{}
         \includegraphics[width=\textwidth]{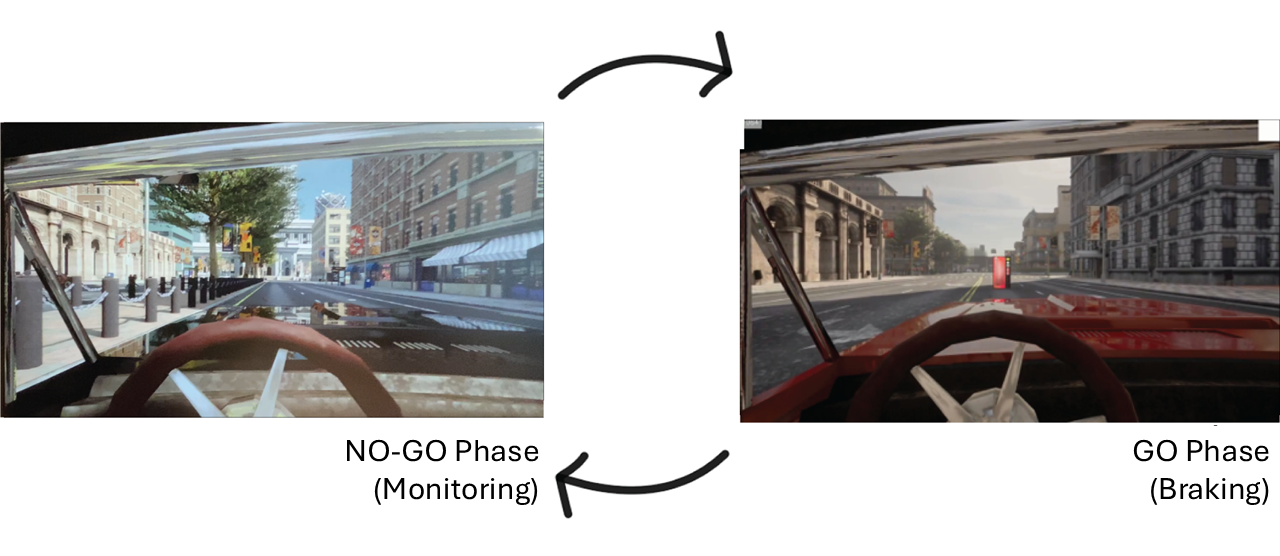}
         \label{fig:brakingtask}
     \end{subfigure}
     
     \begin{subfigure}[b]{0.47\textwidth}
         \centering
         \caption{}
         \includegraphics[height=1.45in]{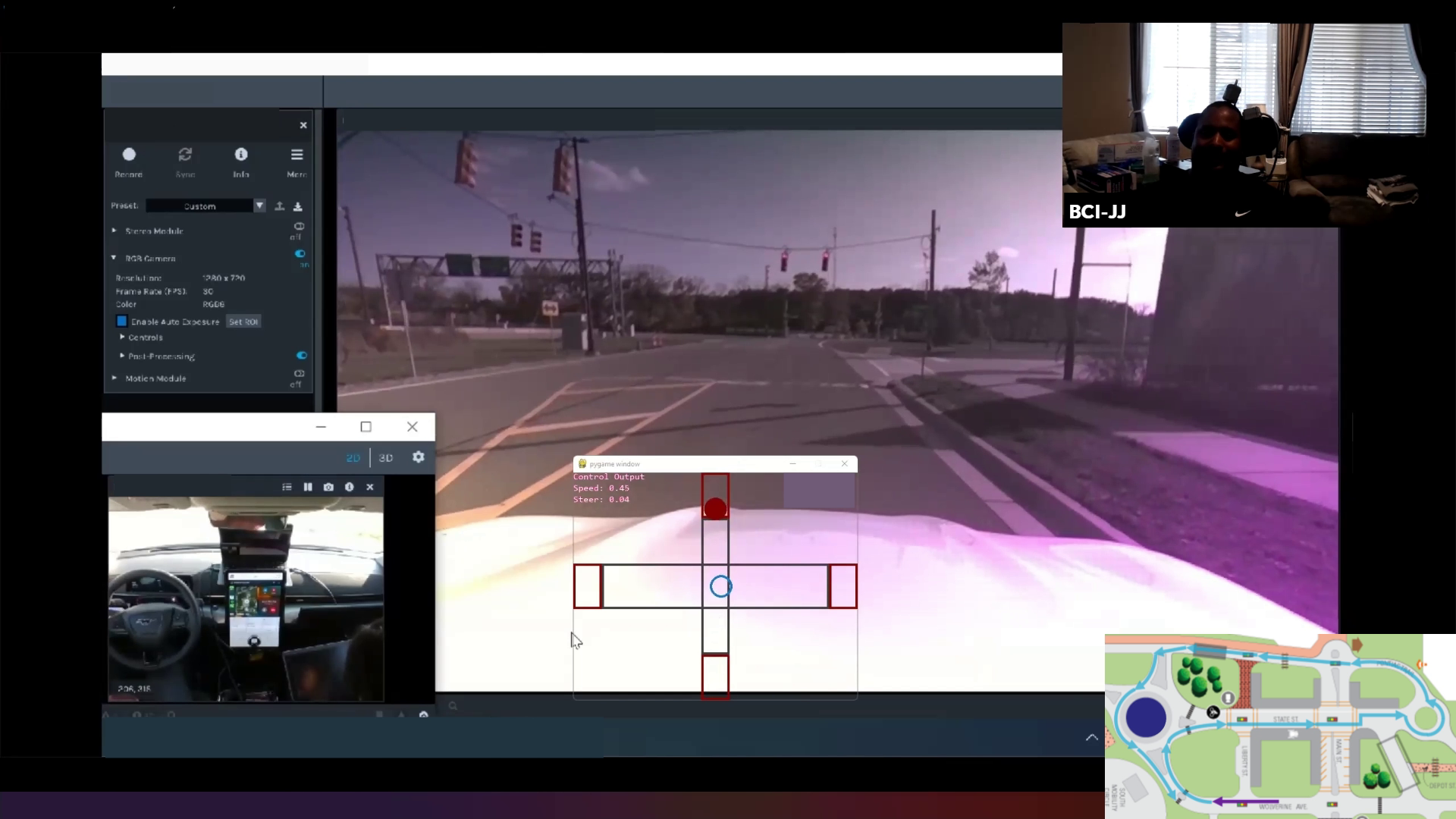}
         \label{fig:mcity}
     \end{subfigure}
     \hfill
     \begin{subfigure}[b]{0.5\textwidth}
         \centering
         \caption{}
         \includegraphics[height=1.45in]{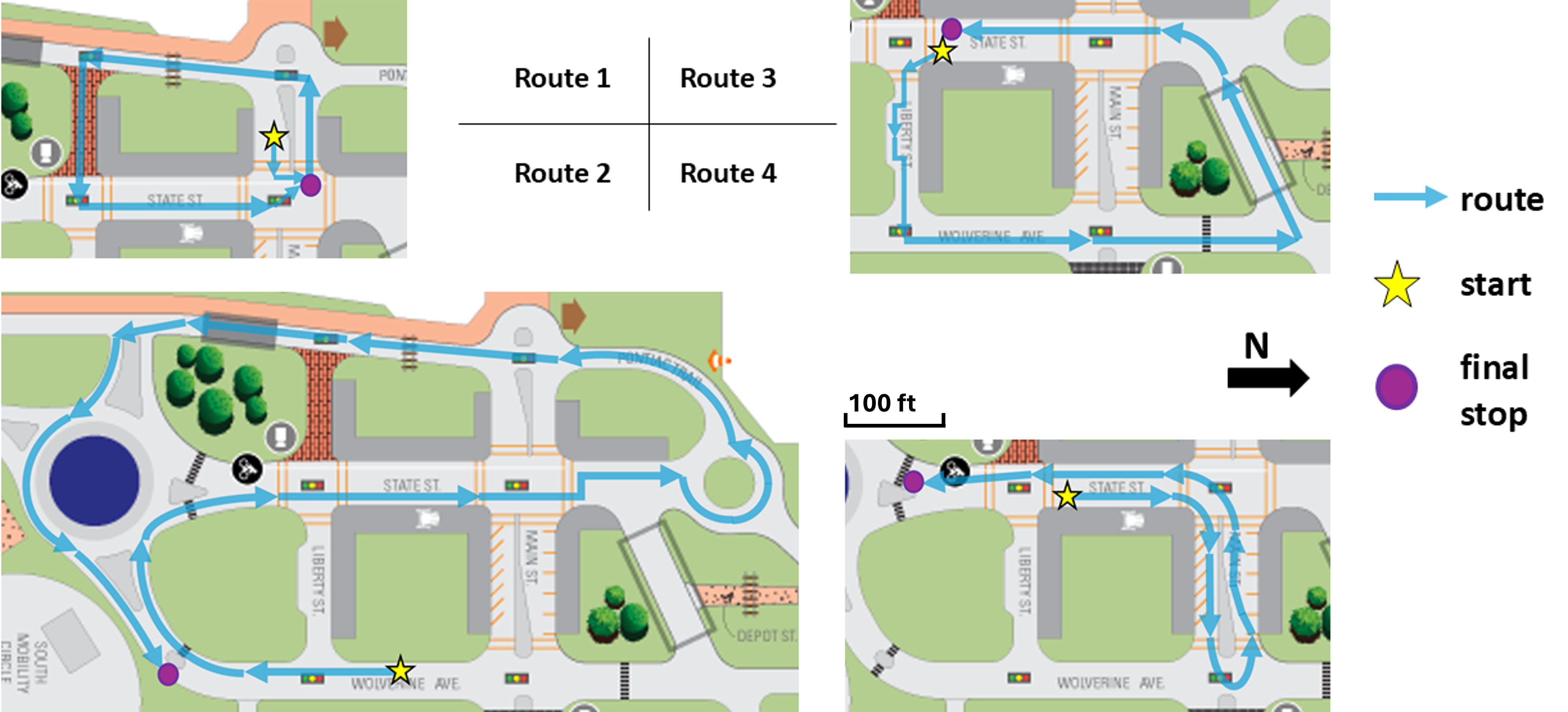}
         \label{fig:route_mcity}
     \end{subfigure}
     \par\smallskip
     \begin{subfigure}[b]{0.47\textwidth}
         \centering
         \caption{}
         \includegraphics[height=1.45in]{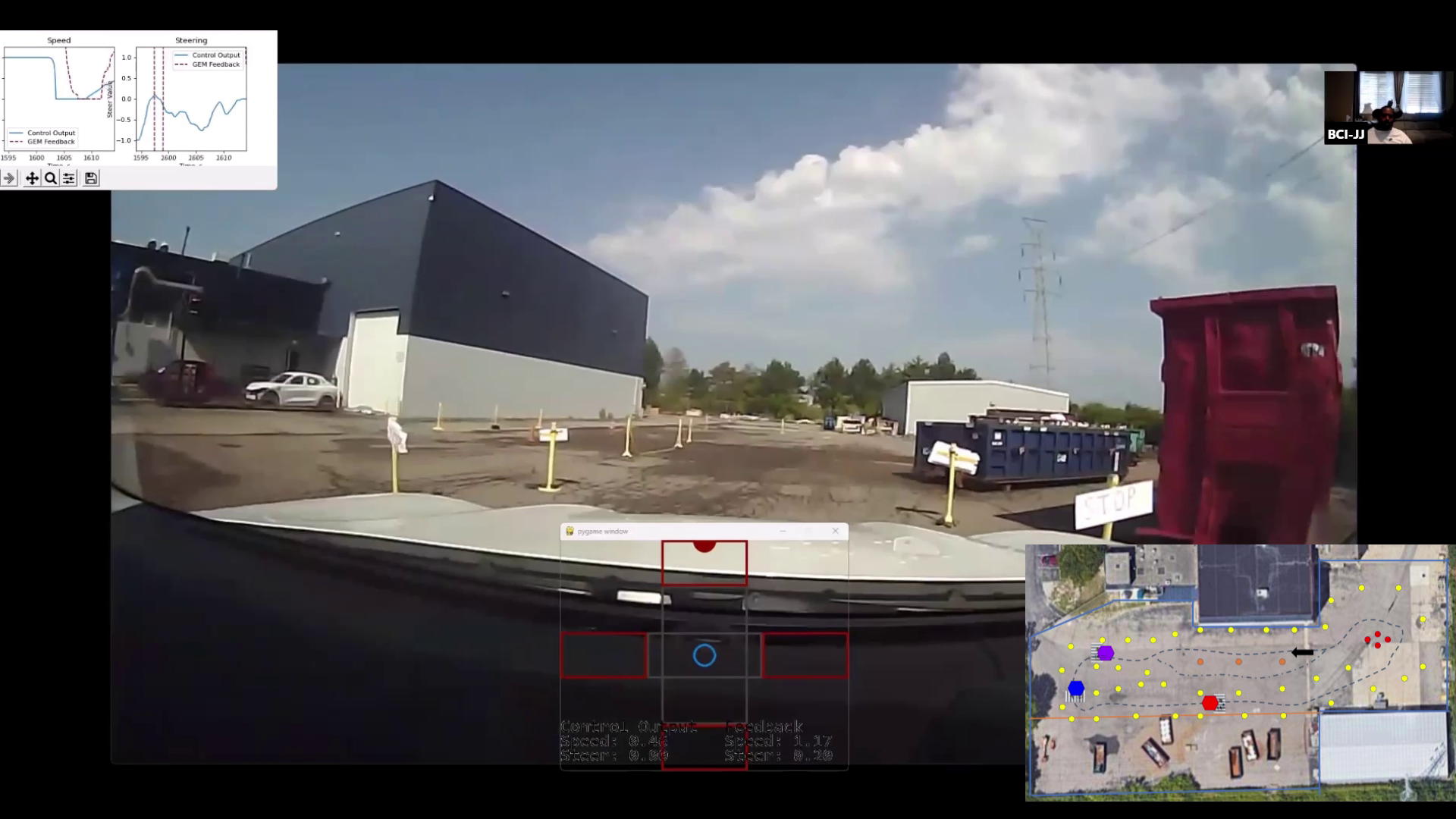}
         \label{fig:parkinglot}
     \end{subfigure}
     \hfill
     \begin{subfigure}[b]{0.5\textwidth}
         \centering
         \caption{}
         \includegraphics[height=1.45in]{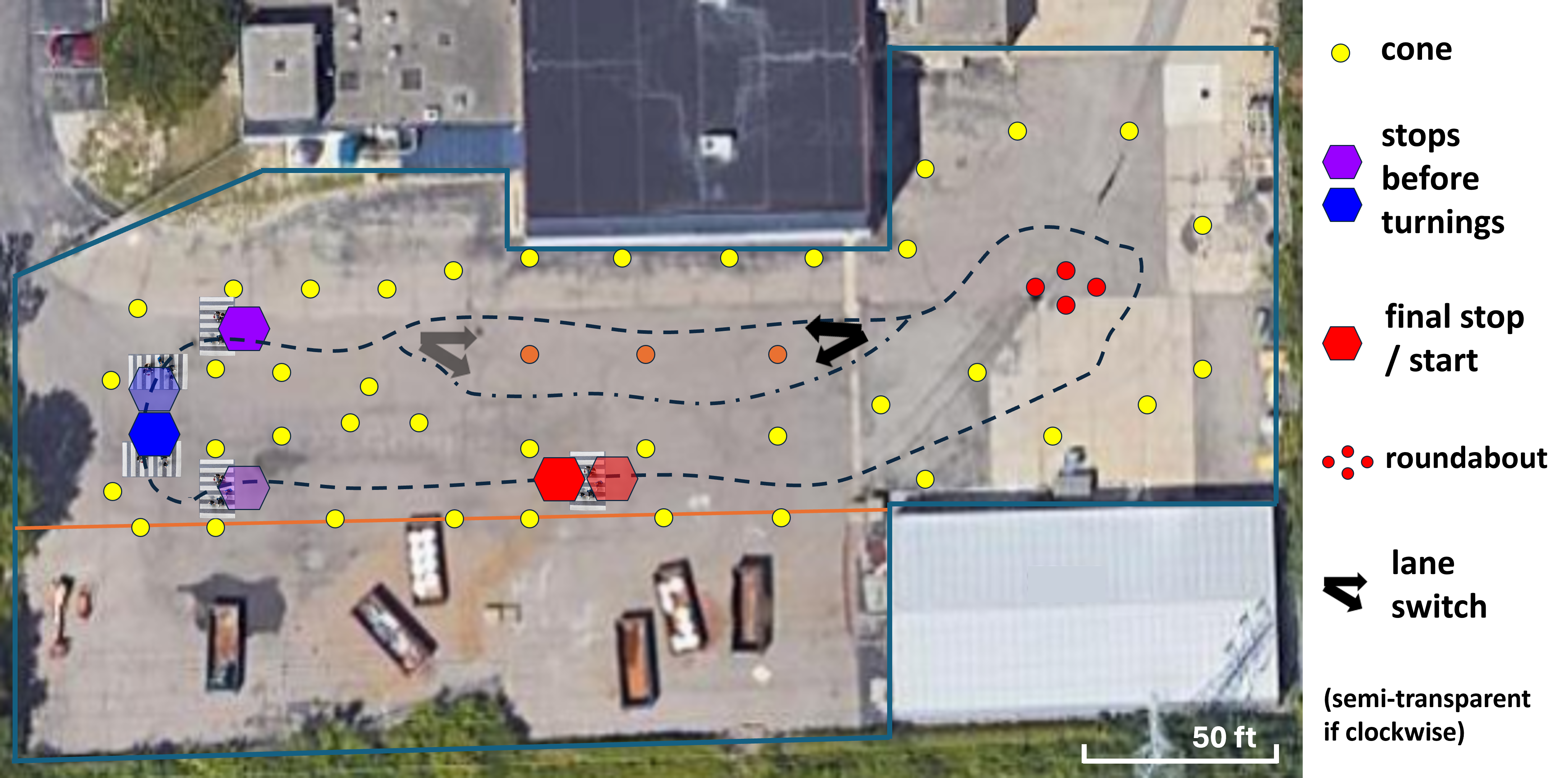}
         \label{fig:route_parkinglot}
     \end{subfigure}
     \par\smallskip
    \begin{subfigure}[b]{0.47\textwidth}
         \centering
         \caption{}
         \includegraphics[height=1.36in]{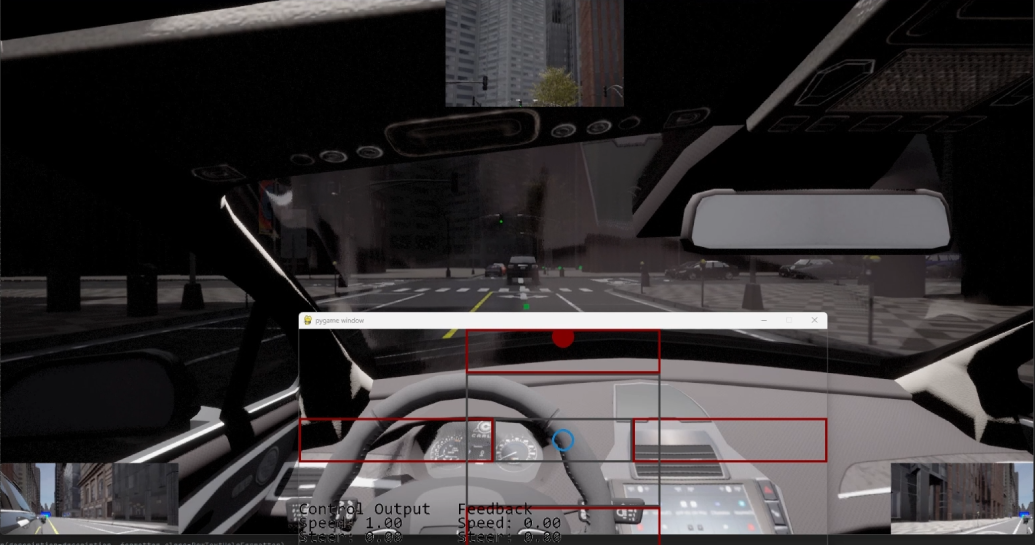}
         \label{fig:leaderboard}
     \end{subfigure}
     \hfill
     \begin{subfigure}[b]{0.5\textwidth}
         \centering
         \caption{}
         \includegraphics[height=1.36in]{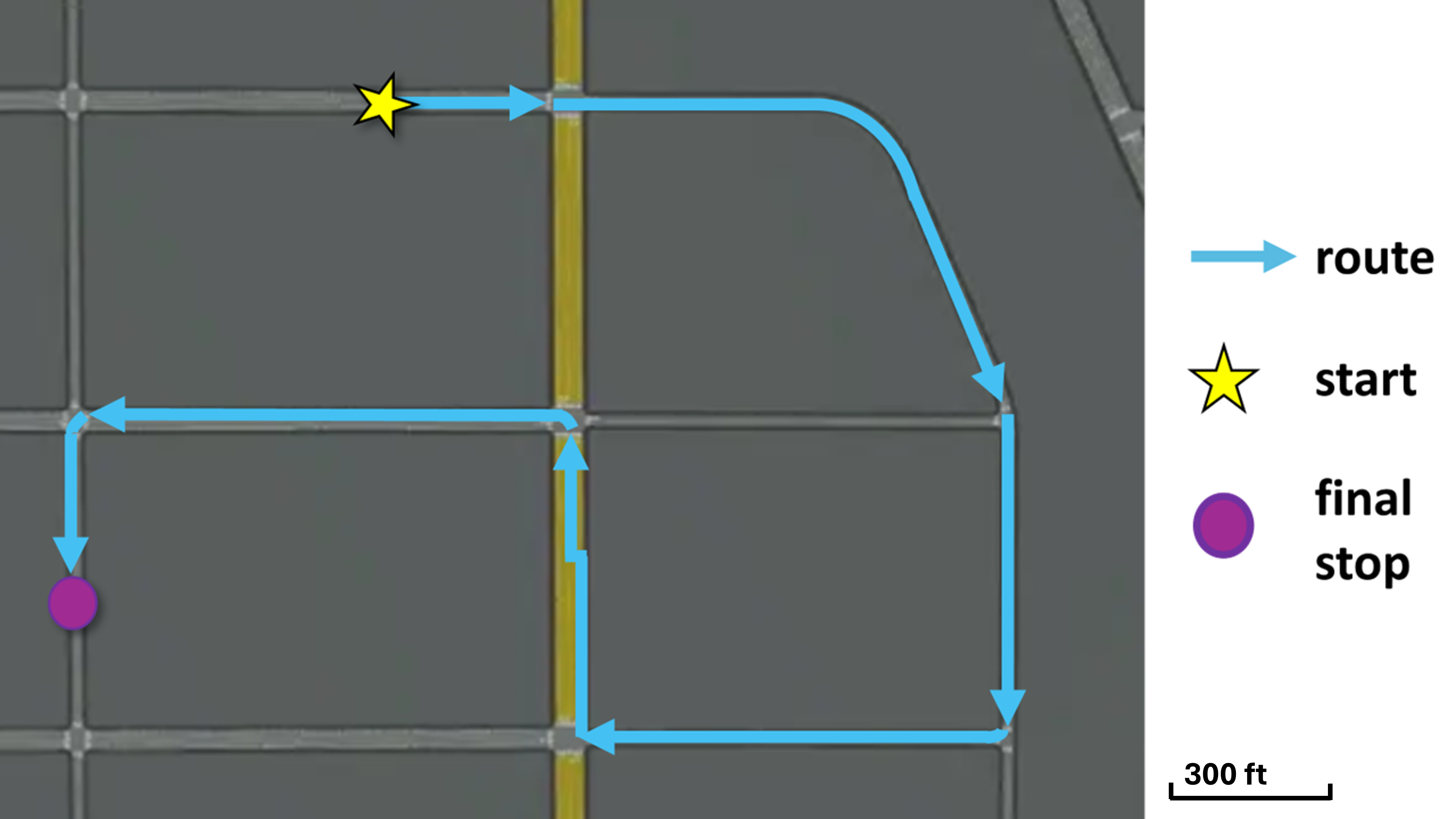}
         \label{fig:route_leaderboard}
     \end{subfigure}
        \caption{\textbf{Display of five tasks and the driving routes.} (\textbf{A}) Simple reaction time task. (\textbf{B}) Simulated braking reaction time task in CARLA 0.9.13. (\textbf{C}) BCI-controlled teledriving of a Mach-E without traffic in the Mcity test facility. (\textbf{D}) Four teledriving routes in Mcity. (\textbf{E}) BCI-controlled teledriving of a Mach-E without traffic on an obstacle course. (\textbf{F}) Four teledriving route options on the loop-shaped obstacle course. (\textbf{G}) Simulated town driving with traffic in the CARLA Leaderboard 2.0. (\textbf{H}) One simulated driving route in Town 12 of CARLA.}
        \label{fig:fivetasks}
\end{figure}

To evaluate the capability of our BCI system to control a vehicle, we conducted a series of five tasks, progressing from in-lab to more complex and practical applications (see Table~\ref{tab:tasksummary}). First, we compared the speed and proficiency of BCI-JJ to a control group of motor intact individuals during two reaction time tasks. Task 1 tested the reaction times in a simple task \cite{gordon_stimulus_1967,fieandt_personal_1956}, in which each participant made an attempted or actual click, via our BCI click decoder (see the ``\nameref{sec:click}'' section) or a computer mouse, in response to a target appearing on a computer screen (see Figure~\ref{fig:simpletask}). Task 2 tested the braking reaction times during simulated semi-autonomous driving in CARLA 0.9.13 \cite{dosovitskiy_carla_2017} in which each participant was asked only to brake (again via a click) the virtual vehicle in response to each random obstacle appearing in front of it driving in a virtual town environment with random settings of sunlight and shadows (see Figure~\ref{fig:brakingtask}). We evaluated the reaction times of the motor intact control group compared to BCI-JJ in both reaction time tasks, finding that BCI-JJ had similar reaction performance measures and significantly faster reaction times than most of the motor intact participants.

As a proof-of-concept, we then tested the ability of our BCI system to remotely control the speed and steering of a Ford Mustang Mach-E vehicle in real time (see Figures~\ref{fig:bcisys} and~\ref{fig:bcimechanism}). Using a single-effector cursor-movement-based BCI decoder (see the ``\nameref{sec:cursor}'' section) and control system, BCI-JJ navigated the Mach-E vehicle in Mcity, as free driving through randomly mapped routes without following any traffic lights or stop signs, in Task 3 (see Figures~\ref{fig:mcity} and~\ref{fig:route_mcity}) and on a predefined obstacle course, as testing many components of a standard driving task, in Task 4 (see Figures~\ref{fig:parkinglot} and~\ref{fig:route_parkinglot}). We set the maximum speed of the Ford Mach-E vehicle to 4 mph. Such low-speed vehicle control could ensure safety of the commercially available vehicle and safety drivers, stable control on a real-world obstacle course with narrow lanes, and smooth display of the video feedback. We demonstrated the first implementation of intracortical BCI control of a full-sized commercially available vehicle in a real-world driving environment. 

Once the basic safety and feasibility of vehicle control was established, we implemented a more complex simulated town driving task modified from the CARLA Leaderboard 2.0, using the bimanual cursor-and-click control, to evaluate our BCI control system in more realistic traffic situations. Task 5 mimicked the downtown driving scenarios in a big city with skyscrapers and heavy traffic (see Figures~\ref{fig:leaderboard} and~\ref{fig:route_leaderboard}). Similar to the low-speed teledriving tasks, we kept the maximum speed of the virtual agent vehicle at 5 mph, the same as the traffic flow speed, for more stable control and smoother display of the video feed. We compared the simulated driving performance of BCI-JJ to that of the motor intact control group, both with bimanual control which included click movement with the left index finger for full-stop braking and cursor movement with the right thumb for speed and steering adjustments. Using simulated driving proficiency scores, calculated based on the occurrence of traffic infractions and the completeness of the route in Task 5, we were able to show that BCI-JJ had similar driving proficiency compared to the motor intact control group, indicating that the BCI control system can achieve essential functions of a vehicle while driving.

\subsection*{BCI reaction times and performance measures}
The main purpose of our study is to apply our BCI system to real-world applications beyond the laboratory as a proof-of-concept. The prerequisite for achieving this aim is to prove, in a controlled laboratory environment, that a participant using our BCI system can react as fast and precisely as individuals with intact motor functions, while controlling potential biases related to age and gender. Therefore, we first designed two reaction time tasks with various target and scenario complexities, using the same concept of responding with a decoded or actual click as soon as a target appeared. The simple reaction time task had a target presented at varying time intervals on an otherwise black computer screen, and the participant was asked to click with either our BCI click decoder or a computer mouse as quickly as possible each time the target appeared in each GO trial (see Figure~\ref{fig:simpletask}). The braking reaction time task had obstacles appeared at random on a predetermined path in the CARLA 0.9.13 driving simulator \cite{dosovitskiy_carla_2017}, and the participant was asked to click for full-stop braking of an otherwise autonomous driving vehicle whenever an obstacle appeared in its path in the GO phase (see Figure~\ref{fig:brakingtask}). The ``\nameref{sec:rt-tests}'' section has a full description of the experimental settings for the two reaction time tasks. 

To set up a comparison study of BCI-JJ with a control group, we recruited 20 motor intact participants (M01\textendash M20) with no gender bias and their ages normally distributed between 35 and 65. The average age of this motor intact control group was 50, the same as the age of BCI-JJ at the time of testing. For both the simple and braking reaction time tasks described above, all 20 motor intact participants in this control group were asked to click on the left button of our computer mouse with their right index finger, in consistent with their daily habit, to react to targets. Meanwhile, BCI-JJ was asked to test six hand effectors to make attempted clicks via our BCI click decoder in the simple reaction time task, with the two best effectors chosen for the braking reaction time task. 

We assessed the reaction times, accuracy, sensitivity, and specificity of participants' responses over each run in both reaction time tasks. A reaction time was considered valid in a GO trial/phase if the participant's response occurred between 50 ms and 1000 ms, after which time the target disappeared from the screen. A true positive case would refer to a GO trial/phase with only one click between 50 ms and 1000 ms. A true negative case would refer to a NO-GO trial/phase with no click. A false positive case would refer to either a NO-GO trial/phase with any click, a GO trial/phase with any click before 50 ms, or a GO trial/phase with any click after 1000 ms from the start of the Target Phase. A false negative case would refer to a GO trial/phase with no click. The accuracy measures the total number of correct cases (true positives + true negatives) divided by the total number of cases (true positives + true negatives + false positives + false negatives). The sensitivity measures the number of true positives divided by the total number of true positives and false negatives. The specificity measures the number of true negatives divided by the total number of true negatives and false positives. 

\begin{figure}
     \centering
     \begin{subfigure}[b]{0.48\textwidth}
         \centering
         \caption{}
         \includegraphics[width=\textwidth]{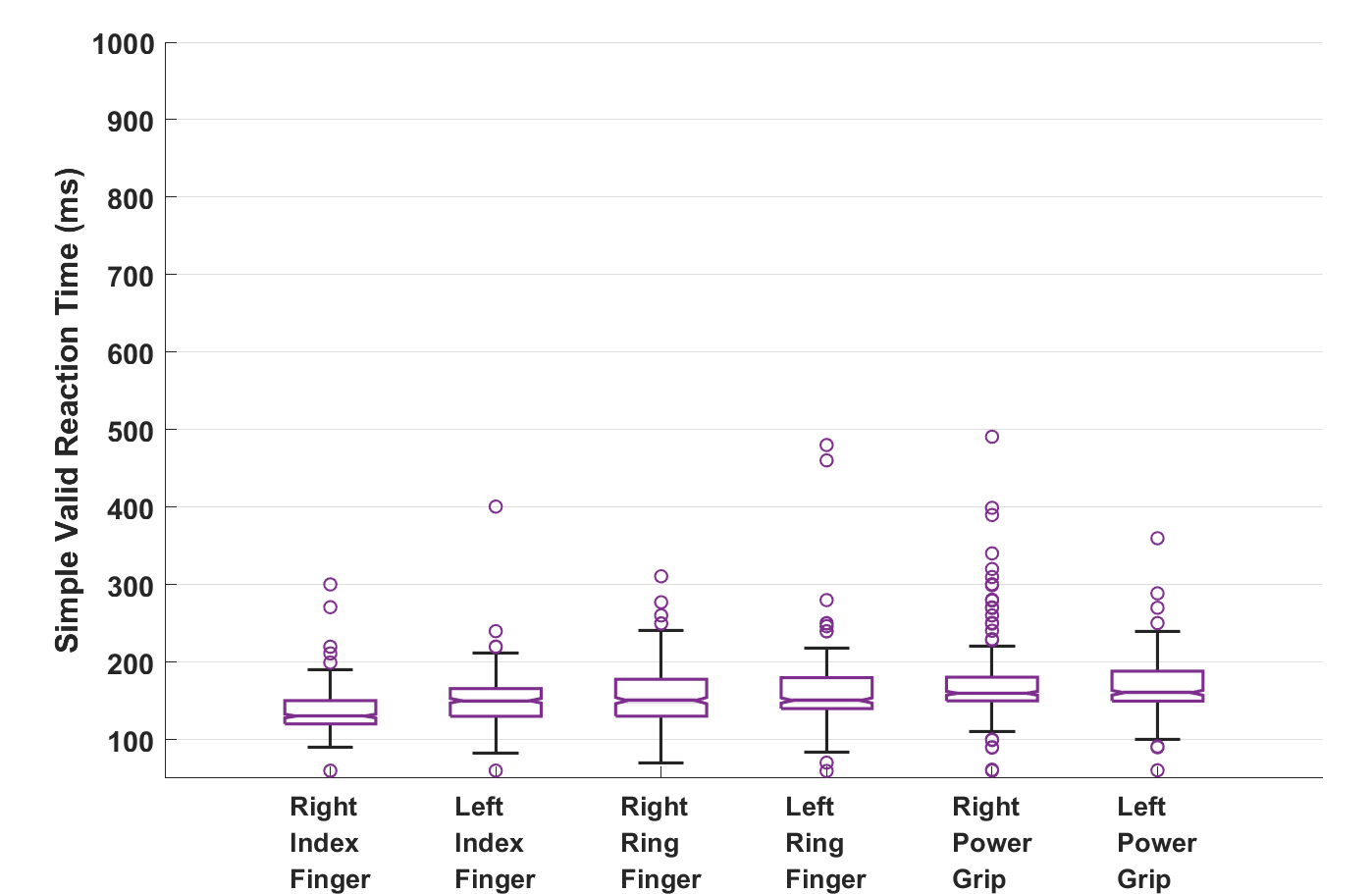}
         \label{fig:JJsimpleValidRT}
     \end{subfigure}
     \hfill
     \begin{subfigure}[b]{0.48\textwidth}
         \centering
         \caption{}
         \includegraphics[width=\textwidth]{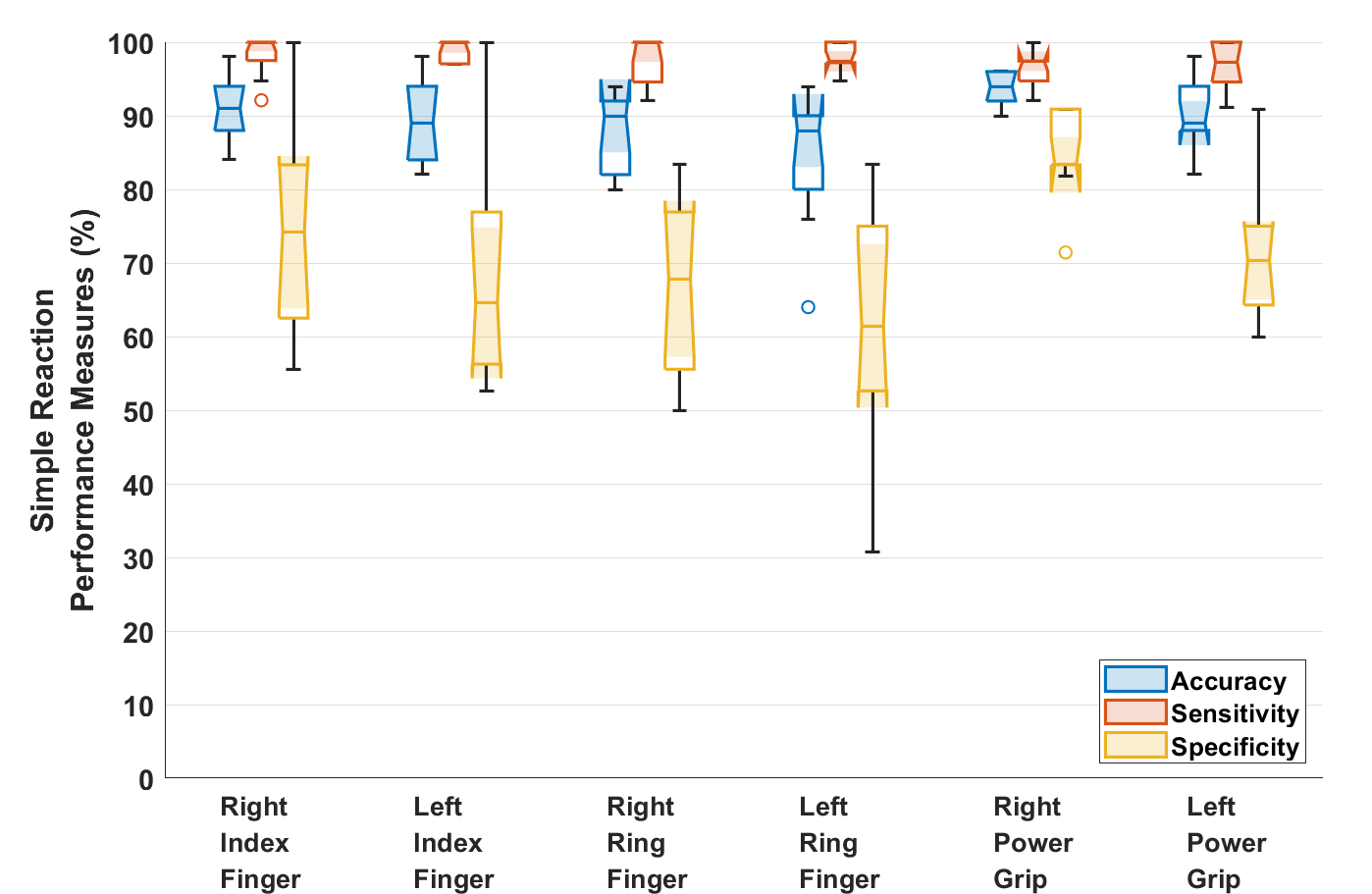}
         \label{fig:JJsimplecost}
     \end{subfigure}
    
    \begin{subfigure}[b]{0.48\textwidth}
         \centering
         \caption{}
         \includegraphics[width=\textwidth] {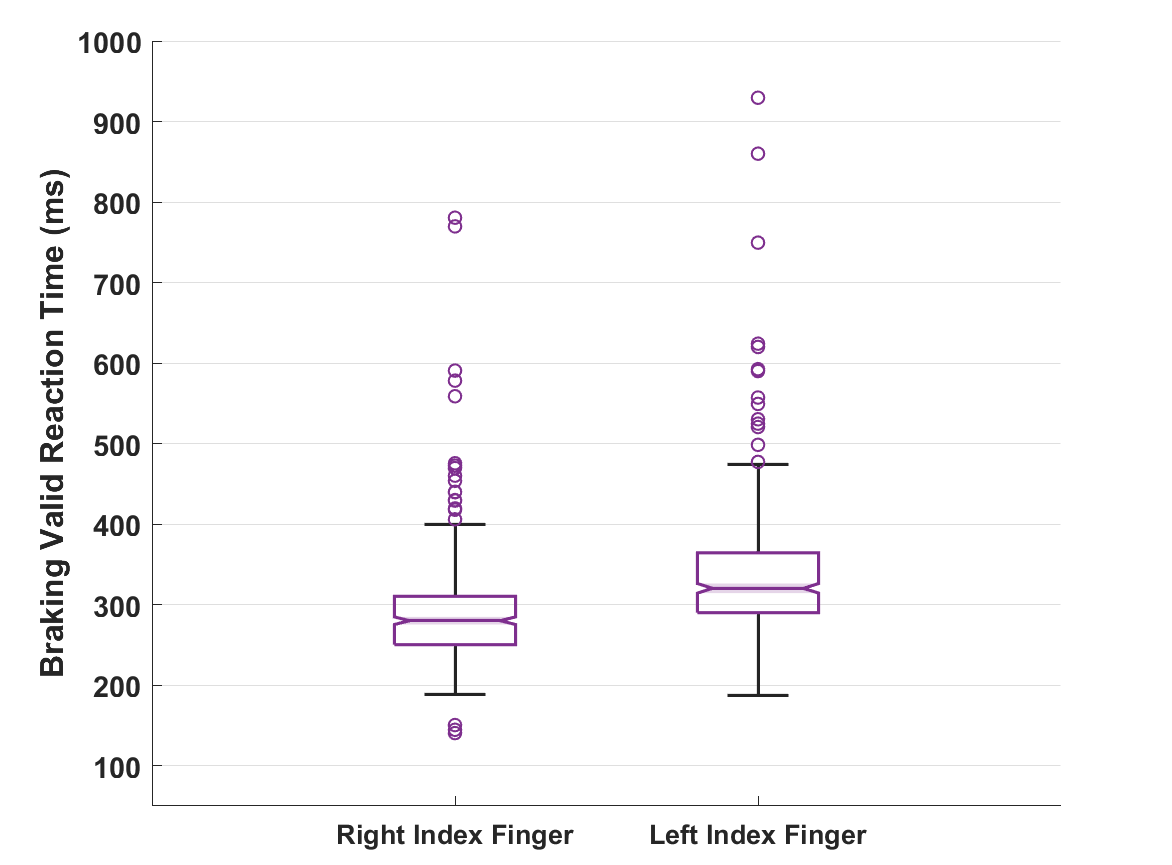}
         \label{fig:JJbrakingValidRT}
     \end{subfigure}
     \hfill
     \begin{subfigure}[b]{0.48\textwidth}
         \centering
         \caption{}
         \includegraphics[width=\textwidth]{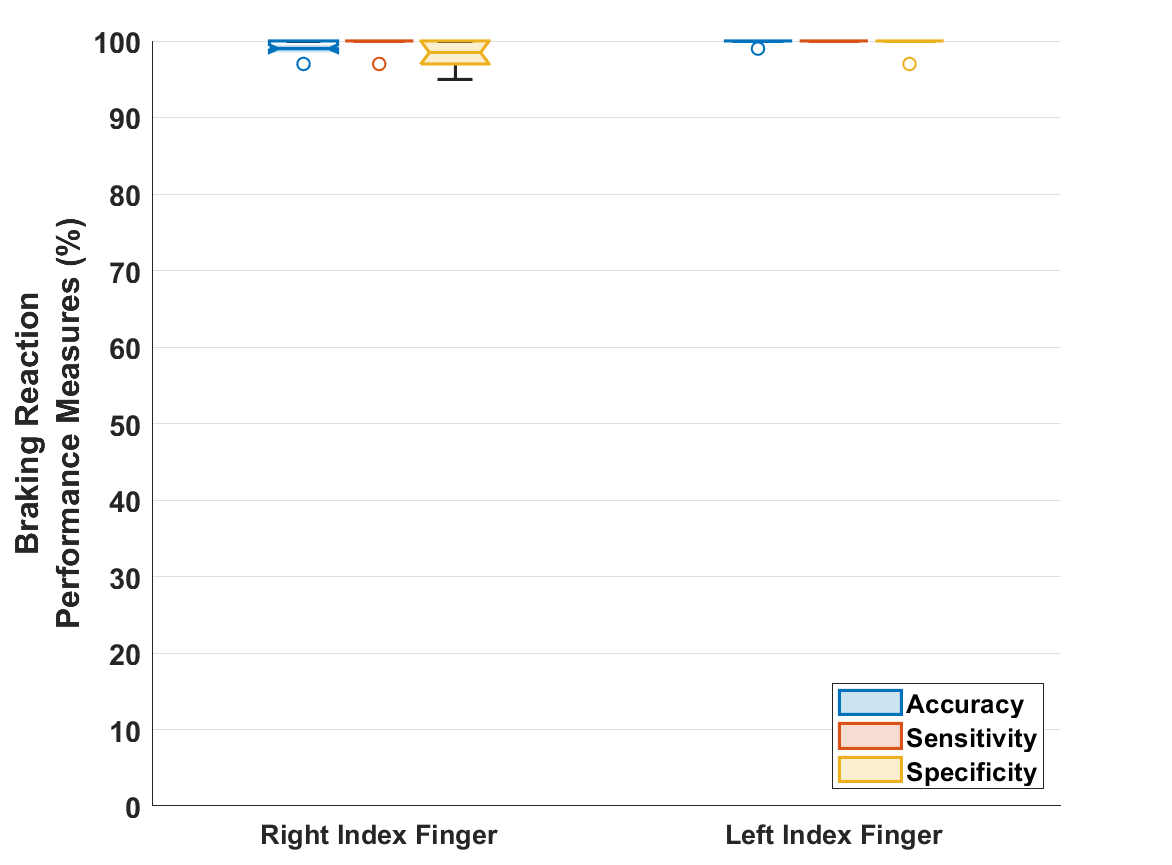}
         \label{fig:JJbrakingcost}
     \end{subfigure}
    \captionsetup{aboveskip=0pt}
     \caption{\textbf{BCI trial-based simple and braking reaction time task results among different hand effectors of BCI-JJ.} (\textbf{A} and \textbf{B}) For the simple reaction time task, we collected 10 runs for each of BCI-JJ's six effectors (i.e., right and left index fingers, ring fingers, and power grips) consisting of 40 GO trials and 10 randomly interleaved NO-GO catch trials per run. We compared simple reaction times within the valid range from 50 ms to 1000 ms in GO trials/phases among the six effectors. We also conducted comparisons of simple reaction performance measures (i.e., accuracy, sensitivity, specificity) among these effectors. (\textbf{C} and \textbf{D}) For the braking reaction time task, we collected 10 runs per right or left index finger consisting of 40 trials per run, with one NO-GO phase and one GO phase in each trial. We compared valid braking reaction times and reaction performance measures between these two effectors. Box charts whose shaded notches around the median lines do not overlap have different medians at the corrected $5\%$ significance level. }
        \label{fig:JJreactResults}
\end{figure}

\subsubsection*{\itshape BCI simple reaction performance among different hand effectors}
To assess the basic reaction times of our BCI control system, we conducted a simple reaction time task in which BCI-JJ attempted to click with one of six different effectors: left and right index fingers, left and right ring fingers, and left and right power grips to complete this action. Decoded click movement intentions for each effector were digitally executed as mouse clicks to complete the task. BCI-JJ was asked to complete 10 runs consisting of 40 GO trials and 10 randomly interleaved NO-GO catch trials (indicated by a low-pitch auditory stimulus during the ITI phase) per run. BCI-JJ repeated these 10 runs for each effector separately.  

We compared the valid simple reaction times, accuracy, sensitivity, and specificity across all six effectors for BCI-JJ using a one-way ANOVA test followed by pairwise comparisons with all 10 runs for each effector combined. The reaction times for BCI-JJ's right (contralateral to the recording hemisphere) index finger performed significantly faster than all other effectors ($p<0.05$, Bonferroni corrected) with an average reaction time of 137 ms (26.8 ms SD). Using his right index finger led to not significantly different performance measures compared to using the rest effectors, with an average accuracy at 90.8$\%$ (4.13$\%$ SD), an average sensitivity at 98.4$\%$ (2.82$\%$ SD), and an average specificity at 73.3$\%$ (13.8$\%$ SD) (see Figures~\ref{fig:JJsimpleValidRT} and~\ref{fig:JJsimplecost}, tables~\ref{Stab:JJsimple} and~\ref{Stab:JJsimpleANOVA}).

\subsubsection*{\itshape BCI braking reaction performance with index finger effectors}
To evaluate the reaction times of our BCI control system in a more realistic driving scenario, we conducted a braking reaction time task in a virtual town environment using the CARLA 0.9.13 driving simulator \cite{dosovitskiy_carla_2017}. BCI-JJ completed 10 runs per right or left index finger, consisting of 40 trials per run, with one NO-GO phase and one GO phase in each trial. The effectors we evaluated were limited to the index fingers for this task, because these two effectors provided him with the best performance among all the six hand effectors tested during the previous simple reaction time task. 

We compared the valid braking reaction times, accuracy, sensitivity, and specificity between the left and right index finger effectors via two-sample t-tests for BCI-JJ (see tables~\ref{Stab:JJbraking} and~\ref{Stab:JJbrakingTTests}). Like the simple reaction time task, BCI-JJ had significantly faster reaction times using the right index finger, compared to the left, showing 290 ms (64.9 ms SD) versus 338 ms (82.2 ms SD) with $p < 0.05$ (see Figure~\ref{fig:JJbrakingValidRT}). BCI-JJ showed greater than $90\%$ accuracy, sensitivity, and specificity across both effectors throughout all trials for this simulated braking task (see Figure~\ref{fig:JJbrakingcost}). Comparing each performance measure (i.e., accuracy, sensitivity, and specificity) of BCI-JJ's right and left index fingers, their differences are either at the edge of the corrected $5\%$ significance level or clearly not significant (see table~\ref{Stab:JJbrakingTTests}).

\subsubsection*{\itshape BCI versus motor intact reaction performance}

Each of the 20 motor intact control participants completed 4 runs of the simple reaction time task and 5 runs of the braking reaction time task, both with the same number of trials per run and layout as the same tasks for BCI-JJ. 

\begin{figure}
     \centering
     \begin{subfigure}[b]{0.48\textwidth}
         \centering
         \caption{}
         \includegraphics[width=0.9\textwidth]{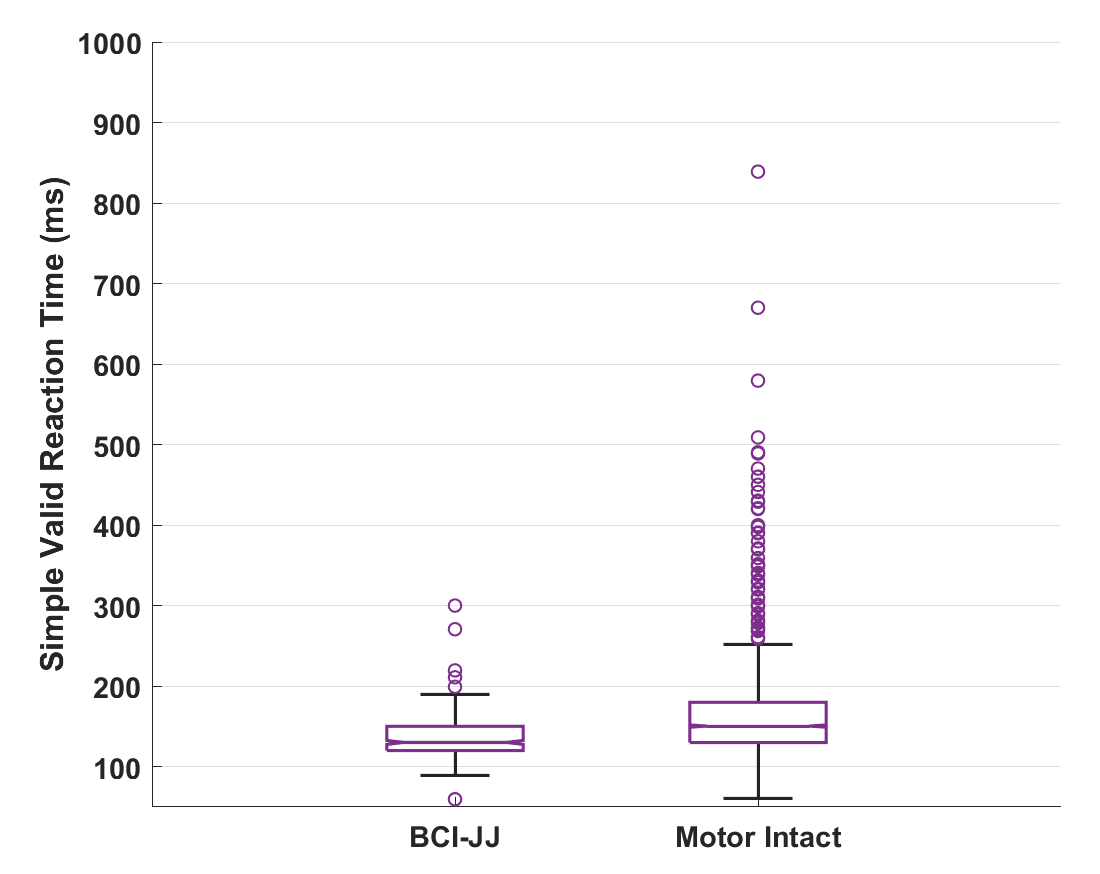}
         \label{fig:AGsimpleValidRT}
     \end{subfigure}
     \hfill
     \begin{subfigure}[b]{0.48\textwidth}
         \centering
         \caption{}
         \includegraphics[width=0.9\textwidth]{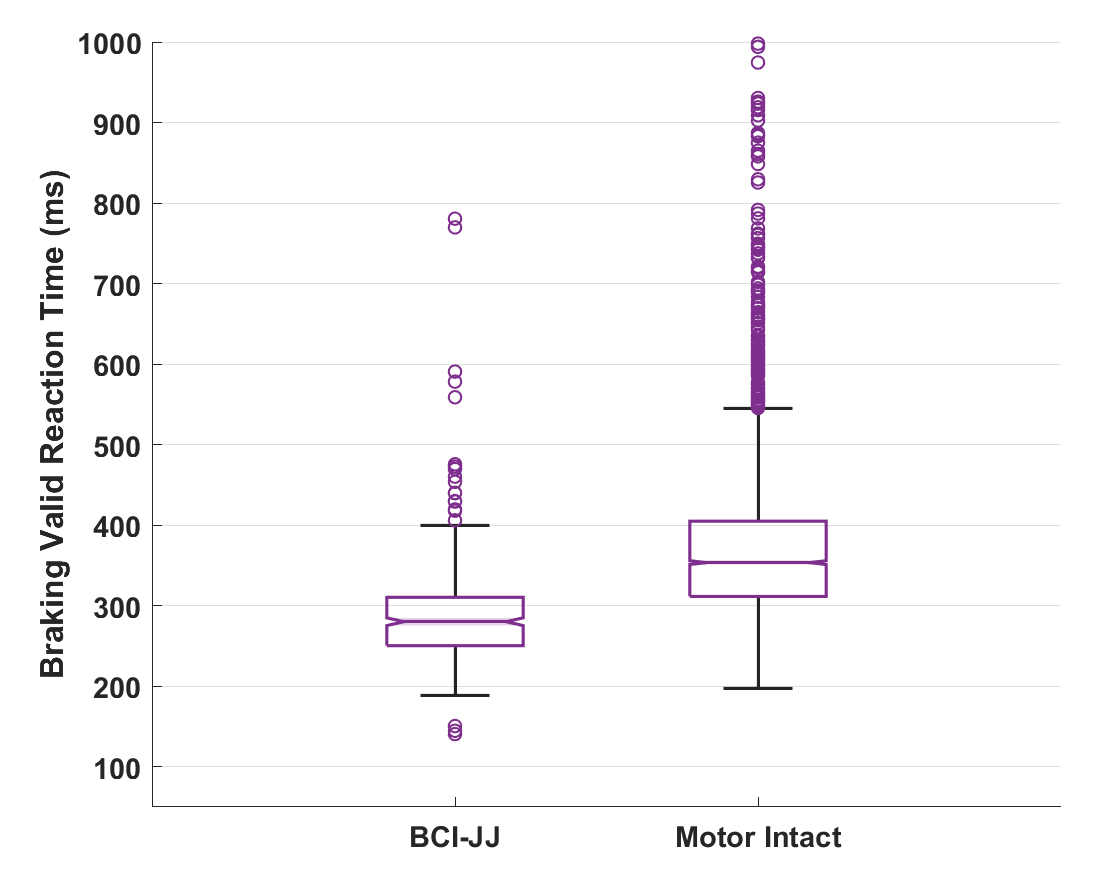}
         \label{fig:AGbrakingValidRT}
     \end{subfigure}
     
     \begin{subfigure}[b]{0.48\textwidth}
         \centering
         \caption{}
         \includegraphics[width=0.9\textwidth]{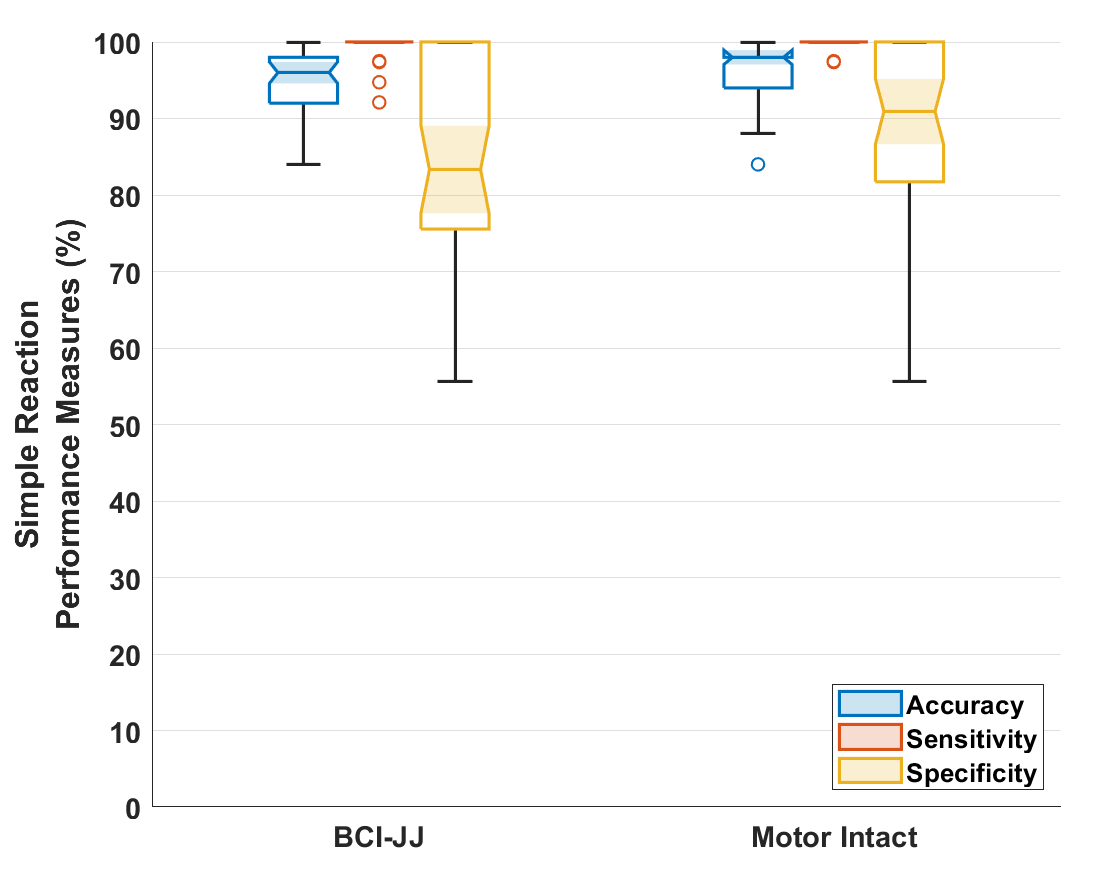}
         \label{fig:AGsimplecost}
     \end{subfigure}
     \hfill
     \begin{subfigure}[b]{0.48\textwidth}
         \centering
         \caption{}
         \includegraphics[width=0.9\textwidth]{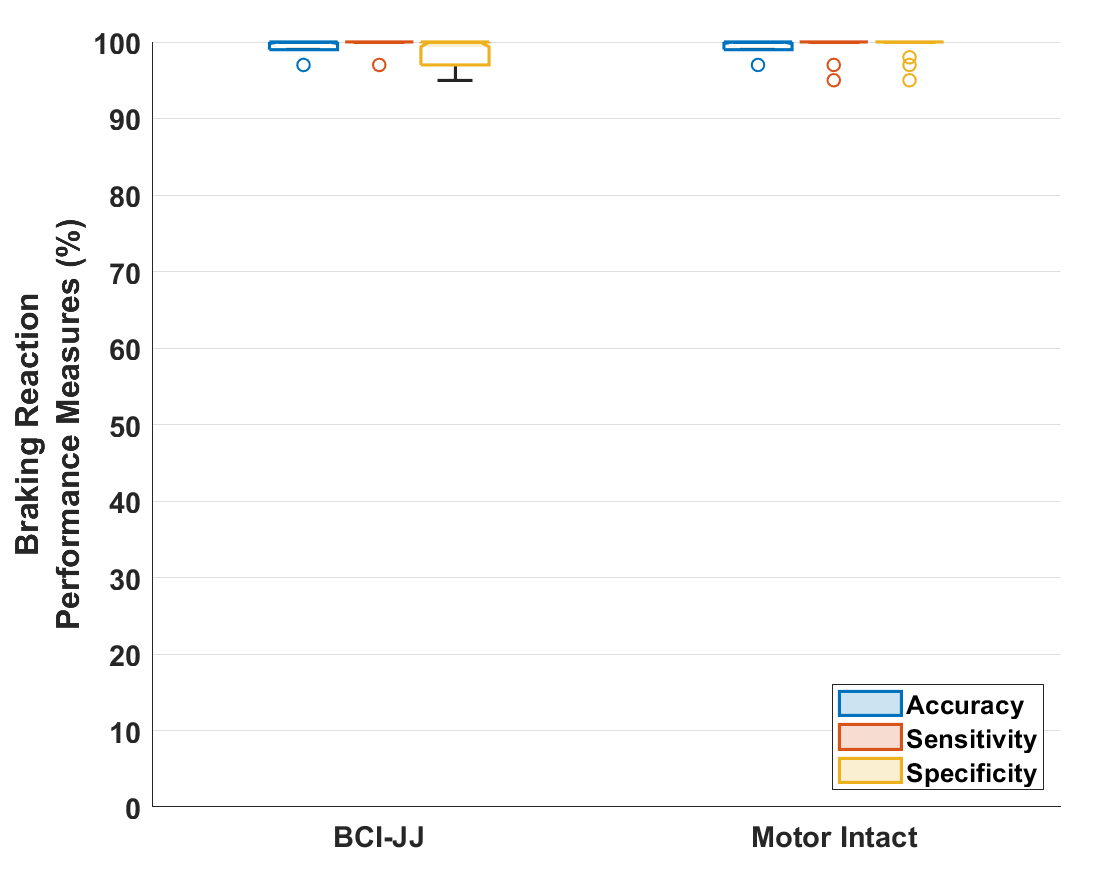}
         \label{fig:AGbrakingcost}
     \end{subfigure}
     
        \caption{\textbf{Trial-based simple and braking reaction time task results between BCI-JJ and the motor intact control group with the same right index finger effector.} We recruited 20 motor intact participants with no gender bias and an average age of $50 \pm 15$, the same as the age of BCI-JJ at the time of testing. For the simple reaction time task, we collected 10 runs from BCI-JJ and 4 runs from each of the 20 motor intact participants, with each run consisting of 40 GO trials and 10 randomly interleaved NO-GO catch trials. For the braking reaction time task, we collected 10 runs from BCI-JJ and 5 runs from each motor intact participant, consisting of 40 trials per run, with one NO-GO phase and one GO phase in each trial. We compared the valid reaction times between 50 ms and 1000 ms in GO trials/phases between BCI-JJ and the motor intact control group for (\textbf{A}) the simple reaction time task and (\textbf{B}) the braking reaction time task. We also conducted group comparisons of reaction accuracy, sensitivity and specificity for (\textbf{C}) the simple reaction time task and (\textbf{D}) the braking reaction time task. Box charts whose shaded notches around the median lines do not overlap have different medians at the corrected $5\%$ significance level. }
        \label{fig:AGreactResults}
\end{figure}

Between BCI-JJ and the motor intact control group, we applied two-sample t-tests to compare the valid reaction times, accuracy, sensitivity, and specificity using their right index finger effector (see Figure~\ref{fig:AGreactResults} and tables~\ref{Stab:GroupSimple},~\ref{Stab:GroupSimpleTTest2},~\ref{Stab:AGSimpleANOVA},~\ref{Stab:GroupBraking},~\ref{Stab:GroupBrakingTTest2}, and~\ref{Stab:AGBrakingANOVA}). For both tasks, BCI-JJ had significantly faster valid reaction times when compared to the motor intact control group ($p<0.05$). However, both BCI-JJ and the control group had slower reaction times on average in the more complex braking reaction time task compared to the simple reaction time task (see Figures~\ref{fig:AGsimpleValidRT} and~\ref{fig:AGbrakingValidRT}). The motor intact control group had significantly higher accuracy and specificity ($p<0.05$) when compared to BCI-JJ during the simple reaction time task (see Figure~\ref{fig:AGsimplecost}). There were no significant differences in accuracy, sensitivity, or specificity performance between the two groups during the braking reaction time task (see Figure~\ref{fig:AGbrakingcost}). This indicates that the reaction performance of the BCI control system in BCI-JJ was, on average, faster than and as reliable as the motor intact control group of comparable ages, no matter of the complexities of the two reaction time tasks. 

Among BCI-JJ and all 20 motor intact participants (M01\textendash M20) individually, we also compared the reaction performance using a one-way ANOVA test followed by pairwise comparisons (see figure~\ref{Sfig:21reactValidRT} for the valid reaction times and figures~\ref{Sfig:21reactAcc},~\ref{Sfig:21reactSens},~\ref{Sfig:21reactSpec} for the reaction accuracy, sensitivity, and specificity). During the simple reaction time task, BCI-JJ had significantly faster valid reaction times than 12 out of 20 motor intact participants, and significantly lower accuracy and specificity compared to 7 motor intact participants ($p<0.05$, Bonferroni corrected), which were not necessarily the same participants between metrics. There were no significant differences in sensitivity between BCI-JJ and any motor intact participants (see tables~\ref{Stab:GroupSimple} and~\ref{Stab:AGSimpleANOVA}). During the braking reaction time task, BCI-JJ had significantly faster valid reaction times than 18 out of 20 motor intact individuals, and significantly higher sensitivity compared to 1 motor intact participant ($p<0.05$, Bonferroni corrected). There were no significant differences in accuracy and specificity between BCI-JJ and any motor intact participants (see tables~\ref{Stab:GroupBraking} and~\ref{Stab:AGBrakingANOVA}). This indicates that BCI-JJ, via our BCI click control, reacted at least as fast and precisely as most of the motor intact control participants, no matter of the complexities of the two reaction time tasks.

\subsection*{BCI driving performance}
Via the two reaction time tasks, we established that the reaction performance of BCI-JJ with index finger clicks decoded by our BCI system was at least as fast and reliable as the average performance of the motor intact participants clicking a computer mouse. In order to verify the robustness of our BCI system in dealing with more realistic applications, such as driving a vehicle, we outfitted a Ford Mustang Mach-E vehicle to be remotely controllable in real-world environments without traffic. Via teledriving, we first tested the speed and steering control with cursor movement decoded by our BCI system in a free-driving scenario without following any traffic lights or stop signs inside an urban test facility called Mcity (see Figure~\ref{fig:mcity}). Mcity is a closed test facility in Michigan made to represent an urban driving environment without traffic, whereas the obstacle course included components of a typical driving test for a motor intact person (e.g., full stops, lane switches, turnings, roundabouts, and obstacle avoidance). As shown in Figure~\ref{fig:bcimechanism}, the overlay module obtained the decoded x and y cursor movement values from the effector controller module and displayed them as positions of the blue and red circles, which would be post-processed for steering and speed control of the vehicle, respectively (see the ``\nameref{sec:driving-tests}'' section for a full description). Then we designed an obstacle course and applied teledriving with the same Mach-E vehicle which could verify the same BCI driving system in a more applicable real-world setting mimicking many components of a standard driving test (see Figure~\ref{fig:parkinglot}). Finally, we increased the complexity of the BCI decoder from the single-effector cursor control for steering and speed to the bimanual cursor-and-click control, which enabled more precise full-stop braking via clicks with the left index finger effector and continuous steering and speed adjustments via the horizontal and vertical cursor movement, respectively, with the right thumb. To ensure the safety, precision, and complexity of our driving evaluations with bimanual control, we temporarily switched back to a laboratory setting and compared the driving proficiency of BCI-JJ to the motor intact control group using a virtual vehicle in a simulated town environment with heavy traffic and a more sophisticated route modified from the CARLA Leaderboard 2.0 (see Figure~\ref{fig:leaderboard}). The ``\nameref{sec:driving-tests}'' section has full descriptions of the experimental settings for the three driving tasks. 

To ensure safe driving in accordance to many components of a standard driving test, we set up safety criteria (see the ``\nameref{sec:safety-criteria}'' section) and evaluated the following infractions, including the proportion of the completed route distance ($C$), the number of collisions ($N_c$), the number of lane deviations ($N_l$), and the number of running red traffic lights or stop signs if available ($N_s$). The infraction metric aggregated all these infractions using the following equation: 
\begin{equation}
\text{Driving Score} = C \times 0.8^{N_c} \times 0.9^{N_l} \times 0.9^{N_s}.
\label{eq:driving}
\end{equation}
It started with an ideal 1.0 base score, which was reduced each time an infraction was committed. Here, $C$ could be a decimal between 0 and 1.0 with 1.0 as full completion of the route. $N_c$, $N_l$ and $N_s$ were all nonnegative numbers whose increase would cause a decrease in the driving score. 

For both teledriving tasks in Mcity and through the obstacle course, BCI teledriving infractions were evaluated based on recorded videos by three independent human evaluators from the California Institute of Technology, Blackrock Neurotech, and Ford Motor Company. Please note that if the vehicle was about to collide with some dangerous obstacle, $N_c$ would increase by 1, and the safety driver in the test vehicle would take over the vehicle control for a few seconds to move it back to the right track before the BCI teledriving could resume. If the vehicle was about to turn left at an intersection where a left-turn lane was available, it should merge to that left-turn lane before its left turn during BCI teledriving; otherwise, $N_l$ would increase by 1. If an evaluator was uncertain about the occurrence of any infraction component ($N_c$, $N_l$ or $N_s$) (e.g., whether there was a lane deviation at $t=45$ seconds in Run 2), he/she would mark it with a quantity increase of 0.5 instead of 1 (e.g., $N_l \mathrel{+}= 0.5$). 

For the simulated town driving task, the infractions of BCI-JJ and the motor intact participants were automatically measured by the CARLA Leaderboard 2.0 at the end of each run to compute the driving score for the simulated route. So in this case, $N_c$, $N_l$ and $N_s$ were all nonnegative integers without the possibility of increasing by 0.5.

\begin{figure}
     \centering
     
     \begin{subfigure}[b]{0.49\textwidth}
         \centering
         \caption{}
         \includegraphics[width=0.72\textwidth]{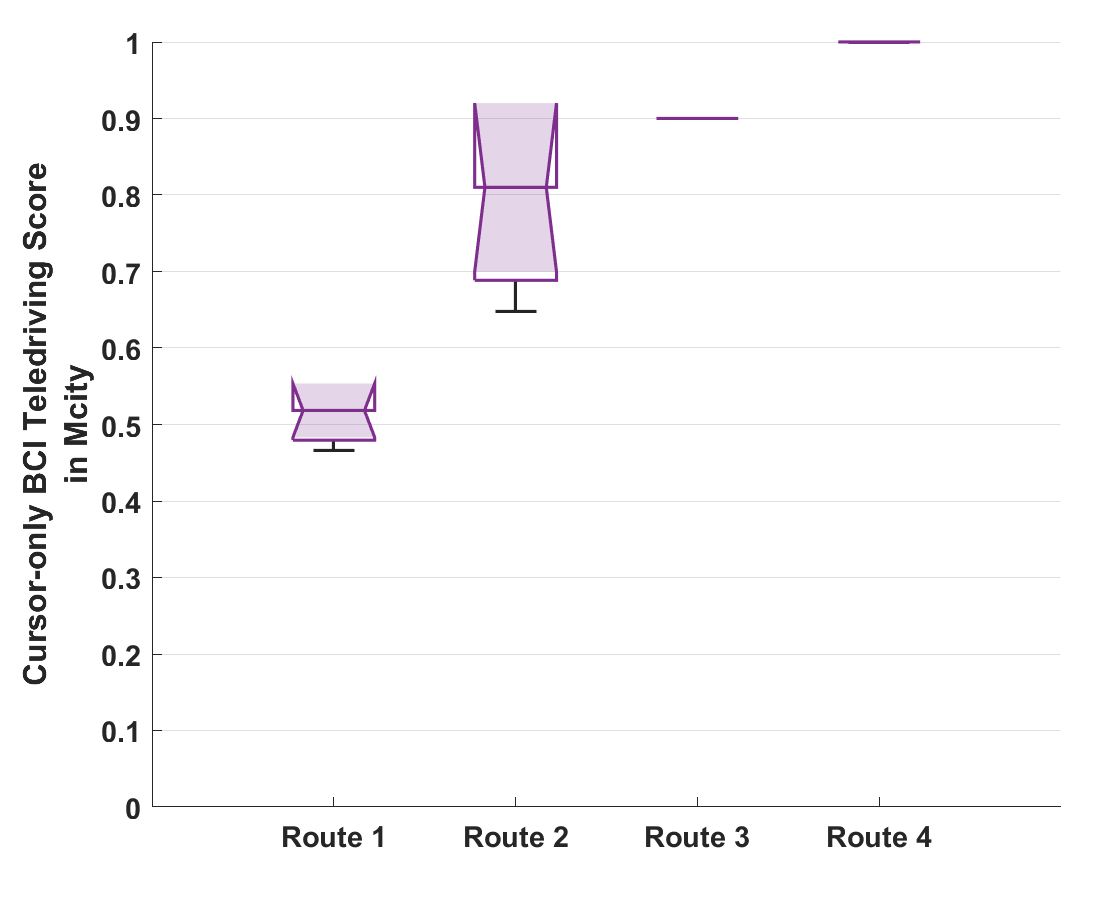}
         \label{fig:JJmcity_score}
     \end{subfigure}
     \hfill
     \begin{subfigure}[b]{0.49\textwidth}
         \centering
         \caption{}
         \includegraphics[width=0.72\textwidth]{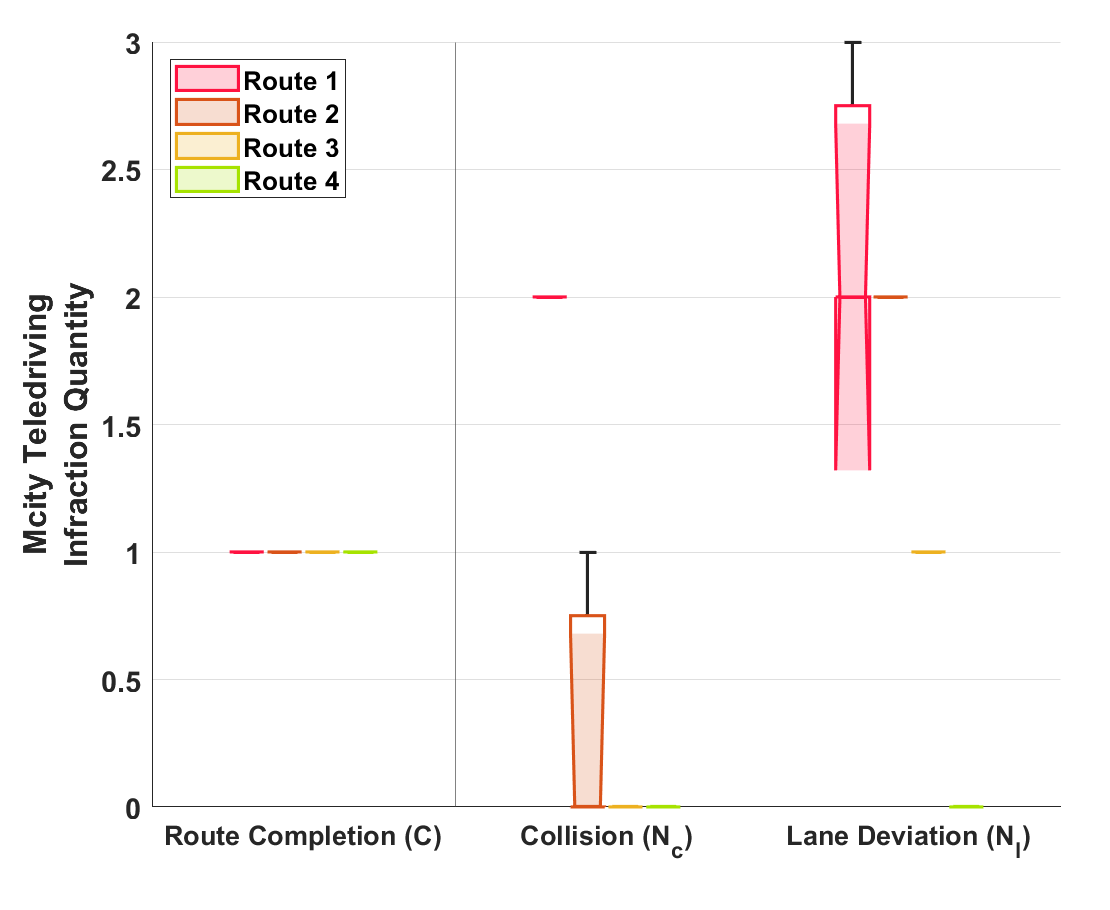}
         \label{fig:JJmcity_infrac}
     \end{subfigure}

     \begin{subfigure}[b]{0.49\textwidth}
         \centering
         \caption{}
         \includegraphics[width=0.72\textwidth]{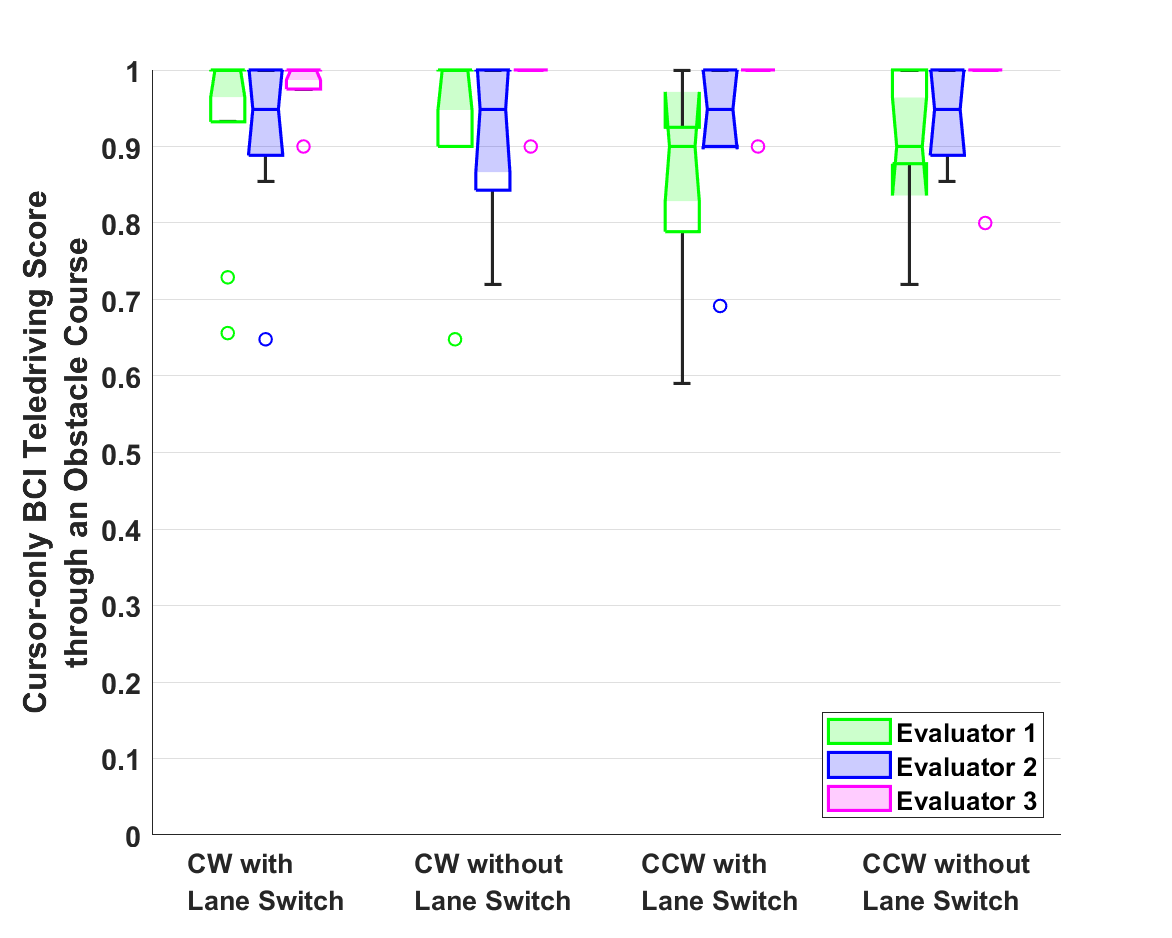}
         \label{fig:JJparkinglot_score}
     \end{subfigure}
     \hfill
     \begin{subfigure}[b]{0.49\textwidth}
         \centering
         \caption{}
         \includegraphics[width=0.72\textwidth]{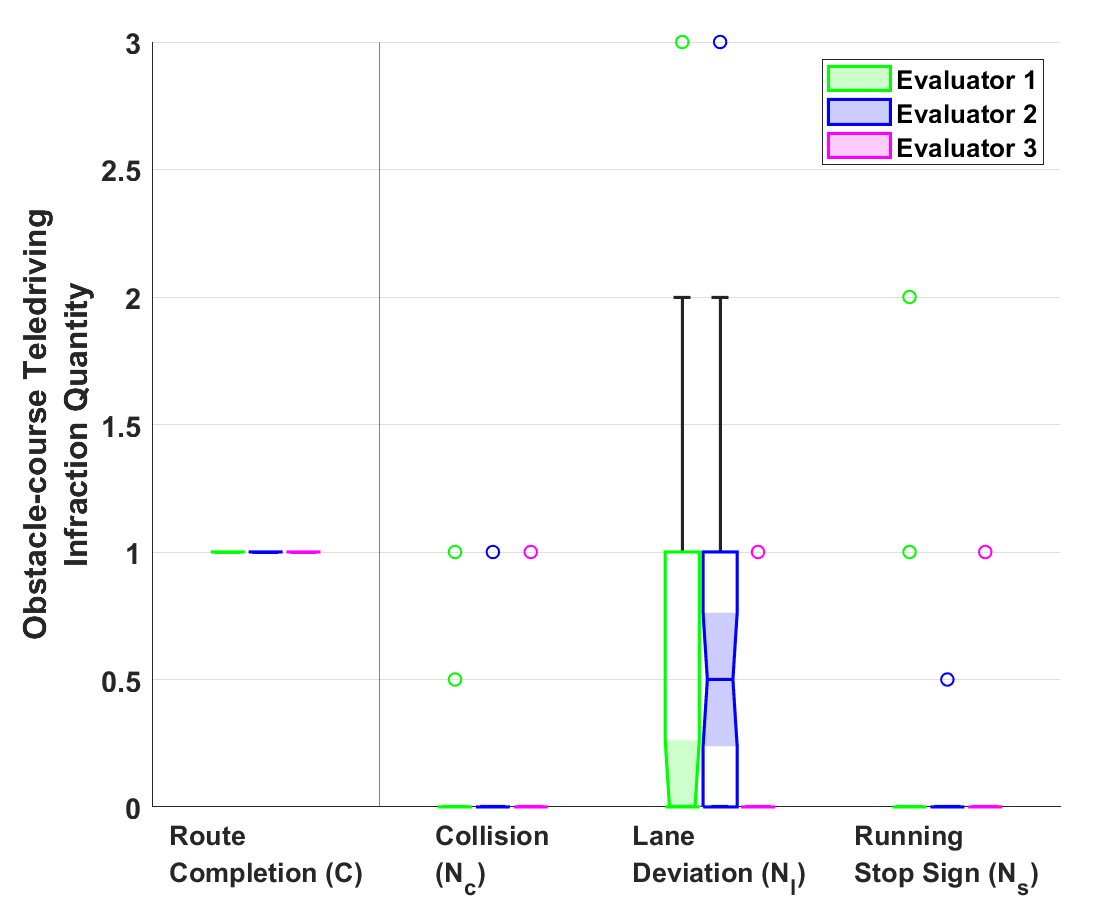}
         \label{fig:JJparkinglot_infrac}
     \end{subfigure}
    
    \begin{subfigure}[b]{0.49\textwidth}
         \centering
         \caption{}
         \includegraphics[width=0.72\textwidth]{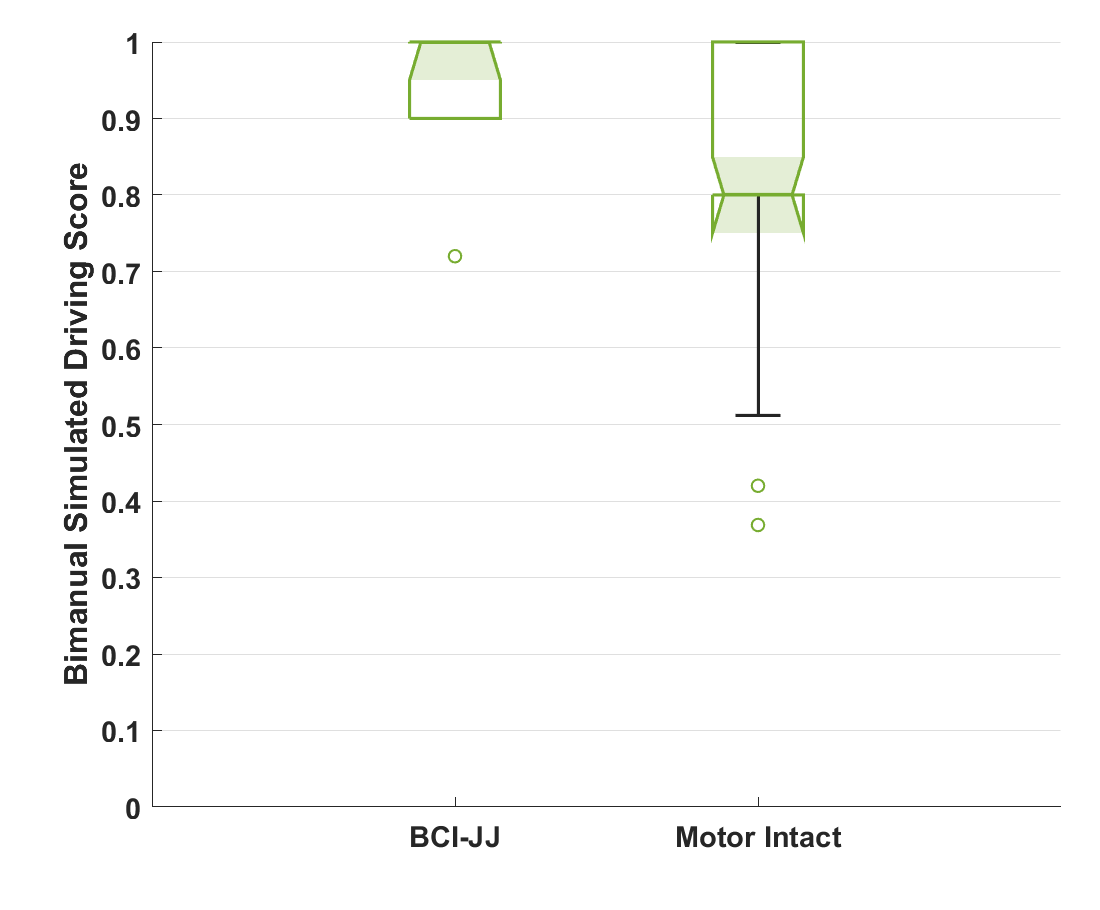}
         \label{fig:AGleaderboard_score}
     \end{subfigure}
     \hfill
     \begin{subfigure}[b]{0.49\textwidth}
         \centering
         \caption{}
         \includegraphics[width=0.72\textwidth]{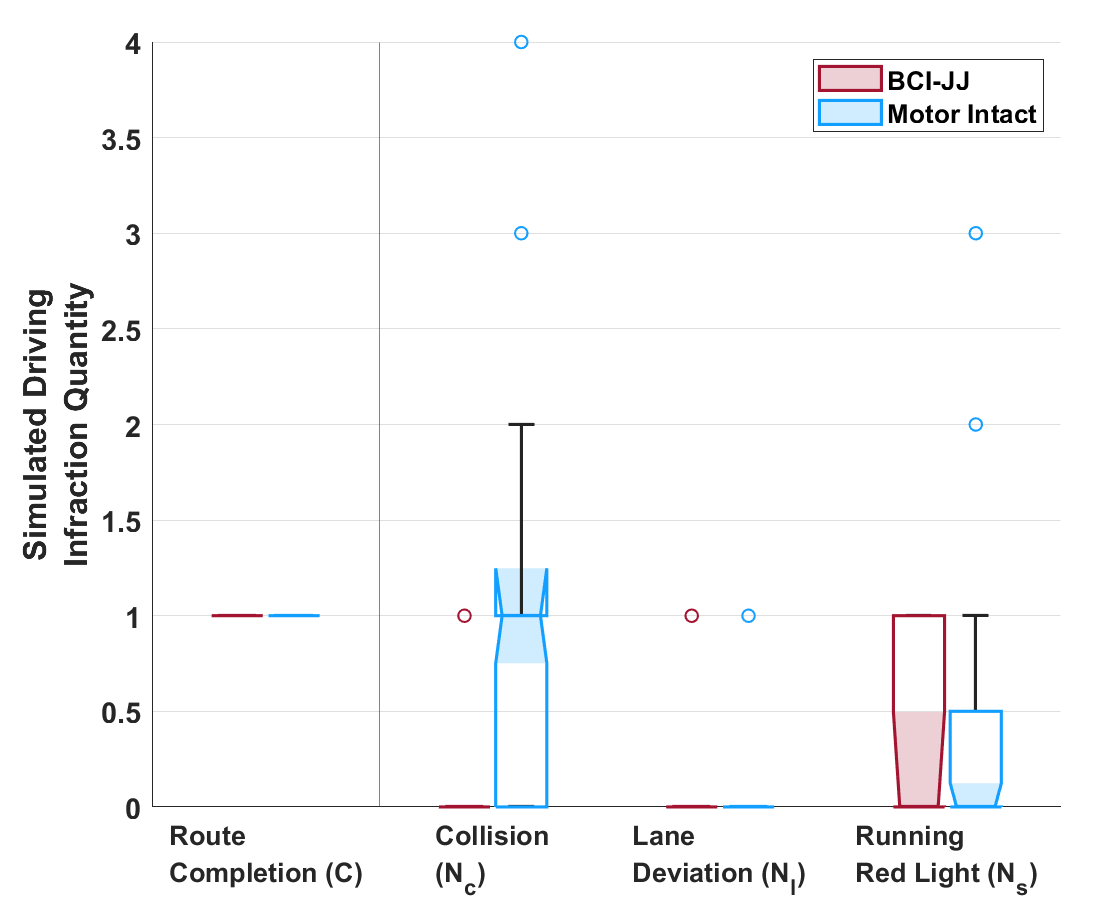}
         \label{fig:AGleaderboard_infrac}
     \end{subfigure}
     
    \caption{\textbf{BCI driving scores and infractions.} It started with an ideal 1.0 base score in each run, which was reduced each time an infraction was committed. (\textbf{A} and \textbf{B}) The teledriving scores and infractions of BCI-JJ using the single-effector cursor control for steering and speed of a Ford Mustang Mach-E vehicle through four random routes in the Mcity urban test facility without traffic. (\textbf{C} and \textbf{D}) The teledriving scores and infractions of BCI-JJ using the same cursor-movement control through an obstacle course. CW: clockwise. CCW: counterclockwise. (\textbf{E} and \textbf{F}) The simulated driving scores and infractions of BCI-JJ and the motor intact control group with the bimanual cursor-and-click control for steering, speed, and braking of a virtual vehicle in a busy town. }
    \label{fig:drivingResults}
\end{figure}

\subsubsection*{\itshape Teledriving performance in Mcity}
As a proof-of-concept, we started the evaluations of our BCI-enabled driving system with the single-effector cursor control for steering and speed of a commercially available vehicle through a real-world urban environment without traffic. BCI-JJ, who was living in California, remotely controlled a Ford Mustang Mach-E vehicle through Mcity, a mock urban road environment located in Michigan. The goal of this Mcity teledriving task was to evaluate the continuousness and proficiency of BCI driving control in speed and steering, rather than to repeatedly evaluate reaction times or braking as we had done in the previous two reaction time tasks. Therefore, BCI-JJ was asked to ignore all traffic lights and stop signs in this urban test facility, focusing only on the cursor control for steering and speed with his right thumb. BCI-JJ completed four randomly mapped routes through Mcity, with one run per route across two sessions (see movies~\ref{Smovie:mcity-route1},~\ref{Smovie:mcity-route2},~\ref{Smovie:mcity-route3} and~\ref{Smovie:mcity-route4}). Each route included various driving elements (e.g., turnings at intersections, roundabouts, U-turns, lane merges, etc., see Figure~\ref{fig:route_mcity} and the ``\nameref{sec:driving-tests}'' section for a full description). 

With the BCI teledriving infractions in Mcity evaluated by the three independent human evaluators, we applied the infraction metric to compute the cursor-only BCI teledriving score in Mcity using Eq.~\ref{eq:driving} in which $N_s$ was set to zero, because in this task no traffic light or stop sign was considered. Using BCI-decoded cursor movement for speed and steering control of the Mach-E vehicle, BCI-JJ completed each of the four random routes through Mcity in 2.5 to 7.5 minutes and obtained an average driving score of 0.780 (0.211 SD) across the four routes (see Figure~\ref{fig:JJmcity_score}), with his driving performance improving over the course of each run (see Figure~\ref{fig:JJmcity_infrac}).

\subsubsection*{\itshape Teledriving performance on an obstacle course}

After establishing the free-driving capability of our single-effector cursor-movement-based BCI control system for a commercially available vehicle in the Mcity closed test facility, we then demonstrated the capability of the same BCI control system to navigate the same vehicle on a more complex, predefined obstacle course. To do this, BCI-JJ remotely controlled the same Mach-E vehicle through a loop-shaped driving obstacle course which included many components of a standard driving test (e.g., lane switches, turnings, full stops, roundabouts, obstacle avoidance, etc.) (see Figure~\ref{fig:route_parkinglot} and the ``\nameref{sec:driving-tests}'' section for a full description). In this obstacle-course teledriving task, BCI-JJ completed 9 runs per route for a total of four different route options (i.e., a clockwise or counterclockwise loop with or without a lane-switch segment) across three sessions (see movies~\ref{Smovie:obstacle-course-cw-ls},~\ref{Smovie:obstacle-course-cw-nols},~\ref{Smovie:obstacle-course-ccw-ls} and~\ref{Smovie:obstacle-course-ccw-nols} for one example run per route). 

With the BCI teledriving infractions through the obstacle course evaluated by the three independent evaluators, we applied the infraction metric to compute the BCI teledriving score through the obstacle course using Eq.~\ref{eq:driving} in which $N_s$ represented the number of running stop signs. During all 36 runs through the obstacle course with the single-effector cursor control for steering and speed of the Mach-E vehicle, BCI-JJ received a composite driving score of 0.939 (0.0957 SD), where 1.0 is considered a perfect score (see Figure~\ref{fig:JJparkinglot_score}). There were no significant differences in scores among the four different routes taken through the course (see Figure~\ref{fig:JJparkinglot_infrac}), indicating that our BCI control system has the capability to remotely operate the basic controls of a commercially available vehicle. To our knowledge, this is the first successful demonstration of full BCI control of a full-sized commercially available vehicle in a real-world driving scenario. 

\subsubsection*{\itshape Simulated driving performance in a virtual town}
After establishing that BCI-JJ could remotely drive a physical vehicle with continuous control of its speed and steering, we upgraded our BCI teledriving system to include bimanual control for cursor and click movement. BCI-JJ's right thumb was used as before for cursor movement on the overlay for continuous changes in speed and steering; meanwhile, his left index finger was used for clicks on the overlay to achieve fast and accurate full-stop braking action. In order to more realistically test the BCI driving proficiency while ensuring the safety and consistency of measurement comparable to the motor intact behavior, we switched the driving test scenario to a more complex simulated town environment. Using the CARLA Autonomous Driving Leaderboard 2.0 simulator, we applied an approximately 20-minute route in a sophisticated virtual environment that mimicked traffic in a big city. In this simulated town driving task, the route consisted of four-way intersections with traffic lights, curves, lane changes, and an equal amount of left and right turns (see Figure~\ref{fig:route_mcity} and the ``\nameref{sec:driving-tests}'' section for a full description). BCI-JJ and the same control group of 20 motor intact participants were asked to navigate the simulated urban route with a virtual vehicle in which they had to maintain control of the vehicle's speed, steering, and braking, while adhering to standard traffic guidelines. For motor intact participants, the bimanual cursor-and-click control of the virtual vehicle was programmed using the right thumbstick and the top left button on a joystick (see figure~\ref{Sfig:joystick}). Each participant received 15 minutes of practice time before their runs were recorded. BCI-JJ completed 10 runs in four sessions, and each motor intact participant completed two runs in two sessions. With the infractions automatically measured by the CARLA Leaderboard 2.0 at the end of each run, we applied the infraction metric to compute the bimanual simulated driving score using Eq.~\ref{eq:driving} in which $N_s$ represented the number of running red traffic lights. 

Using a right-tailed two sample t-test, we compared the composite simulated driving scores for BCI-JJ versus the motor intact control group. All participants applied the bimanual cursor-and-click control for steering, speed, and full-stop braking of the virtual vehicle. Across all runs, BCI-JJ, via our BCI control system, performed significantly better than the motor intact control group ($p<0.05$), with an average driving score of 0.924 (0.115 SD), compared to an average score of 0.823 (0.160 SD) for the motor intact control group, where a score of 1.0 is perfect (see Figure~\ref{fig:AGleaderboard_score}, table~\ref{Stab:GroupLeaderboard}, and movie~\ref{Smovie:simulated-driving} for a side-by-side comparison of the average simulated driving performance between BCI-JJ and a motor intact participant). Additionally, we used two sample t-tests to compare the number of average traffic violations between BCI-JJ and the motor intact control group. BCI-JJ had significantly fewer collisions per run, on average, than the control group ($p<0.05$). There were no significant differences between these two groups in the number of red light violations or lane deviations (see Figure~\ref{fig:AGleaderboard_infrac} and table~\ref{Stab:GroupLeaderboard}). These results reflect the previous findings of two reaction time tasks in which BCI-JJ reacted at least as fast and precisely as the motor intact control group. We also compared the simulated driving scores between BCI-JJ and all 20 motor intact participants individually, using a one-way ANOVA test followed by pairwise comparisons. We found BCI-JJ had his simulated town driving performance similar to 19 motor intact participants and only significantly better than 1 motor intact participant ($p<0.05$ Bonferroni corrected, see figure~\ref{Sfig:21drivingScore}). These results indicate that the BCI control system can be used to control the essential functions of a vehicle with the same proficiency level as motor intact people in a realistic driving environment.

\begin{figure}
     \centering
     \begin{subfigure}[b]{0.65\textwidth}
         \centering
         \caption{}
         \includegraphics[height=2.5in]{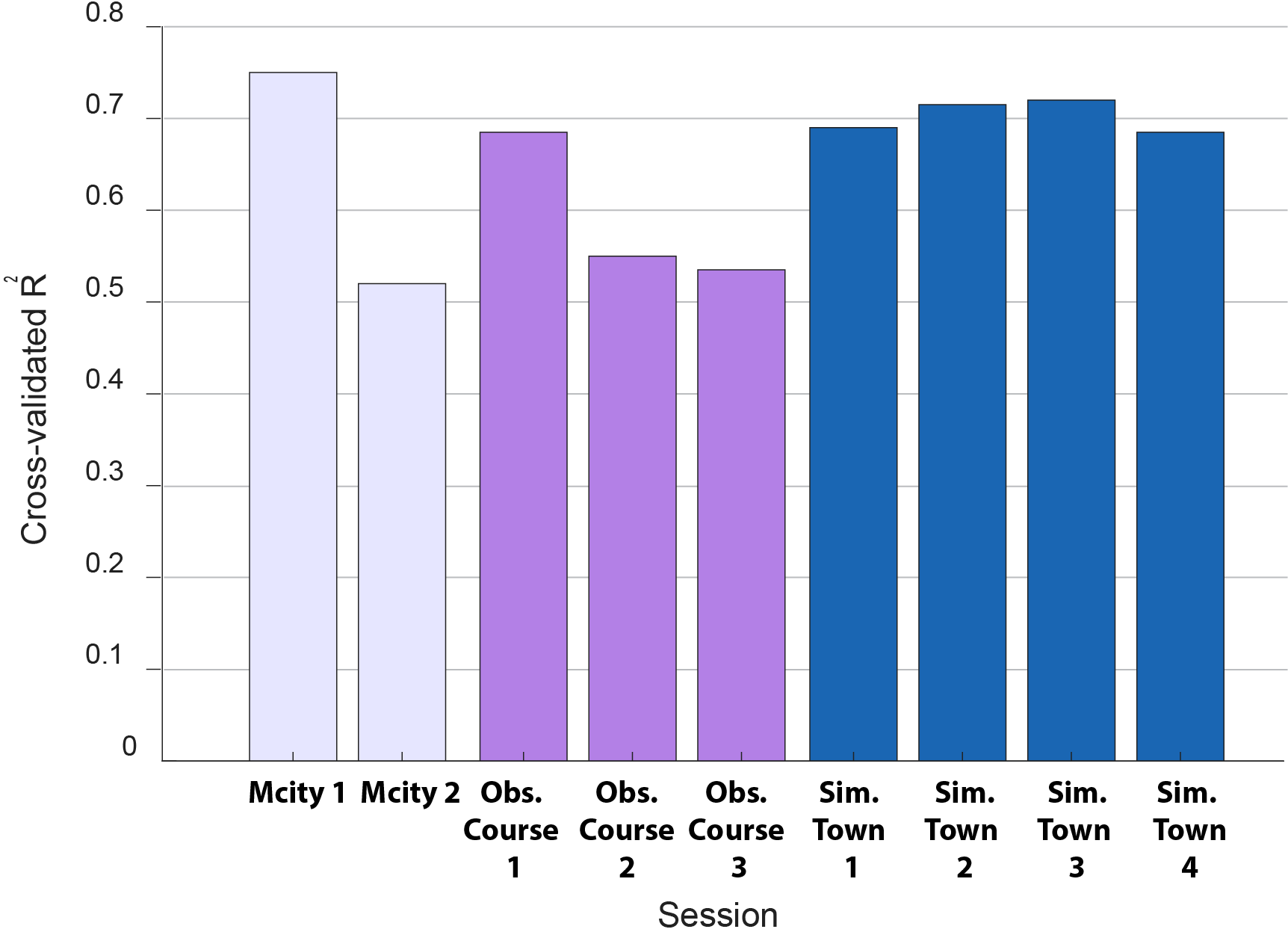}
         \label{fig:drivingR2}
     \end{subfigure}
     \hfill
     \begin{subfigure}[b]{0.33\textwidth}
         \centering
         \caption{}
         \includegraphics[height=2.5in]{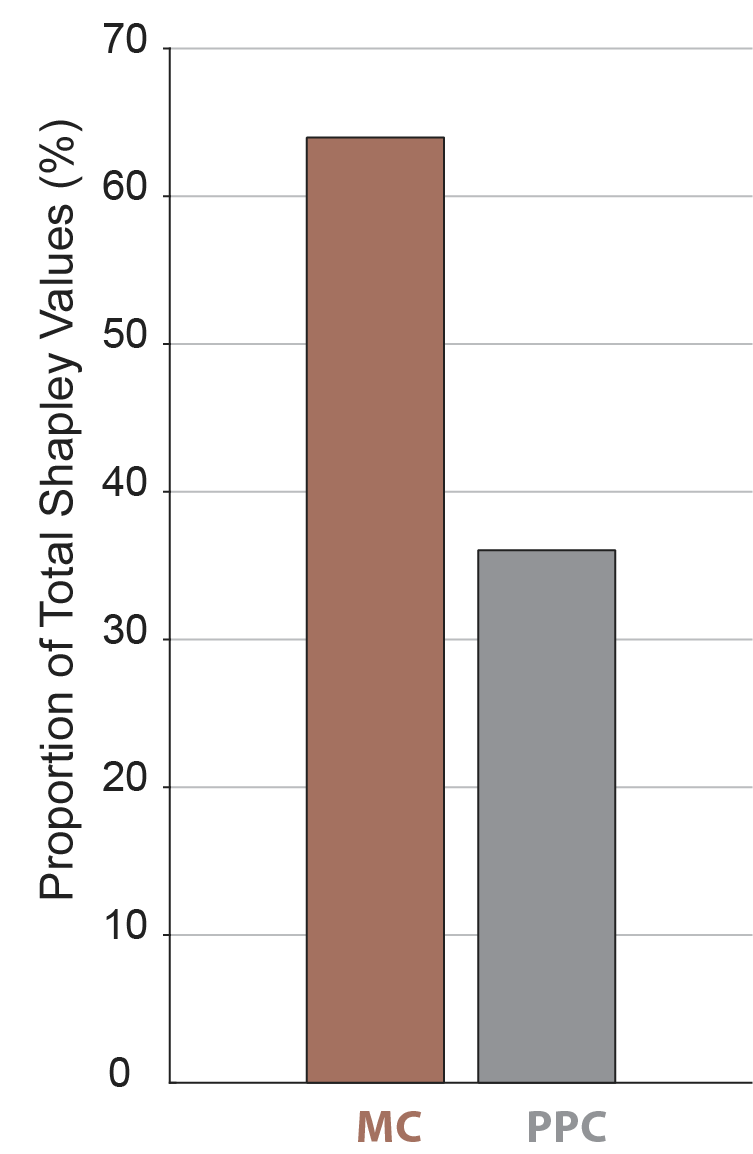}
         \label{fig:drivingMcPpcContrib}
     \end{subfigure}
     
     \begin{subfigure}[b]{0.32\textwidth}
         \centering
         \caption{}
         \includegraphics[height=2.0in]{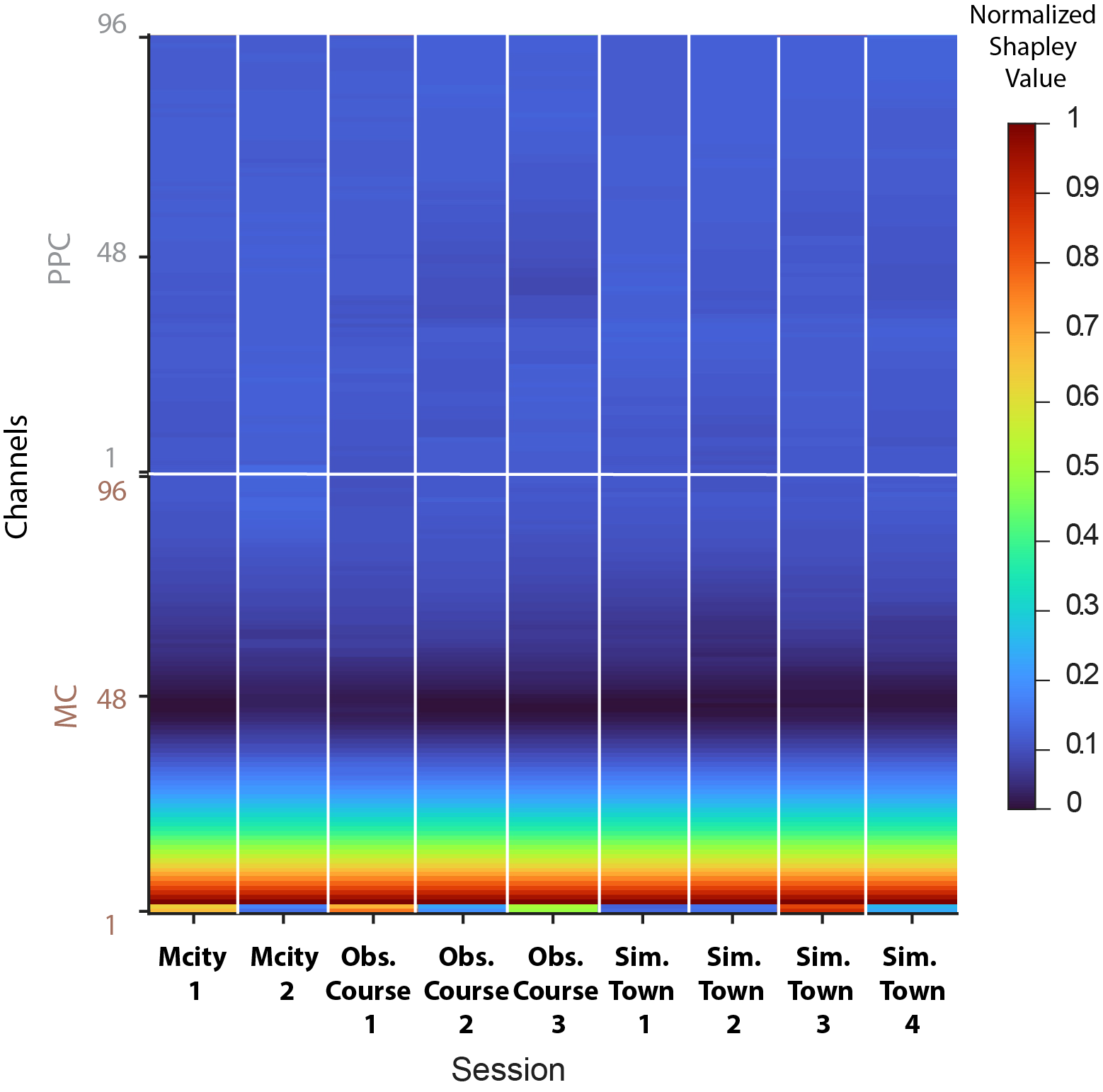}
         \label{fig:drivingChannelContrib}
     \end{subfigure}
    \hfill
    \begin{subfigure}[b]{0.32\textwidth}
         \centering
         \caption{}
         \includegraphics[height=2.0in] {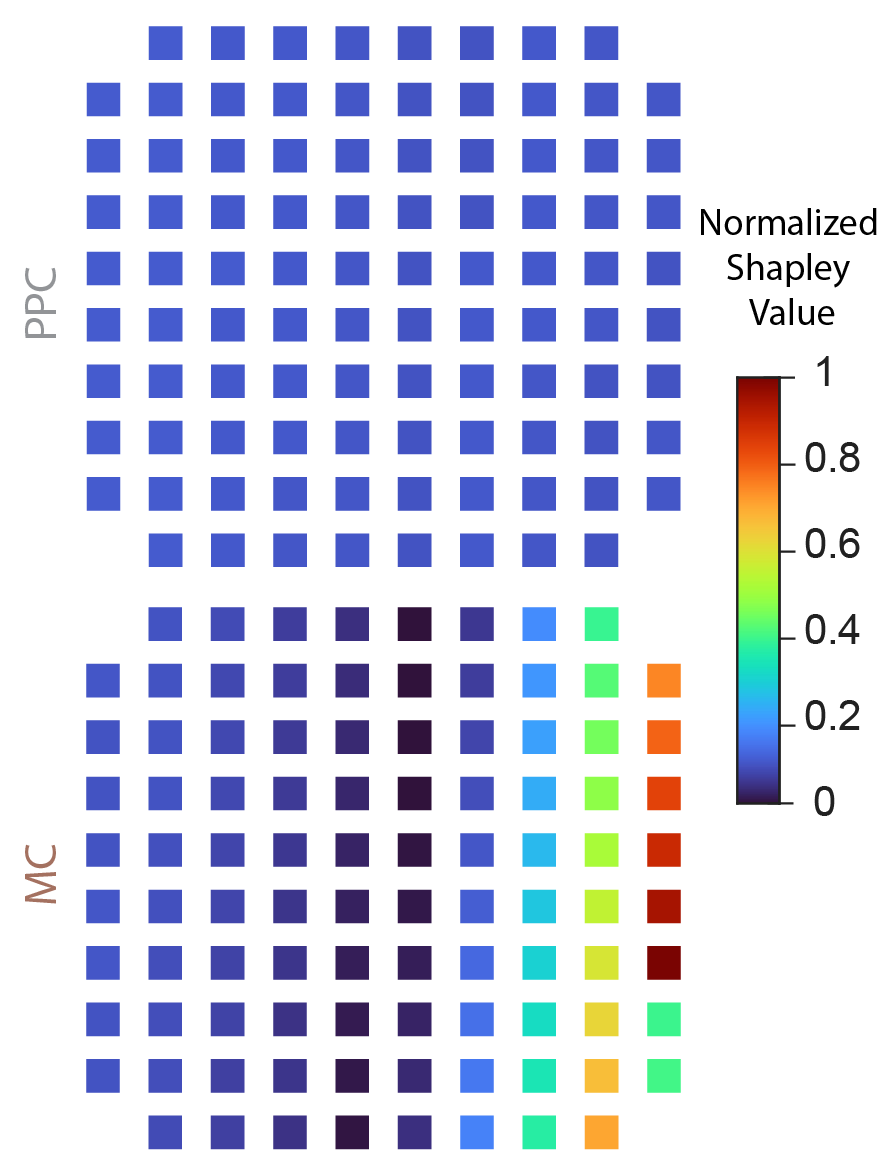}
         \label{fig:drivingTopography}
     \end{subfigure}
     \hfill
     \begin{subfigure}[b]{0.32\textwidth}
         \centering
         \caption{}
         \includegraphics[height=1.63in]{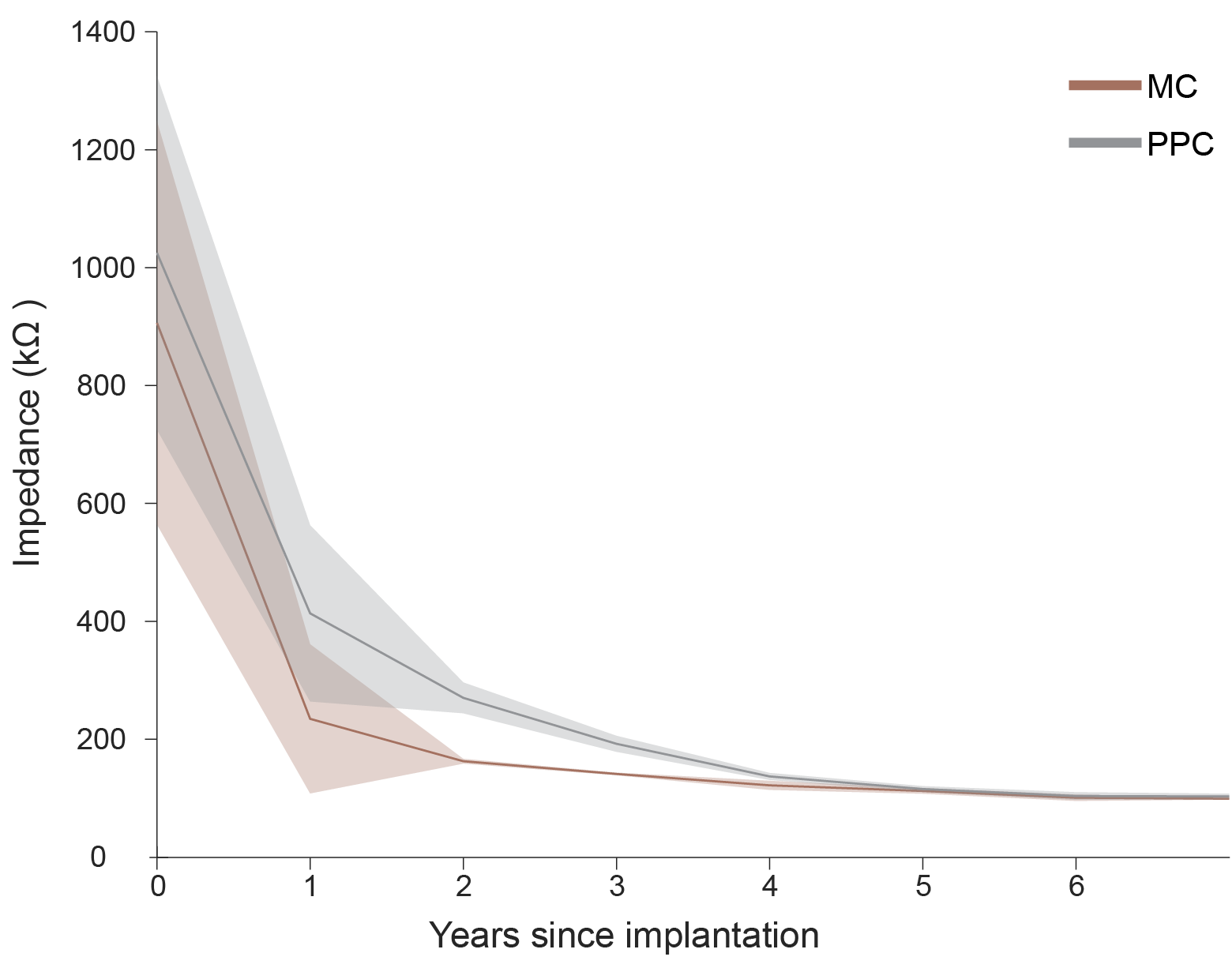}
         \label{fig:impedance}
     \end{subfigure}
    
     \caption{\textbf{Decoder stability and signal contributions.} (\textbf{A}) Cross-validated $R^2$ values for the trained decoder on center-out data on each session of the Mcity teledriving task, the obstacle-course teledriving task, and the simulated town driving task. (\textbf{B}) Proportion of average Shapley values for BCI-JJ's motor cortex (MC) and posterior parietal cortex (PPC), illustrating the contribution of each brain area to the decoder. (\textbf{C}) Normalized Shapley values for each channel across sessions, demonstrating the contribution of each channel to the decoder trained in each session. (\textbf{D}) Average normalized Shapley values for each channel plotted topographically for each electrode array implanted in BCI-JJ. (\textbf{E}) Average array impedance over years since implantation. Solid lines show the mean impedance of all channels for each electrode array across all sessions in which impedance was measured in a year; shaded areas represent one standard deviation (SD). }
        \label{fig:decoderTraining}
\end{figure}

\subsubsection*{\itshape Decoder stability for driving}
Each session began with a series of center-out tasks in which the computer progressively reduced its control over the cursor, ultimately giving control to the participant (see the ``\nameref{sec:decoder-training}'' section). The $R^2$ of the trained decoders was stable across the three years of recordings (see Figure~\ref{fig:drivingR2}) done for this study, with the linear regression not significantly different from a constant ($p=0.47$). We also calculated Shapley values for each driving session across the Mcity teledriving task, the obstacle-course teledriving task, and the simulated town driving task. Shapley values estimate each feature's contribution to a decoder. By calculating the contributions of all channels for each implanted electrode (see Figure~\ref{fig:drivingMcPpcContrib}), we found that MC contributed a mean of $64\%$ and PPC contributed a mean of $36\%$ ($0.8\%$ SD for both). Furthermore, the 32 channels that contributed the most to the decoder were the same across recorded sessions over the years of recordings (see Figure~\ref{fig:drivingChannelContrib}). The locations of these high-contributing channels followed a clear topographic pattern in MC (see Figure~\ref{fig:drivingTopography}), but PPC did not show this topographic distribution.

\subsubsection*{\itshape Teledriving and simulated driving lags}
In teleoperation tasks that require real-time feedback, a short latency between command generation and visual feedback is crucial for achieving precise control. To estimate the lag between BCI-JJ's intentions and the video feedback of the vehicle's movements, we measured the temporal difference between the overlay state and the vehicle's movement direction as inferred from optic flow analysis of the videos on all three BCI-enabled driving tasks (see the ``\nameref{sec:lag-estimate}'' section). This method accounted for delays caused by digital communication, video recording, and mechanical execution of commands. 

A one-way ANOVA revealed a significant difference ($p<0.001$) in the estimated lags of the three tasks (see figure~\ref{Sfig:drivinglags}). A Tukey's HSD Test for multiple comparisons showed that the average lags among the three groups were significantly different. The Mcity teledriving task had the longest mean lag of -1333 ms (530 ms SD), the obstacle-course teledriving task had a mean lag of -750 ms (70 ms SD), and the simulated town driving task had the least mean lag of -200 ms (40 ms SD). We also found a strong negative Pearson correlation coefficient of -0.953 ($p<0.001$) between the magnitude of the lags and the scores adjusted for the route lengths (see the ``\nameref{sec:lag-estimate}'' section).

\section*{DISCUSSION}
\subsection*{Reaction times and robustness in tasks with various complexities}
We conducted two reaction time tasks (i.e., the simple reaction time task and the braking reaction time task) by using different hand effectors of BCI-JJ with attempted clicks decoded by our BCI system and the right index finger of 20 motor intact participants with actual clicks via a computer mouse. Among tests of six hand effectors (i.e., right and left index fingers, ring fingers, and power grips) of BCI-JJ in the simple reaction time task, his right index finger had significantly faster reaction times and competitive performance measures. Compared with his left index finger in the braking reaction time task, BCI-JJ's right index finger also had significantly faster braking reaction times; however, both index finger effectors had stable enough reaction performance in order to allow the vehicle's braking in the face of emergencies. While the right thumb was selected for BCI-JJ to control the cursor movement for steering and speed changes among all three of our BCI driving tasks, his left index finger was selected to control the clicks for full-stop braking in the simulated town driving task. The differences in effector control are likely due to the placement of the arrays within MC, which has a rough somatotopic map for effectors, and degree of mixed selectivity of effectors, which is greater in PPC than MC \cite{zhang_partially_2017,kadlec_distinct_2025}. For fair comparisons, the same two effectors (i.e., the left index finger for clicks and the right thumb for cursor movement) were selected for motor intact participants during the simulated town driving task. Our reaction time task results have shown that the reaction times increase with task complexity for both BCI-JJ and the motor intact control group \cite{boisgontier_complexity_2014}. These results also indicate that reading human brain signals directly from the PPC and MC regions can provide reaction performance at least as fast and reliable as the motor intact behavior. This feature of our BCI system can be further applied to other time-sensitive devices.

\subsection*{Bimanual cursor-and-click control with BCI}
Compared with largely focused single-effector control by many existing BCI systems \cite{aflalo_decoding_2015,afshar_single-trial_2011,patrick-krueger_state_2024}, multi-effector movement allows greater functionality for people with paralysis. Other groups have done bimanual BCI studies in the lab such as a delayed-movement task \cite{deo_brain_2024} or virtual quadcopter control to reach specific points without a designated route, any other vehicles or any real-world traffic elements \cite{willsey_high-performance_2025}, whereas our bimanually controlled BCI system has been successfully applied to driving a virtual vehicle to follow a designated route through heavy traffic in a simulated town environment with our BCI performance comparable to a motor intact control group. Our bimanual control involves both the click movement via the left index finger and the cursor movement via the right thumb simultaneously and accurately using the partial least squares regression (PLSR) decoder with a feature extraction algorithm called FENet \cite{haghi_enhanced_2024} (see the ``\nameref{sec:FENet}'' section for a full description). We first documented the stability and accuracy of cursor and click control separately via two reaction time tasks and two teledriving tasks with a commercially available vehicle, respectively. The first teledriving task occurred in the Mcity closed test facility, whereas the second teledriving task took place on an obstacle course which contained many components of a standard driving test. Then we bimanually combined click control for braking and cursor control for speed and steering of a vehicle. In order to minimize the bias of comparisons between our BCI system and motor intact behavior for bimanual driving control, we asked all participants to navigate a virtual agent vehicle through a simulated busy urban town. As judged by the simulator itself at the end of each run, the bimanual driving performance decoded by our BCI system from the PPC and MC regions of BCI-JJ turned out to have the same level of stability and proficiency as using a joystick by the motor intact control group. We achieved stable bimanual decoding, performing in trials that lasted nearly two orders of magnitude longer than typical cursor control tests, which has not been demonstrated before. Beyond this, most BCI experiments report results over a few days or weeks, whereas here we show that our system's performance remained stable for several years.

\subsection*{BCI beyond the lab: real-world driving applications}
Most existing intracortical BCI systems have been applied to controlled tasks within the laboratory or home settings \cite{hochberg_reach_2012,willett_high-performance_2023,aflalo_decoding_2015,nuyujukian_cortical_2018,andersen_preserved_2022}. We show that current advances in intracortical BCIs extend the use of this technology to real-world scenarios, potentially improving the mobility of people with paralysis. Previous BCI systems for vehicle control suffered from temporal and spatial limitations \cite{haufe_electrophysiology-based_2014,zhou_novel_2021,zhuang_motion_2017,zhang_brain-controlled_2023}. We developed a modular system that included neural signal processing, bimanual cursor-and-click decoders, a graphical user interface, communication protocols, and engineering modules, enabling vehicle control. Our intracortical BCI system can provide first-of-its-kind driving experiences for a tetraplegic participant in the real-world environments by remotely controlling the steering and speed of a commercially available vehicle. BCI-JJ has performed single-effector cursor-movement-based BCI teledriving stably with a Ford Mach-E vehicle both in the Mcity closed test facility encompassing an urban environment via randomly mapped routes and on a predefined obstacle course including many components of a standard driving test, according to the independent human evaluations based on the recorded videos. Our BCI system also allows BCI-JJ to drive in a simulated town environment with heavy traffic by controlling the speed, steering, and full-stop braking of a virtual vehicle. Using the more advanced bimanual cursor-and-click control via our BCI decoder, BCI-JJ has had his BCI-controlled driving performance in the simulated town via CARLA Leaderboard 2.0 comparable to the joystick-controlled performance of 20 motor intact individuals with their average age the same as BCI-JJ's, according to the infraction components automatically recorded by the CARLA Leaderboard 2.0 at the end of each run. It is also important to note that the NeuroPort Electrodes had been implanted in BCI-JJ for three years when this study started, whereas the performance of our decoder remained stable across three years of recordings for this study, contributing to reliable translatability from neuroscience to BCI real-world applications in the long term. 

\subsection*{Limitations}
\subsubsection*{\itshape Driving test settings}
The teledriving test settings were less practical than the driving experiences that would normally occur in our daily lives. For the Mcity teledriving task, we asked BCI-JJ to complete teledriving of the Ford Mustang Mach-E vehicle on randomly mapped routes, meaning he could ignore all the traffic lights and stop signs in this closed test facility and instead focus on keeping the vehicle in the lane and avoiding collisions. For the obstacle-course teledriving task, we added stop signs and narrower lanes for the same vehicle to follow. To evaluate whether our teledriving setup served as a proof-of-concept for our BCI system's ability to reliably control the movement of a commercially available vehicle in real-world scenarios, we used domain-specific metrics designed to assess driving performance rather than general BCI metrics. This includes the CARLA Leaderboard 2.0 infraction measures and relevant parts of the California DMV driving test checklist \cite{department_of_motor_vehicles_california_2025} (see the ``\nameref{sec:safety-criteria}'' section). 

CARLA Leaderboard 2.0 was designed for autonomous driving tests, incorporating various in-vehicle sensors and artificial intelligence algorithms. Limitations of the simulation environment prevented either BCI-JJ or the motor intact control group from achieving perfect driving performance in the simulated town driving task with bimanual cursor-and-click control on the overlay. Some of the limitations are intrinsic to the CARLA simulator, such as a restricted and fixed field of view for the virtual agent vehicle, no yielding rule for other autonomous vehicles, and partial incoherence for the physics between the virtual agent vehicle and a physical vehicle. Moreover, some limitations are related to the experimental setup, such as the joystick controller and the monitor. However, the simulator was a valuable testing framework, as it provided objective evaluations of many practical features of real-world driving environments, including heavy traffic, functioning traffic lights, and computer-measurable infraction metrics.

\subsubsection*{\itshape Potential biases on the BCI versus motor intact control comparisons}
Various factors can lead to biased comparisons, such as prior experience with driving simulators, joystick usage, and the level of involvement in the tasks between BCI-JJ and 20 motor intact participants. However, we made sure that the motor intact participants were selected with no gender bias and with their average age the same as BCI-JJ's. We only had the opportunity to test the BCI reaction time tasks and driving tasks with one tetraplegic participant who might already have faster reaction times before his injury than the average performance of the motor intact control group. BCI-JJ also had longer practice time than motor intact participants, although we tried to make sure the tasks were straightforward with a steep learning curve. We also do not have a comparison between our BCI system and the motor intact control group for driving a commercially available vehicle in the real-world traffic environment. 

\subsubsection*{\itshape Driving lags and restrictions}
We found that the teledriving system we used experienced significant lags during both the Mcity and the obstacle-course tasks. Humans are capable of adjusting their motor actions to visual feedback lags greater than 200 ms, but longer lags tend to increase the perceived complexity of the task \cite{sheridan_remote_1963,fitts_information_1954}. These long lags between motor intention and visual feedback during the teledriving tasks made driving the vehicle more challenging, as the participant did not receive timely visual information on the effect of his actions. This delayed sensorimotor integration can result in slower motor responses and increased overcorrections, negatively affecting control precision.

As a result, we set the maximum speed of the simulated vehicle to 5 mph and the Ford Mach-E vehicle to 4 mph. Such low-speed vehicle control could ensure safety of the commercially available vehicle and safety drivers, stable control in both a simulated busy town environment and a real-world obstacle course with narrow lanes, and smooth display of the video feedback. Completely eliminating all lags in the teleoperation pipeline might prove challenging, as this issue was mainly affected by digital communication delays. Factors such as availability and bandwidth of the wireless network contributed to these lags. Further development of a simulated driving setup with a VR system and/or an in-vehicle BCI driver setting for broader and more realistic views and shorter lags can help improve the driving realism and performance of both BCI and motor intact participants.

\subsubsection*{\itshape Long-term recording quality and decoder stability}
We found that the performance of our decoder was stable over three years of this study (see Figure~\ref{fig:drivingR2}). It is important to note that BCI-JJ had been implanted with the NeuroPort Electrodes for three years by the time the study started, with the impedance on the channels diminishing at a much slower rate than during the initial years after implantation (see Figure~\ref{fig:impedance}). Even with the downgraded recording signal quality, the high stability of the decoder could still benefit our real-world BCI applications.

\subsection*{Future work}
\subsubsection*{\itshape Potential improvement for the decoder}
MC and PPC have traditionally been understood to have a functional organization with distinct areas responsible for the movement of different body parts. In this study, NeuroPort Electrodes were implanted in regions of the cortices that are typically associated with finger movements. However, results from existing studies have shown that neurons in these regions encode motor information of body parts beyond the hand \cite{zhang_partially_2017,kadlec_distinct_2025,willett_hand_2020}. Understanding how simultaneous movements across the body are encoded in these areas remains an open area of study. Improving our decoder to include effectors beyond the hand area would allow our participants to use their feet to control acceleration and braking, providing a driving experience that more closely resembles that of motor intact individuals and enriching the overall user experience of BCI control. 

Early work on simultaneous multi-effector encoding suggests a complex neural representation that requires non-linear decoders for accurate interpretation, making the development of more complex decoders essential for the evolution of BCIs. The implementation of these complex algorithms will demand more processing power. However, for real-world applications, BCIs must be portable and operate with limited access to power sources. As a result, decoding algorithms will need to be transitioned from high-power general-purpose processors to low-energy application-specific integrated circuits (ASICs) \cite{shaeri_246-mm2_2024,bulfer_192-channel_2025}.

\subsubsection*{\itshape In-vehicle driving with multi-effector BCI control}
We applied the cursor-movement-based BCI control with the right thumb effector for steering and speed adjustments when teledriving a commercially available vehicle, and then added the additional click control with the left index finger for full-stop braking in the simulated town driving task. The bimanual setting has been proved to allow BCI-JJ to drive with the same proficiency level as the motor intact control group in a virtual town with heavy traffic, so we would like to apply the same bimanual setting or even more advanced multi-effector control to driving a commercially available vehicle in the future. The lags between between BCI-JJ's intentions and the video feedback of the vehicle's movements in the three driving tasks have a negative correlation with the adjusted driving scores based on the route length. Instead of just remotely driving a commercially available vehicle, we would like to further test our BCI control system with a BCI participant sitting and driving inside the vehicle for enhanced mobility, wider field of view, and reduced lags.

\subsubsection*{\itshape Application with autonomous vehicles and robotics}
Intracortical BCI control with high proficiency can improve safety and personal experiences when driving a semi-autonomous vehicle or using a robot \cite{cai_rules_2023}. By building on current results and improving multi-effector decoding, we can pursue more advanced engineering outcomes for driving with BCI. We can enrich our BCI system with a decoder of cognitive states which could provide a vigilance alert signal, trigger built-in vehicle safety features, and/or switch the vehicle control between autonomous driving and driver's manual control. This advanced vehicle control system might eventually enable BCI participants to drive via their brain control inside the test vehicle instead of remotely with occasionally non-negligible telecommunication delay. In both simulated and physical urban scenarios with a semi-autonomous vehicle, the enhanced BCI system could continuously track the vigilance of each participant. If a participant is detected as having reduced attention, the system will alert them and strengthen the vehicle's safety controls. Unlike an obstacle detection sensor (ODS) that comes with many commercially available vehicles, such continuous BCI vigilance check can remind the driver to remain focused on the traffic at all times, rather than waiting until the vehicle gets too close to an obstacle or another vehicle. Although autonomous vehicles might be able to drive without human intervention in common scenarios, they might not respond adequately under special conditions (e.g., sudden appearances of other cars, pedestrians, obstacles, and unpredictable weather conditions). Therefore, the multi-effector BCI control and the autonomous driving features of the vehicle could be alternatively activated or used in combination under different situations. Such hierarchical BCI application can further improve the mobility of people with paralysis and advance the accuracy and flexibility of vehicle control.

\section*{MATERIALS AND METHODS}
\subsection*{Intracortical BCI technology}
The neural recordings in this study were collected from one participant enrolled in a BCI clinical study (ClinicalTrials.gov Identifier: NCT01958086). The institutional review boards of the California Institute of Technology, Casa Colina Hospital and Centers for Healthcare, and the University of California, Los Angeles, approved study procedures, including informed consent, implant surgery, and experiment design. BCI-JJ is a right-handed male 50 years old with a C4-C5 level spinal cord injury that occurred approximately three years before his enrollment in the study. BCI-JJ was implanted with two arrays of 96-channel NeuroPort Electrodes, including one array in the left superior parietal lobule of his PPC (MNI -34, -58, 67) and the other in his left MC close to the hand knob (MNI -39, -15, 69). Neural signals were acquired, amplified, bandpass filtered (0.3 Hz to 7.5 kHz), and digitized (30 kHz, 16 bits/sample) from electrodes using the neural signal processors called NeuroPort Systems (Blackrock Neurotech, Salt Lake City, UT). 

It is important to note that BCI-JJ's electrodes had been implanted for more than five years at the time of testing. The signal-to-noise ratio (SNR) of the recorded brain signals had decreased to approximately one-third of the value observed in the first year after implantation. Due to this decline, a custom feature extraction algorithm, FENet (see the ``\nameref{sec:FENet}'' section), played a crucial role in enabling BCI-JJ to control the vehicle with the proficiency described in this paper.

\subsection*{Motor intact participants}
We recruited 20 motor intact individuals, including 10 males and 10 females with an average age of $50 \pm 15$, the same as the age of BCI-JJ at the time of testing. The motor intact control group participated in the simple reaction time task, the braking reaction time task, and the simulated town driving task. Each motor intact participant attended two sessions on their own to complete these three tasks. All participants were informed about the purpose of the study and provided their written informed consent prior to participating. They were compensated at the end of each session based on their task performance. 11 out of 20 motor intact participants filled out an online post-session questionnaire after they completed the simulated town driving task. All those 11 people are right-handed. They all got their regular driver's licenses before participating. The rest of the questionnaire's results are shown in figure~\ref{Sfig:questionnaireResults}. Although $36.4\%$ of them would experience motion sickness when they take a ride or even drive a physical vehicle (see figure~\ref{Sfig:motionsickness}), none of them experienced motion sickness when they controlled the virtual vehicle in our simulated setting (see figure~\ref{Sfig:sicknessInSimulation}). Regarding their past experiences in simulated driving or other simulated games, $63.6\%$ of them never tried simulated driving, but $54.5\%$ had some experiences in other simulated games (see figure~\ref{Sfig:simulationExperiences}). $90.9\%$ of them had gaming joystick control experience(s) at least once or twice before participating in this task (see figure~\ref{Sfig:joystickExperiences}) and $54.6\%$ of them found the gaming joystick control in this task relatively easy (see figure~\ref{Sfig:joystickInSimulation}). $72.7\%$ of them did not feel fatigued during this task, and $18.2\%$ only felt occasionally or towards the end, with the rest $9.1\%$ could not recall their fatigue level (see figure~\ref{Sfig:fatigue}).

\subsection*{BCI decoder system}
\label{sec:decoder}

\subsubsection*{\itshape Feature extraction} 
\label{sec:FENet}

Our feature extraction procedure was based on a custom algorithm named FENet \cite{haghi_enhanced_2024}. In summary, this algorithm uses a convolutional neural network (CNN) which was trained to generate neural features from the broadband neural data. FENet consists of a set of consecutive feature engineering modules. Each module comprises 1-D convolutional filters, nonlinear activation functions, and pooling layers. The input signal is passed into M-1 back-to-back feature engineering modules. Each module receives the input data of the $i$th feature engineering module, and the data passed through the two separate temporal 1-D convolutional filters. The output of the upper filter is downsampled by 2 and is passed through a leaky ReLU nonlinear activation function. Then the output of the current filter is passed through an adaptive average pooling layer to summarize extracted temporal patterns into a single feature. The output of the lower filter is passed to the next feature engineering module. This process is repeated to find the output feature vector. A PLSR approach is applied to the output feature matrix for the electrodes to reduce the number of features (see the top part of Figure~\ref{fig:bcimechanism}).

\subsubsection*{\itshape Cursor movement decoding}
\label{sec:cursor}
We used PLSR to analyze the relationship between FENet features and movement directions. PLSR projects the predictive and observed variables into a new space by finding pairs of weight vectors that maximize the covariance between the two projections \cite{wold_pls-regression_2001}. Specifically, for each pair of neural activity data ($N$) and corresponding movement directions ($M$), we computed the FENet features from the neural activity in every 30 ms and transformed these values into z-scores. After determining the weight matrix from the training dataset using PLSR, we used it to predict $M$ from $N$ on untrained data. Finally, we applied exponential smoothing to the predicted $M$ values.

\subsubsection*{\itshape Click classification}
\label{sec:click}
Our click classifier involved a two-step process. In the first step, a Linear Discriminant Analysis (LDA) model calculated the probability that the current population firing rate was related to the intention of clicking. The second step corresponded to a Hidden Markov Model (HMM) that used the LDA probability as the input to set the state of the click between on and off. The classifiers were trained on individual trial data from a center-out task that required BCI-JJ to attempt to click (i.e., moving his index finger as if activating a trigger) when a target appeared in a reaction time task or when full-stop braking was needed in a driving task. 

\subsubsection*{\itshape Cursor-and-click training}
\label{sec:decoder-training}
BCI-JJ was asked to perform a series of center-out task runs to train our cursor-movement decoder and our click classifier in each session. The center-out task involved a central fixation point and eight radial targets spaced at 45-degree intervals around a circle. Each run consisted of 48 trials. In each trial, one of the targets would be randomly selected to change its color from gray to red, signaling the start of the trial. A cursor would then move from the center to the indicated target. After the cursor reached the target, it would return to the center. 

BCI-JJ completed three runs of the center-out task to train the cursor-movement decoder for each driving session. In this case, the participant was instructed to attempt to move his right thumb following the cursor movement. After each run, the decoder was improved using data from all previous runs. Each subsequent center-out task had decreasing assistance levels: 1.0, 0.4, and 0.2. An assistance level of 1.0 indicated that the computer was fully controlling the cursor with no input from the participant, while an assistance level of 0 meant the participant had complete control over the cursor in a closed-loop test run. During the training and testing process, there sometimes would appear a large decoder bias of cursor movement pulling in the horizontal or vertical direction, which was different from human control error. BCI-JJ would let us know when it happened so that we could adjust the anti-bias setting of the decoder accordingly. 

To evaluate the decoder performance, we used the cross-validated coefficient of determination ($R^2$) as a measure of the strength of the linear association between the predicted and intended kinematics:
\begin{equation}
R^2 = \frac{\sum_{i=1}^{n} (y_i-\hat{y_i})^2}{\sum_{i=1}^{n} (y_i-\bar{y})^2}.
\label{eq:R2}
\end{equation}
Here $n$ was the number of observations. For each observation $i$, $y_i$ was the observed kinematics, $\hat{y_i}$ was the predicted kinematics. $\bar{y}$ was the mean of all observed kinematics. The $R^2$ values for the horizontal and vertical dimensions were computed independently and then averaged for reporting.

BCI-JJ performed five runs of a modified center-out click task in the open loop to train the click classifier for each reaction-time or simulated town driving session. In this task, the cursor would briefly change color when reaching a target on $70\%$ of trials. BCI-JJ was asked to attempt clicking, such as flexing his left index finger, when the cursor changed color. For each reaction-time session that only relied on clicks, the assistance level of the center-out click task remained at 1.0 for computer-controlled cursor movement. For each simulated driving session that relied on bimanual cursor-and-click control, the click classifier was trained after completing the cursor-movement decoder training, with the cursor assistance level set to 0.2. The trained click classifier was tested during a subsequent closed-loop run of the same center-out click task, in which an attempted click would cause the cursor to change color as feedback, while BCI-JJ maintained complete control over cursor movement if the cursor-movement decoder was trained earlier in the same session.

\subsubsection*{\itshape Lag estimation}
\label{sec:lag-estimate}
For each driving task, we estimated the system's lag from the recorded videos of each session by calculating the maximum lag between the participant's intended movement and the actual visual feedback of the vehicle moving in the intended direction (see figure~\ref{Sfig:drivinglags}). We identified the participant's intention by calculating the steering cursor's position on the overlay for each video frame, by locating blobs of blue pixels and averaging their positions. We estimated the vehicle movement by calculating optical flow from the video using the RAFT deep learning algorithm \cite{vedaldi_raft_2020}. This method t+akes two images as input and computes the displacement of each pixel in the first image to its corresponding pixel in the second, generating an array of vectors representing the movement of each pixel. The analyzed video had a resolution of $1920\times980$ and was converted to grayscale for the analysis. The optic flow method was applied to a smaller area of $800\times141$ pixels from the original video, which had an unobstructed view of the road and horizon. The vehicle movement velocity vector was calculated across frames every 66 ms as the mean of the magnitudes of all individual optic flow vectors whose magnitudes exceeded a predefined threshold. We quantified the difference between video frames by calculating the structural similarity index measure (SSIM) between two consecutive frames. Both the participant's intention and the video feedback of the vehicle movement were smoothed to remove noise. Finally, we calculated the cross-correlation between the participant's intention and the video feedback of the vehicle movement and identified the maximum lag (see figures~\ref{Sfig:drivinglags-eg1} and \ref{Sfig:drivinglags-eg2} as an example from the Mcity teledriving task). The cross-correlation was calculated for frames with an SSIM less than 0.99.

We calculated an adjusted score due to the varying lengths of the runs. This adjusted score was determined using the following equation:
\begin{equation}
\text{score}_{adj} = \text{score}_{org}\times\text{length\_rate}.
\label{eq:score}
\end{equation}
Here score$_{org}$ was the original score of a run, and length\_rate was computed as the route length of the current run divided by the mean of the route lengths of all runs (see table~\ref{Stab:route-length} for a summary of route lengths for three driving tasks). Then we could calculate the Pearson correlation between the magnitude of the lags and the adjusted scores. 

This method of measuring lags accounted for all lags between the participant's decoded intention and the visual feedback in our system during execution of that motor plan, except for those in the decoder process and in the transmission of that information to the display PC. The uncomputed lags were estimated to be negligible compared to the previously analyzed lag, because the decoder processed the recordings in 30-ms bins, and the digital communication delay between the decoder and the display PCs was calculated to be less than 1 ms using the ping utility to measure LAN latency.

\subsubsection*{\itshape Channel contributions to decoder}
\label{sec:shapley}
We calculated the Shapley values for each channel to determine their contributions to the decoding process. The Shapley value of a feature is determined by the deviation in the decoder's output at a specific query point from its average output when the feature is excluded \cite{lundberg_unified_2017}. Features with higher Shapley values are considered more important because they have a greater impact on the decoder's output. 

To calculate the Shapley values, we first computed the mean of the FENet features for each channel and time point, and trained our decoder using a procedure similar to that described in the ``\nameref{sec:decoder-training}'' section. Afterwards, we determined the mean of the Shapley values across 120 randomly selected query points for each channel. The contribution of each brain region was calculated as the average of all Shapley values associated with the channels corresponding to that brain area, divided by the total sum of the Shapley values for all channels.

\subsection*{Statistics}

To assess significant differences between populations, we employed two-sample t-tests (between two groups, using ttest2 in MATLAB R2021b) or one-way ANOVA (among more than two groups) followed by a Bonferroni post hoc test to adjust for multiple comparisons (using anova1 and multcompare in MATLAB R2021b). With the Bonferroni correction, a $5\%$ significance level was utilized for all tests, unless otherwise specified for a particular analysis.

We also performed a power analysis to establish the required sample size of individuals with intact motor function for comparisons between two independent means. The analysis was based on an effect size of 0.8, an alpha level of 0.05, a power of 0.95, considering a dropout rate of $25\%$.

\subsection*{Reaction time tests}
\label{sec:rt-tests}
Participants performed two reaction time tasks of increasing complexity (see Figures~\ref{fig:simpletask} and~\ref{fig:brakingtask}). For both tasks, participants were required to react when they saw a visual stimulus. The tetraplegic participant BCI-JJ attempted a click movement as if pushing a button with one of his hand effectors via the BCI system, whereas each motor intact participant performed a click with their right index finger using a computer mouse. In both tasks, the timing of the stimulus was corroborated using a photodiode, and the click signal was recorded directly from the voltage of the mouse button switch for the motor intact individuals. The participants sat 150 cm away from a computer monitor with a diagonal length of 54 cm.

The simple reaction time task required the participants to click as soon as they saw a white circular target with a 5-cm radius appear on a black screen during the Target phase (see Figure~\ref{fig:simpletask}). The onset of the target was randomly taken from a uniform distribution between 1000 and 3000 ms from the start of a trial. The target appeared on the screen for 60 ms. A reaction time comparison to the previous trial would appear on the screen after 1000 ms from the target onset for the current trial; therefore, a valid click must have a reaction time within 1000 ms. However, a click within the first 50 ms from the target onset would be counted as a too-soon false positive click, because such a click might reflect a target prediction from the participant according to the pace of the trials instead of an actual reaction to the new target. Each run of this task had 50 trials, with 10 randomly interleaved NO-GO catch trials during which the participants were not supposed to click, indicated by a low-pitch auditory stimulus during the ITI phase. The valid reaction time was measured between 50 ms and 1000 ms for GO trials.

The braking reaction time task involved clicking to activate the brakes of a vehicle whenever an obstacle appeared in front of a vehicle in a virtual environment within the CARLA 0.9.13 driving simulator (see Figure~\ref{fig:brakingtask}). During this task, the vehicle operated under algorithmic directional control at a constant velocity (5 mph). As the vehicle proceeded down a straight road, an obstacle (e.g., a vending machine, a trash can, etc.) would appear in the vehicle's path (20 m in front of the vehicle) for each GO phase of the 40 trials per run. The participants were instructed to brake using the BCI-decoded click signal as soon as possible after seeing an obstacle appear. Without brake application in time, the vehicle would collide with the obstacle in front. Coming to a full stop required a one-time click that is programmed for a 2-second constant brake. The participant had 1 second from the obstacle onset to begin braking to avoid a collision; therefore, a valid reaction time must be within 1000 ms. A trial was considered successful if there was no collision during the GO phase and no manual braking during the NO-GO phase. However, a click within the first 50 ms from the obstacle onset would be counted as a too-soon false positive click even if there was no collision, because such a click might reflect an obstacle prediction from the participant according to the pace of the trials instead of an actual reaction to the new obstacle. The valid reaction time was measured between 50 ms and 1000 ms for GO phases.

\subsection*{BCI-enabled driving}
\label{sec:driving-tests}

\subsubsection*{\itshape From decoded brain signals to vehicle control}
\label{sec:overlay}
The bottom part of Figure~\ref{fig:bcimechanism} depicts the flow of information from the BCI decoder output in the decoder computer to several modules in the display computer and then to the control of a physical or simulated vehicle. The decoder result was consumed by the effector controller, which was responsible for continuously receiving the decoder output UDP datastream comprising positional cursor and click information, transforming it into Python data structures and publishing it to the overlay.

The overlay was a Python \textsc{pygame} instance, which subscribed to the effector controller and transformed the x-value into the horizontal position of the blue circle and the y-value into the vertical position of the red circle. The overlay further post-processed the horizontal and vertical positions into steering and speed, respectively (for all three driving tasks) and the click signals into full-stop braking (for the bimanually controlled driving task in the simulator). The steering value would change counterclockwise or clockwise exponentially when the blue circle moved to the left or right red rectangular hot zone. The speed value would increase (up to the maximum speed limit) or decrease (down to zero) exponentially when the red circle moved to the top or bottom red rectangular hot zone. If a full-stop braking was triggered by a click signal, the speed value would be immediately brought down to zero for 1 second, meaning the red circle would be immediately brought down to the bottom hot zone. For the two teledriving tasks without traffic, we also programmed slight changes of the steering value when the blue circle moved within the gray central cold zone. The final transformed steering and speed (and full-stop braking) values were displayed on screen and transmitted to the data client. The data client also sent the vehicle feedback values to the overlay display during the obstacle-course teledriving task.

The data client established multiple parallel connections with either the physical Ford Mustang Mach-E vehicle remotely or the virtual vehicle in the CARLA simulator locally. From the data client to the vehicle, the control signals including steering, speed, braking and latency were transmitted over TCP, whereas the clock synchronization signals were transmitted over UDP for lower-latency and better estimates of clock offsets. Meanwhile, the vehicle state was transmitted over UDP from the vehicle to the data client. The real-time video feedback from the vehicle was transmitted to the video display window on the display computer for BCI-JJ to view together with the overlay display while driving the vehicle with the BCI control.

\subsubsection*{\itshape Simulated town driving}
\label{sec:simulated-driving-methods}

For the simulated town driving task modified from the CARLA Autonomous Driving Leaderboard 2.0, we asked BCI-JJ and the group of 20 motor intact participants to hit the vehicle's brake by clicking with their left index finger and to adjust the vehicle's speed and steering by moving the cursor with their right thumb. While BCI-JJ had his attempted movement directly decoded from his PPC and MC signals using our BCI system, the group of 20 motor intact participants physically used a gaming joystick to control the agent vehicle's movement (see figure~\ref{Sfig:joystick}). With the overlay instance floating above the vehicle simulation window (see Figure~\ref{fig:leaderboard}), each participant got real-time visual feedback of their cursor and click control, the view ahead of the vehicle, and the side views. In the CARLA Leaderboard 2.0, we tested in Town 12's downtown areas which featured high-rise skyscrapers arranged into blocks on a consistent grid of roads with heavy traffic. The route that we edited from Town 12 included 2 right turns, 2 left runs, 1 right curve, 1 left lane switch, and 6 traffic lights (see Figure~\ref{fig:route_leaderboard}). The correct route was labeled with green dots on the ground for the agent vehicle to follow. 

The display computer recorded videos for all 10 runs in four sessions controlled by BCI-JJ and for both two runs in two sessions controlled by each motor intact participant (see movie~\ref{Smovie:simulated-driving} for a side-by-side comparison of the average simulated town driving performance between BCI-JJ and a motor intact participant). Each run of the same route lasted for about 20 minutes with both the agent vehicle's maximum speed and the traffic flow speed at 5 mph. If the agent vehicle deviated from the correct route for more than 10 seconds, the run would be stopped immediately and marked as incomplete. With the bimanual cursor-and-click control, the simulated driving performance through this route in the virtual town was evaluated based on (1) the proportion of the completed route distance ($C$), (2) the number of collisions with obstacles ($N_c$), (3) the number of lane deviations ($N_l$), and (4) the number of running red traffic lights ($N_s$). All these factors were automatically measured by the CARLA Leaderboard 2.0 by the end of each run and aggregated together for a composite bimanual simulated town driving score as shown in Eq.~\ref{eq:driving}. It started with an ideal 1.0 base score in each run, which was reduced each time an infraction was committed.

\subsubsection*{\itshape Teledriving with a physical vehicle}
\label{sec:teledriving-module-methods}

For this study, a Ford Mustang Mach-E vehicle was instrumented with additional hardware to facilitate remote motion control of the vehicle as follows: an in-vehicle controller computer with a UDP connection facilitating a telematics node to a MicroAutoBox II (MABx) containing the controller node, which provided data downstream to vehicle motion control modules via CAN. Onboard the in-vehicle telematics node in Michigan, the application received the remote teleoperation commands for speed and steering input from the display computer for BCI-JJ in California published in the form of a ROS message from the software located at the display computer's ROS client over TCP. The ROS master located on the in-vehicle telematics node subscribed to the incoming published messages and refactored the data into a UDP input to the vehicle, which was then received via the controller node on the MABx. The MABx code ran a Simulink model which ingested the UDP data and sent these inputs over CAN to the downstream vehicle motion control modules. This software also provided feedback to the display computer in the reverse path, packaging CAN data back over UDP which was then published out on a ROS message subscribed by the display computer (see Figures~\ref{fig:bcisys} and~\ref{fig:teledriving_car}). The speed and steering values that the in-vehicle controller received from the display computer were between 0 and 1, which were mapped to $0\rightarrow 4$ mph for the actual speed of the vehicle and to $-601.5\rightarrow 601.5$ degrees in the steering wheel angle.

With both BCI-enabled teledriving tasks, there were important components added to ensure the safety of test operators while remote teleoperation was occurring. Within the software on the MABx, safety controls were in place to utilize the Electronic Parking Brake (EPB) and ensure successful exit of teleoperation mode. Thus, safety controls for the safety driver were performed via 1) a manual brake press/hold to intervene in intermediate cases where the BCI-controlled vehicle might encounter an obstacle or experience non-negligible lags (i.e., greater than 1 to 2 seconds), and 2) EPB engagement to fully exit teleoperation mode. On the vehicle server ROS node, the test operators also initiated the testing by publishing a specific message to activate the teleoperation mode with a locally published message prior to initiating the remote participant display computer's ROS node. Finally, the vehicle was calibrated, and the speed request was limited to only perform low speed control of less than or equal to 4 mph. 

\subsubsection*{\itshape Teledriving real-time feedback}
As BCI-JJ was living in California while the Ford research team and this Mach-E vehicle were located in Michigan, it was necessary to establish a teleoperation framework in the initial stages of testing and validation. Our experimental setup consisted of a network created between the display computer and a modem in the Ford Mustang Mach-E SUV (see Figure~\ref{fig:teledriving_car}). A TailScale Virtual Private Network was set up to facilitate the generation of a different IP address for each of the different nodes. The display computer was connected to the Internet via a 1 Gbps fiber connection, while the modem on the vehicle was a Verizon MiFi-2100 5G Modem. In addition, a smartphone was securely mounted on the dash inside the vehicle to allow safety drivers to communicate with the team via video chat. This smartphone was not connected to the other devices in the vehicle to minimize network traffic. 

\subsubsection*{\itshape Teledriving evaluations}
\label{sec:teledriving-eval-methods}

For the Mcity teledriving task with single-effector BCI control over the cursor movement, we tested four randomly mapped routes in this mock urban environment without following any traffic lights or stop signs. BCI-JJ navigated the Mach-E vehicle for one run per route across two sessions and finished each route between 2.5 minutes and 7.5 minutes. We used a camera mounted outside the front windshield of the vehicle to record full videos of those 4 runs (see movies~\ref{Smovie:mcity-route1},~\ref{Smovie:mcity-route2},~\ref{Smovie:mcity-route3} and~\ref{Smovie:mcity-route4}), which were sent to three independent human evaluators after each session to evaluate according to the infraction rubrics. The BCI teledriving infraction components in Mcity included (1) the proportion of the completed route distance ($C$), (2) the number of collisions with obstacles ($N_c$), and (3) the number of lane deviations ($N_l$). All these factors were aggregated together for a composite BCI teledriving score in Mcity as shown in Eq.~\ref{eq:driving}. It started with an ideal 1.0 base score in each run, which was reduced each time an infraction was committed. $N_s$ in Eq.~\ref{eq:driving} was set to zero for this task, because no red traffic light or stop sign was considered.

For the obstacle-course teledriving task with the single-effector BCI control over the cursor movement, we tested four loop-shaped route options containing many components of a standard driving test. Each route was clockwise or counterclockwise with or without a lane-switch segment. BCI-JJ navigated the Mach-E vehicle for 9 runs per route across three sessions and usually finished each run on this obstacle course between 2.5 minutes and 3 minutes. We used the camera mounted within the vehicle to record full videos for all 36 runs (see movies~\ref{Smovie:obstacle-course-cw-ls},~\ref{Smovie:obstacle-course-cw-nols},~\ref{Smovie:obstacle-course-ccw-ls} and~\ref{Smovie:obstacle-course-ccw-nols} for one example run per route), which were sent to three independent evaluators after each session to evaluate according to the infraction rubrics. The BCI teledriving infraction components through this obstacle course included (1) the proportion of the completed route distance ($C$), (2) the number of collisions with obstacles ($N_c$), (3) the number of lane deviations ($N_l$), and (4) the number of running stop signs ($N_s$). All these factors were aggregated together for a composite BCI teledriving score through the obstacle course as shown in Eq.~\ref{eq:driving}. It started with an ideal 1.0 base score in each run, which was reduced each time an infraction was committed.

\subsubsection*{\itshape Safety criteria for driving tasks}
\label{sec:safety-criteria}

We set up the safety criteria for all three driving tasks to include basic components of a standard driving test for motor intact individuals, in accordance with part of the evaluation and metrics from CARLA Autonomous Driving Leaderboard 2.0 and part of the checklist for stops, lane use, lane change, and vehicle control in the ``Driving Performance Evaluation Score Sheet'' \cite{department_of_motor_vehicles_california_2025} from the California Department of Motor Vehicles (DMV) (see table~\ref{Stab:safety-components}). Different from the California DMV score sheet, we did not apply any turn signal light before lane changes or turns in any of the three driving tasks, and we did not consider yielding in the traffic scenario in the simulated town driving task. Different from the Leaderboard 2.0 evaluation and metrics, we did not add any emergency vehicles to the scene, so we did not measure failure to yield to an emergency vehicle in the simulated town driving task. Both the commercially available Ford Mustang Mach-E vehicle and the virtual agent vehicle were configured to have a slow maximum speed (i.e., 4 mph and 5 mph, respectively), which was the same as the traffic flow speed in the simulated case. When a participant controlled the cursor's vertical movement on the overlay via either BCI-decoded signals or the gaming joystick, the vehicle speed could always be kept within a safe range. We also did not introduce pedestrians in any of the three driving tasks for easier safety regulations. With steering and speed control for the teledriving tasks plus additional full-stop braking control for the simulated town driving task, both BCI-JJ and the motor intact control group were supposed to complete each driving run with fewer occurrences of the infractions (including collisions, lane deviations and running red traffic lights or stop signs) to obtain safer driving performance as evaluated by the three human evaluators for the teledriving tasks and by the CARLA Leaderboard 2.0 itself for the simulated town driving task.


\section*{Acknowledgments}
We would like to first thank our tetraplegic participant BCI-JJ and 20 motor intact participants for making this research possible. We also thank Viktor Scherbatyuk from the California Institute of Technology for administrative and technical assistance and the safety drivers Kevin Hwang and Jon Zimmerman from Ford Motor Company for safety management during BCI teledriving. Finally, we would like to express our great appreciation to Marcus Gerhardt from Blackrock Neurotech. He provided us with the original idea and challenge of having a BCI pioneer drive and ultimately race a vehicle on the Bonneville Salt Flats in Utah to show that their disability does not need to be a limit to their independence and accomplishments. This challenge has inspired the entire project team to properly study what is needed to have a vehicle controlled through a BCI and what is possible with today's technology.
\paragraph*{Funding:}\mbox{} \\
T\&C Chen Brain-Machine Interface Center (JG, KP, TA, RAA). \\
James G. Boswell Foundation (KP, RAA). \\
Swartz Foundation (JG). \\
Blackrock Neurotech (XZ, JG, CB, LC, JB, SRM, GK, TA, SSK, FS, RAA). \\
Ford Motor Company (MM, PR, AR, DF).
\paragraph*{Author contributions:}\mbox{} \\
Conceptualization: XZ, JG, MM, PR, AR, TA, SSK, DF, FS, RAA. \\
Methodology: XZ, JG, MM, PR, CB, LC, JB, AR, TA, SSK, RAA. \\
Investigation: XZ, JG, MM, PR, CB, JB, AR, TA, SSK, FS. \\
Visualization: XZ, JG, MM, PR, CB, SRM, GK, SSK. \\
Funding acquisition: JG, KP, AR, TA, SSK, DF, FS, RAA. \\
Project administration: XZ, JG, MM, SRM, KP, ERR, AR, TA, SSK, FS, RAA. \\
Clinical support: AAB, ERR. \\
Supervision: RAA, FS, DF, SSK. \\
Writing –- original draft: XZ, JG, MM, PR, CB, SSK. \\
Writing –- review \& editing: XZ, JG, MM, PR, CB, SRM, SSK, FS, RAA. 
\paragraph*{Competing interests:}\mbox{} \\
XZ, JG, KP, ERR, AAB, TA, and RAA declare that they have no competing interests.\\
CB, LC, JB, SRM, GK, SSK, and FS completed this work as employees and/or contractors for Blackrock Neurotech, a commercial company developing and manufacturing brain computer interface technologies. The work done for this project was done independently of any commercial ventures. The contents are those of the authors and do not necessarily represent the official views of nor an endorsement by Blackrock Neurotech. \\
MM, PR, AR, DF completed this work as employees for the Ford Motor Company, a commercial company developing and manufacturing automotive vehicles. The work done for this project was independent of any commercial ventures or individual financial interests. The contents are those of the authors and do not necessarily represent the official views of nor an endorsement by Ford Motor Company. \\
Ford has IP related to the manuscript but cannot disclose any specific content or details at this time.
\paragraph*{Data and materials availability:}\mbox{} \\
All primary behavioral and neurophysiological data will be archived in the Division of Biology and Biological Engineering at the California Institute of Technology and will be available when the manuscript is published.



\newpage

\renewcommand{\thefigure}{S\arabic{figure}}
\renewcommand{\thetable}{S\arabic{table}}
\renewcommand{\theequation}{S\arabic{equation}}
\renewcommand{\thepage}{S\arabic{page}}
\setcounter{figure}{0}
\setcounter{table}{0}
\setcounter{equation}{0}
\setcounter{page}{1}


\begin{center}
\section*{Supplementary Materials for\\ \mytitle}

Xinyun~Zou$^{\ast}$, Jorge~Gamez$^{\ast}$, Meghna~Menon, Phillip~Ring,\\
Chadwick~Boulay, Likhith~Chitneni, Jackson~Brennecke, Shana~R.~Melby,\\
Gracy~Kureel, Kelsie~Pejsa, Emily~R.~Rosario, Ausaf~A.~Bari,\\
Aniruddh~Ravindran, Tyson~Aflalo, Spencer~S.~Kellis,\\
Dimitar~Filev, Florian~Solzbacher, Richard~A.~Andersen\\
\small$^\ast$Corresponding authors. E-mails: xzou@caltech.edu, jgamez@caltech.edu
\end{center}

\subsubsection*{This PDF file includes:}
Figures S1 to S9\\
Tables S1 to S13\\
Captions for Movies S1 to S9

\subsubsection*{Other Supplementary Materials for this manuscript:}
Movies S1 to S9

\newpage


\begin{figure}[ht!]
\centering
\includegraphics[width=0.8\textwidth]{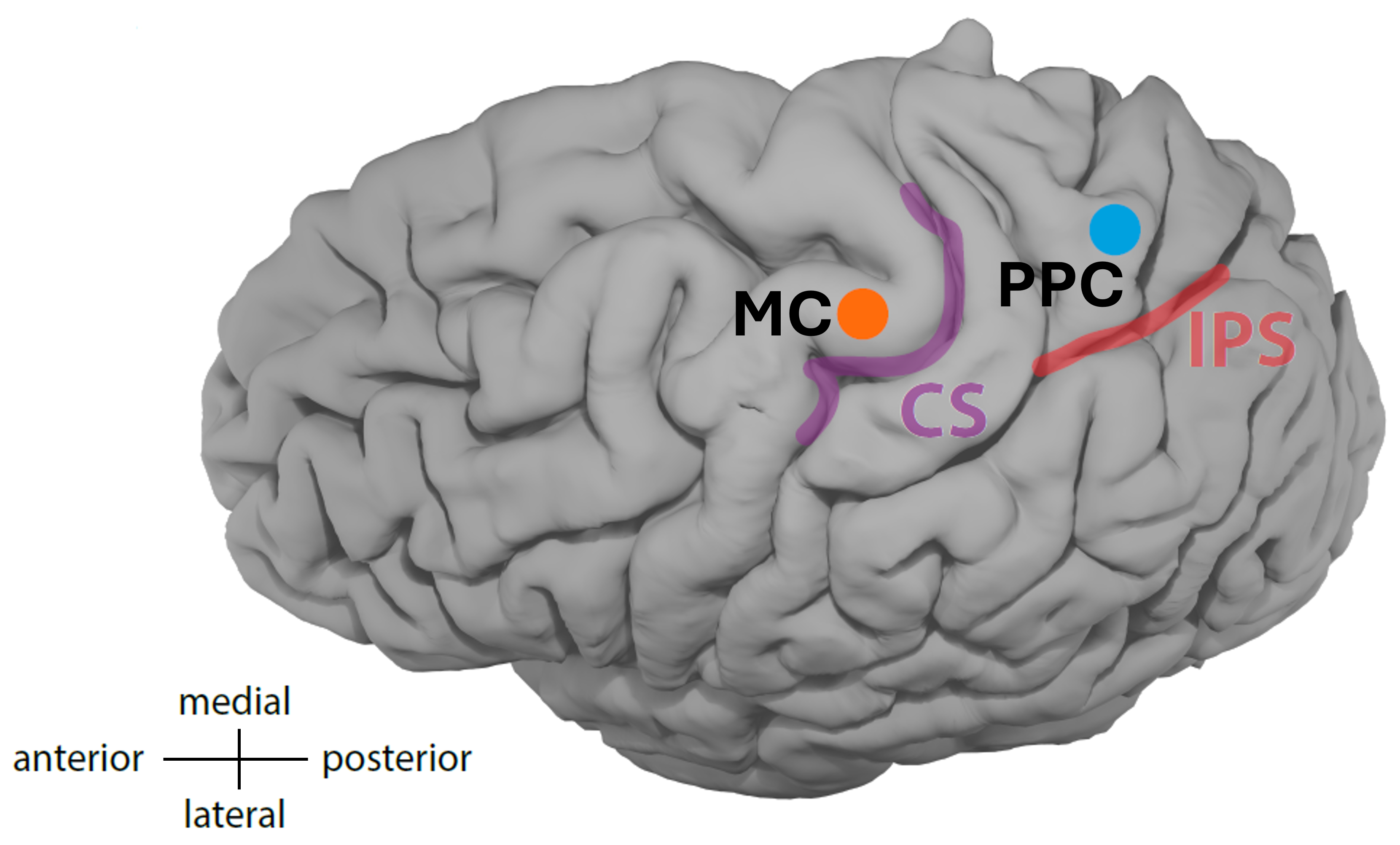}
\caption{\textbf{Brain implant locations of BCI-JJ.} We implanted one array of 96-channel NeuroPort Electrodes \copyright~Blackrock Neurotech in the left superior parietal lobule of his posterior parietal cortex (PPC) (MNI -34, -58, 67) and the other array of 96-channel NeuroPort Electrodes in his left motor cortex (MC) close to the hand knob (MNI -39, -15, 69). CS: central sulcus. IPS: intraparietal sulcus.}
\label{Sfig:brainimplant}
\end{figure}

\begin{figure}
     \centering
     \begin{subfigure}[b]{0.65\textwidth}
         \centering
         \caption{}
         \includegraphics[width=\textwidth]{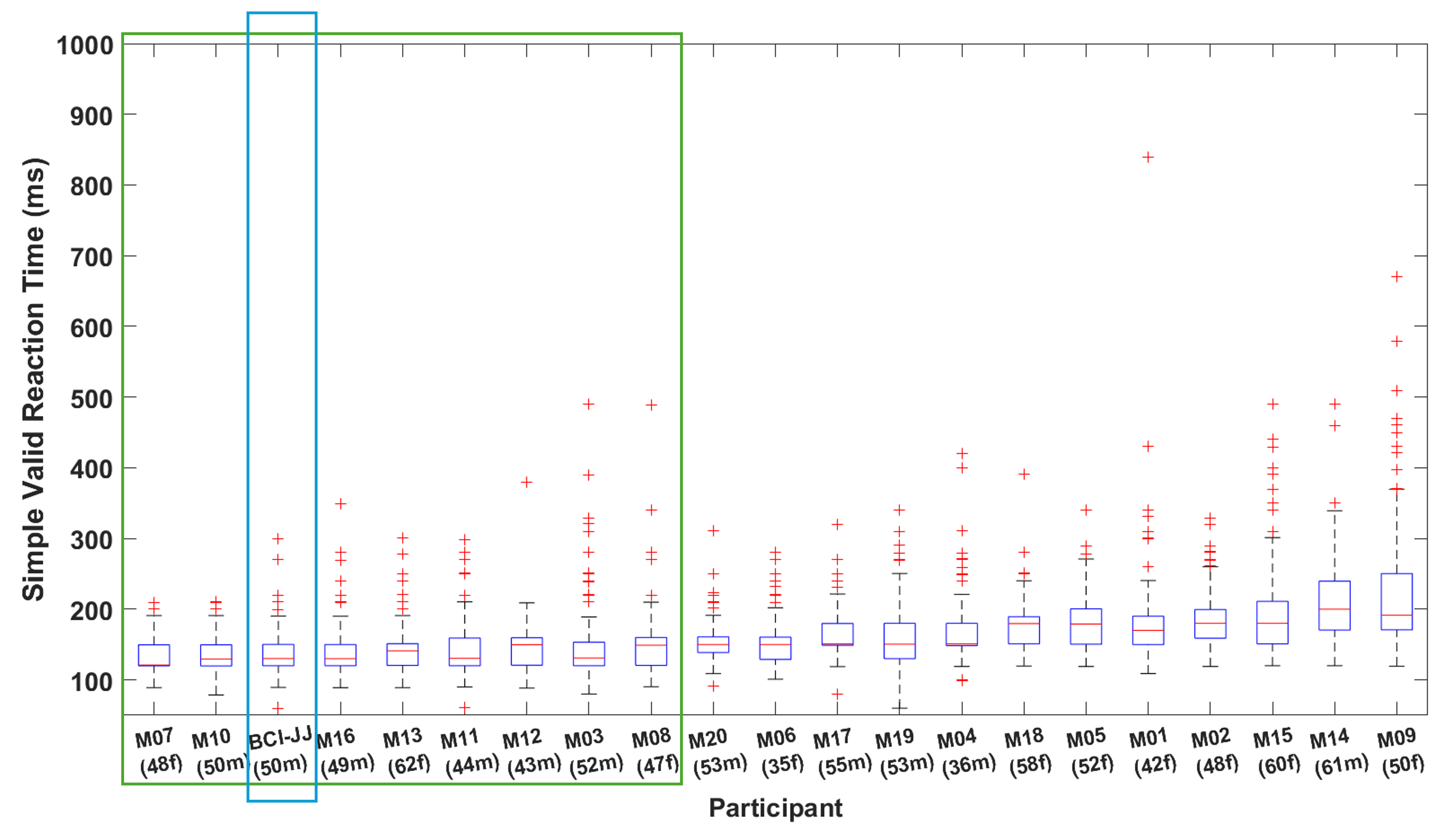}
         \label{Sfig:21simpleValidRT}
     \end{subfigure}
     
     \begin{subfigure}[b]{0.65\textwidth}
         \centering
         \caption{}
         \includegraphics[width=\textwidth]{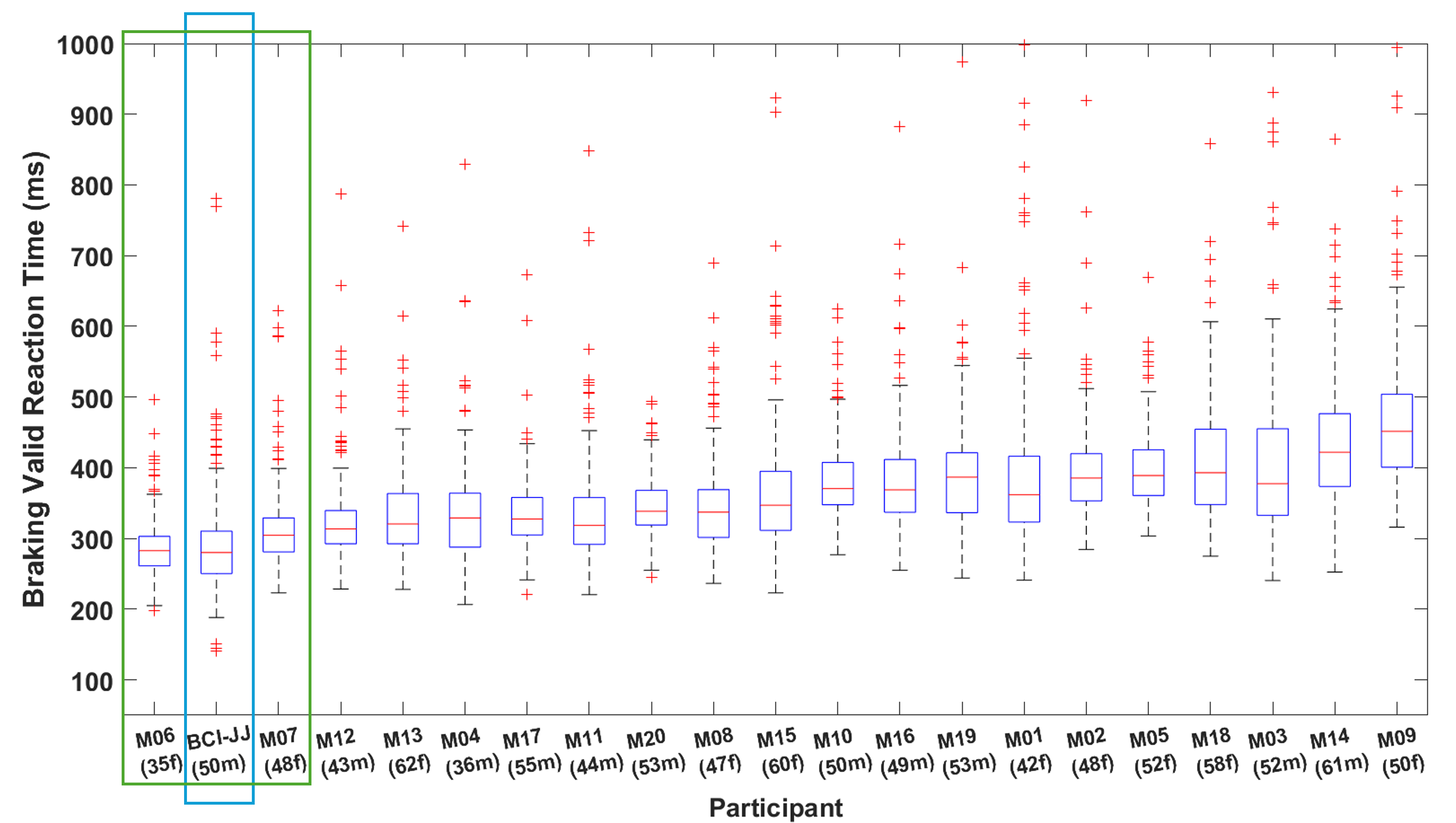}
         \label{Sfig:21brakingValidRT}
     \end{subfigure}
     \captionsetup{aboveskip=0pt}
        \caption{\textbf{Trial-based valid reaction times for simple and braking reaction time tasks among BCI-JJ and all 20 motor intact participants individually with the same right index finger effector.} In the parenthesis under each participant's ID, the number represents their age, and the letter ``f'' or ``m'' represents their gender (female or male). The box plots are sorted according to the average performance of each participant. We conducted pairwise comparisons using a Bonferroni post hoc test following a one-way ANOVA among 21 participants. For the participants whose box plots are outside the green rectangular area, their performance was different from BCI-JJ at the corrected $5\%$ significance level. There are comparisons of the valid reaction times between 50 ms and 1000 ms in GO trials/phases among all 21 participants for (\textbf{A}) the simple reaction time task and (\textbf{B}) the braking reaction time task.}
        \label{Sfig:21reactValidRT}
\end{figure}

\clearpage
\begin{figure}[ht!]
     \centering
     \begin{subfigure}[b]{0.65\textwidth}
         \centering
         \caption{}
         \includegraphics[width=\textwidth]{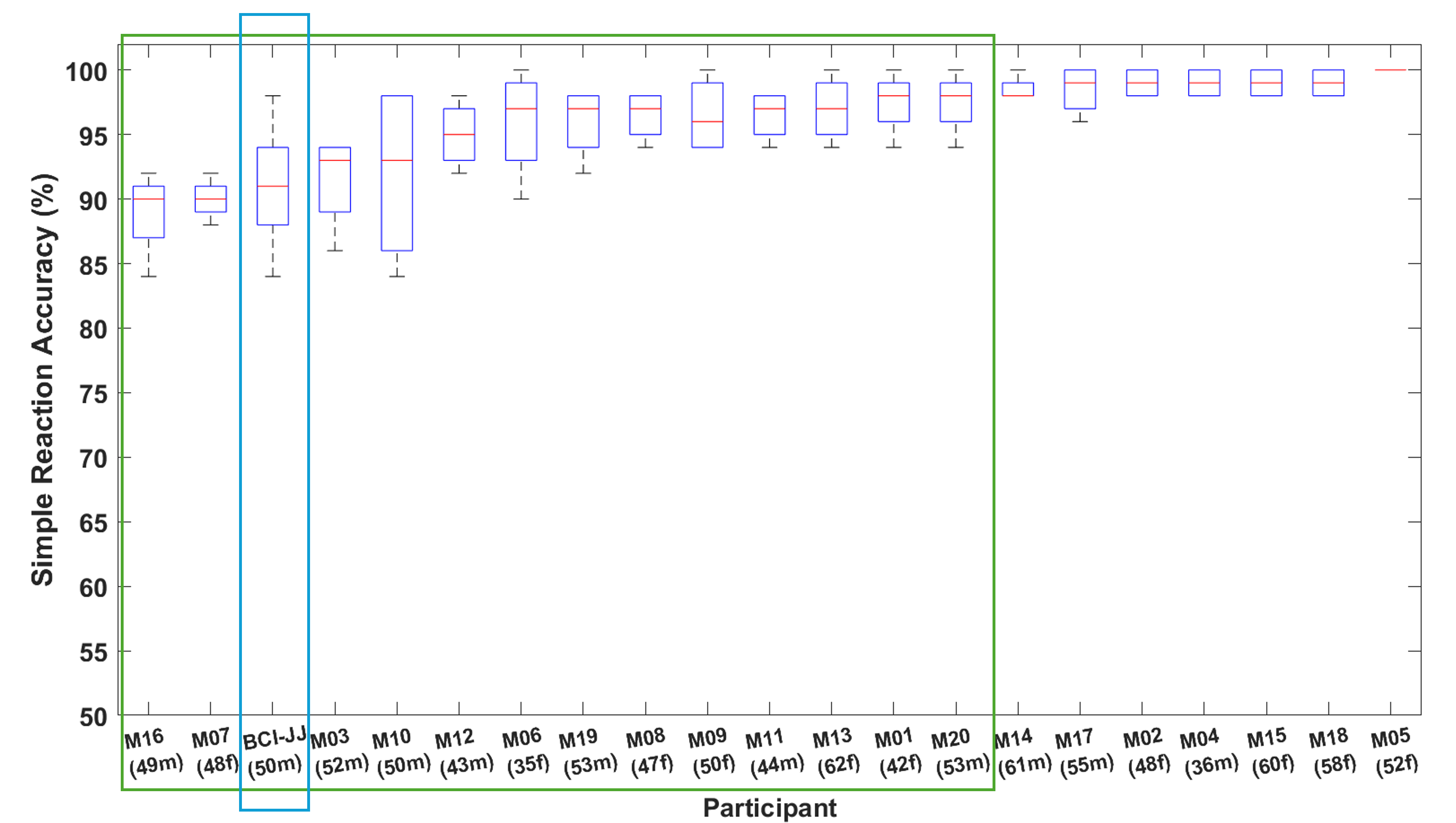}
         \label{Sfig:21simpleacc}
     \end{subfigure}
     
     \begin{subfigure}[b]{0.65\textwidth}
         \centering
         \caption{}
         \includegraphics[width=\textwidth]{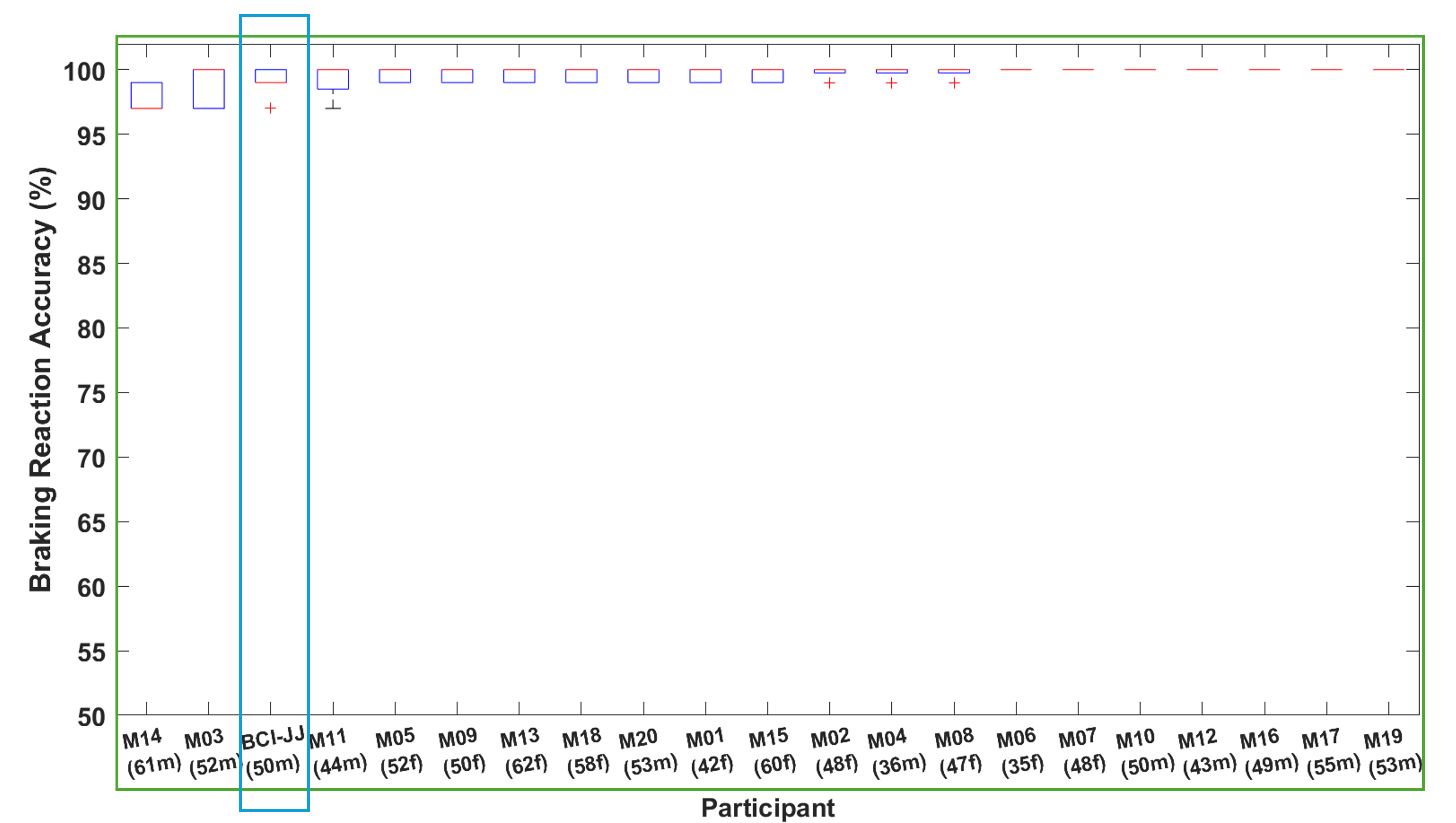}
         \label{Sfig:21brakingacc}
     \end{subfigure}
     \captionsetup{aboveskip=0pt}
        \caption{\textbf{Trial-based reaction accuracy for simple and braking reaction time tasks among BCI-JJ and all 20 motor intact participants individually with the same right index finger effector.} In the parenthesis under each participant's ID, the number represents their age, and the letter ``f'' or ``m'' represents their gender (female or male). The box plots are sorted according to the average performance of each participant. We conducted pairwise comparisons using a Bonferroni post hoc test following a one-way ANOVA among 21 participants. For participants whose box plots are outside the green rectangular area, their performance was different from BCI-JJ at the corrected $5\%$ significance level. There are comparisons of reaction accuracy among all 21 participants for (\textbf{A}) the simple reaction time task and (\textbf{B}) the braking reaction time task.}
        \label{Sfig:21reactAcc}
\end{figure}

\clearpage
\begin{figure}[ht!]
     \centering
     \begin{subfigure}[b]{0.65\textwidth}
         \centering
         \caption{}
         \includegraphics[width=\textwidth]{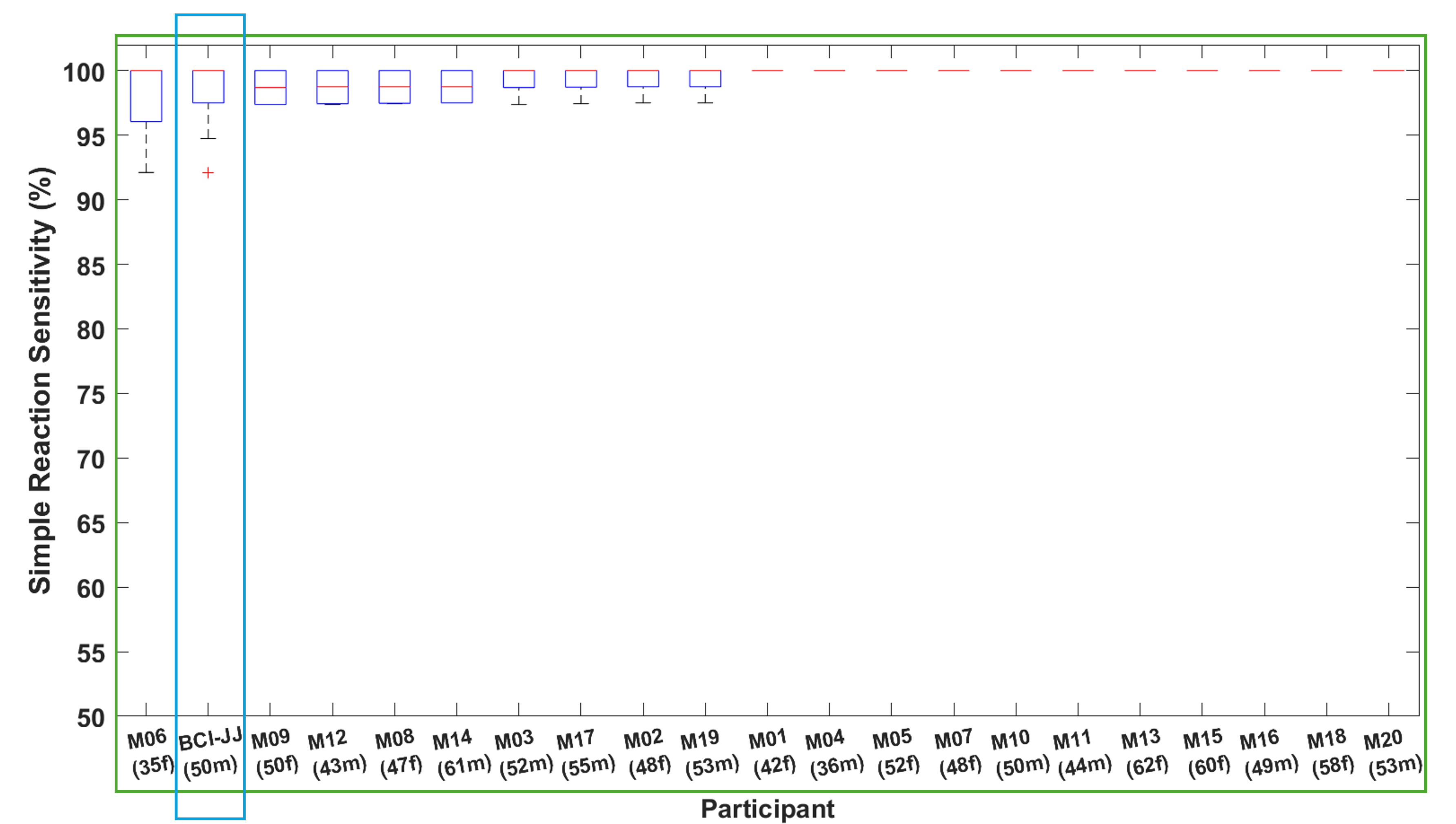}
         \label{Sfig:21simplesens}
     \end{subfigure}
     
     \begin{subfigure}[b]{0.65\textwidth}
         \centering
         \caption{}
         \includegraphics[width=\textwidth]{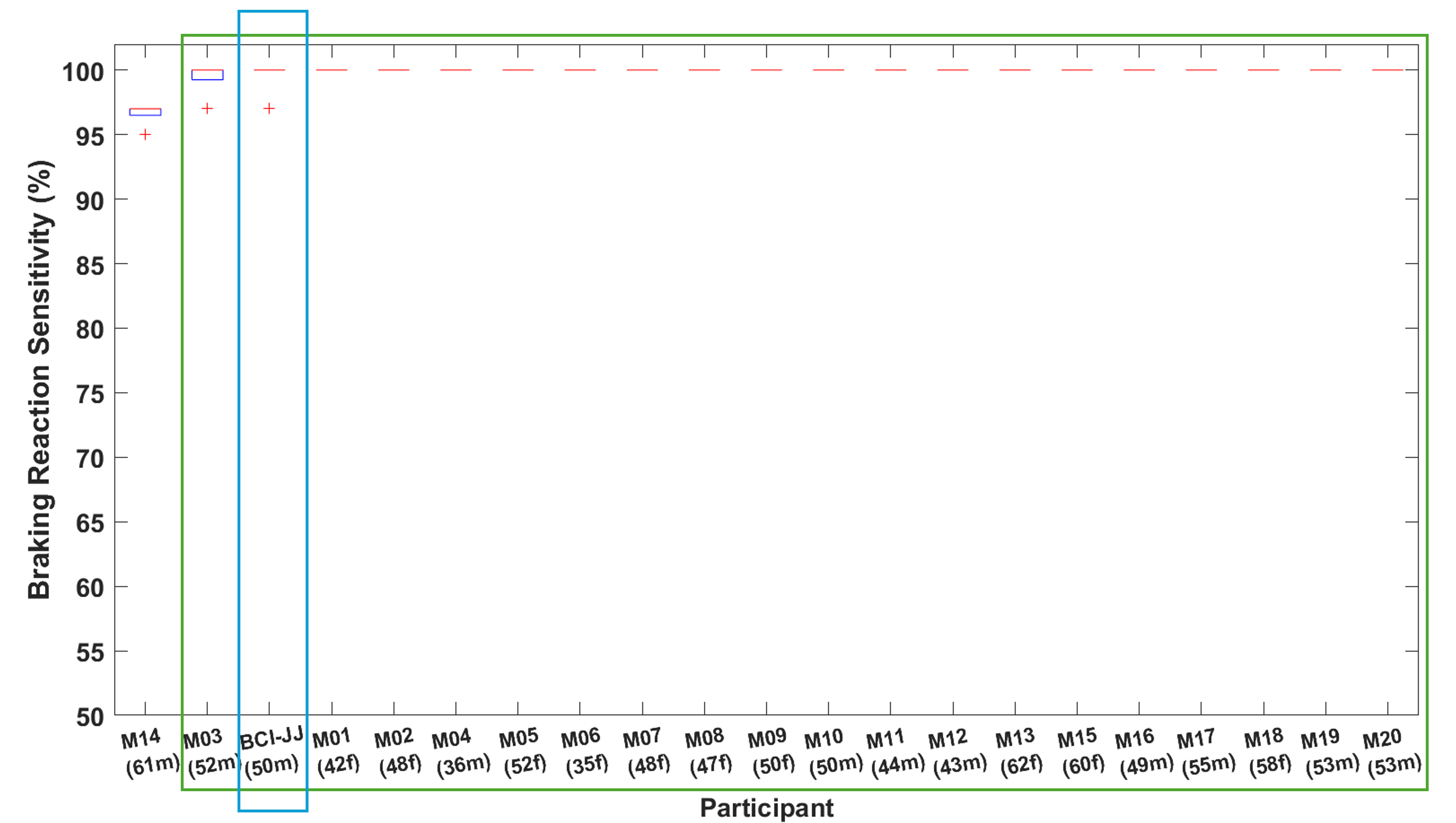}
         \label{Sfig:21brakingsens}
     \end{subfigure}
     \captionsetup{aboveskip=0pt}
        \caption{\textbf{Trial-based reaction sensitivity for simple and braking reaction time tasks among BCI-JJ and all 20 motor intact participants individually with the same right index finger effector.} In the parenthesis under each participant's ID, the number represents their age, and the letter ``f'' or ``m'' represents their gender (female or male). The box plots are sorted according to the average performance of each participant. We conducted pairwise comparisons using a Bonferroni post hoc test following a one-way ANOVA among 21 participants. For participants whose box plots are outside the green rectangular area, their performance was different from BCI-JJ at the corrected $5\%$ significance level. There are comparisons of reaction sensitivity among all 21 participants for (\textbf{A}) the simple reaction time task and (\textbf{B}) the braking reaction time task.}
        \label{Sfig:21reactSens}
\end{figure}

\clearpage
\begin{figure}[ht!]
     \centering
     \begin{subfigure}[b]{0.65\textwidth}
         \centering
         \caption{}
         \includegraphics[width=\textwidth]{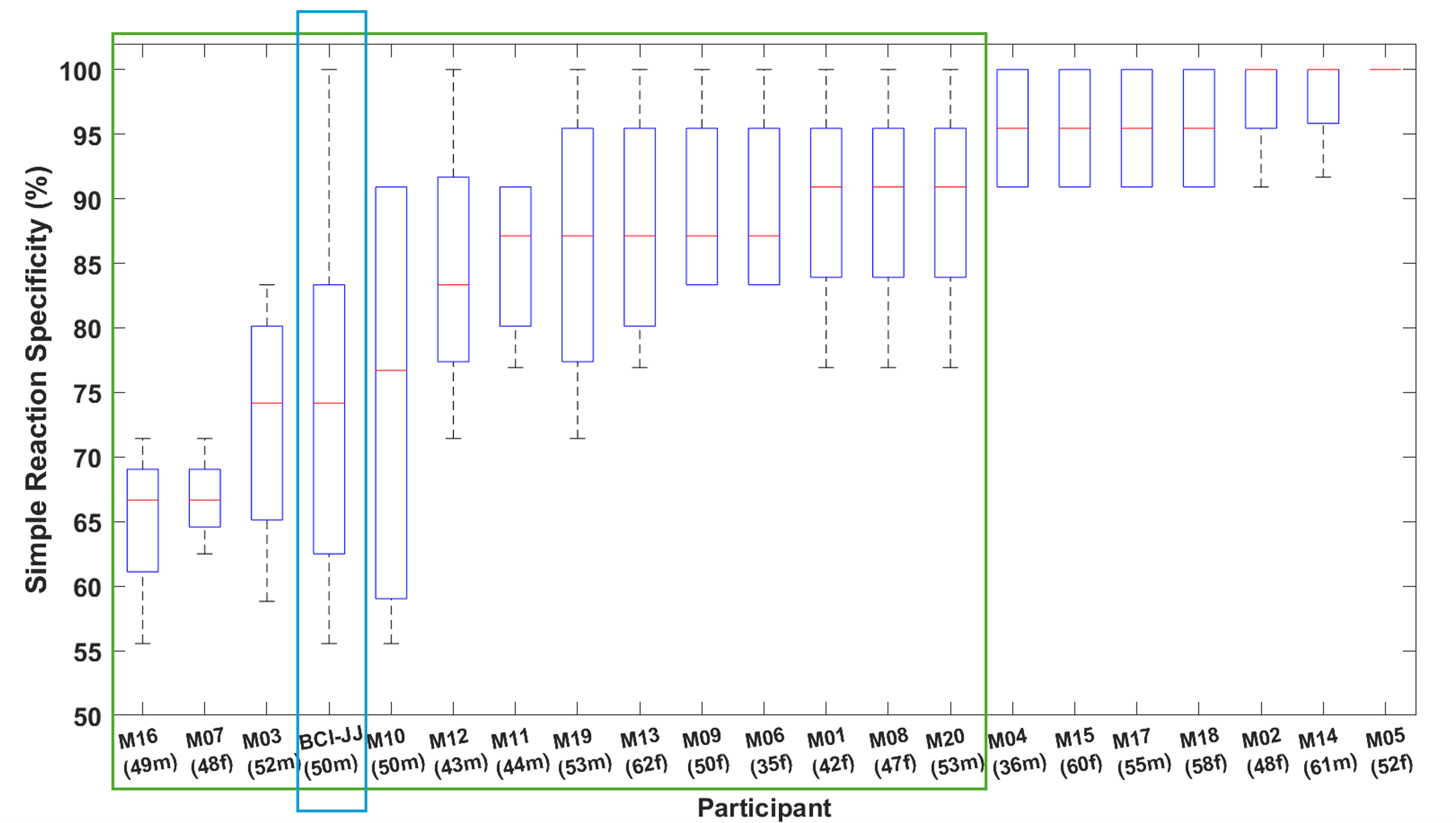}
         \label{Sfig:21simplespec}
     \end{subfigure}
     
     \begin{subfigure}[b]{0.65\textwidth}
         \centering
         \caption{}
         \includegraphics[width=\textwidth]{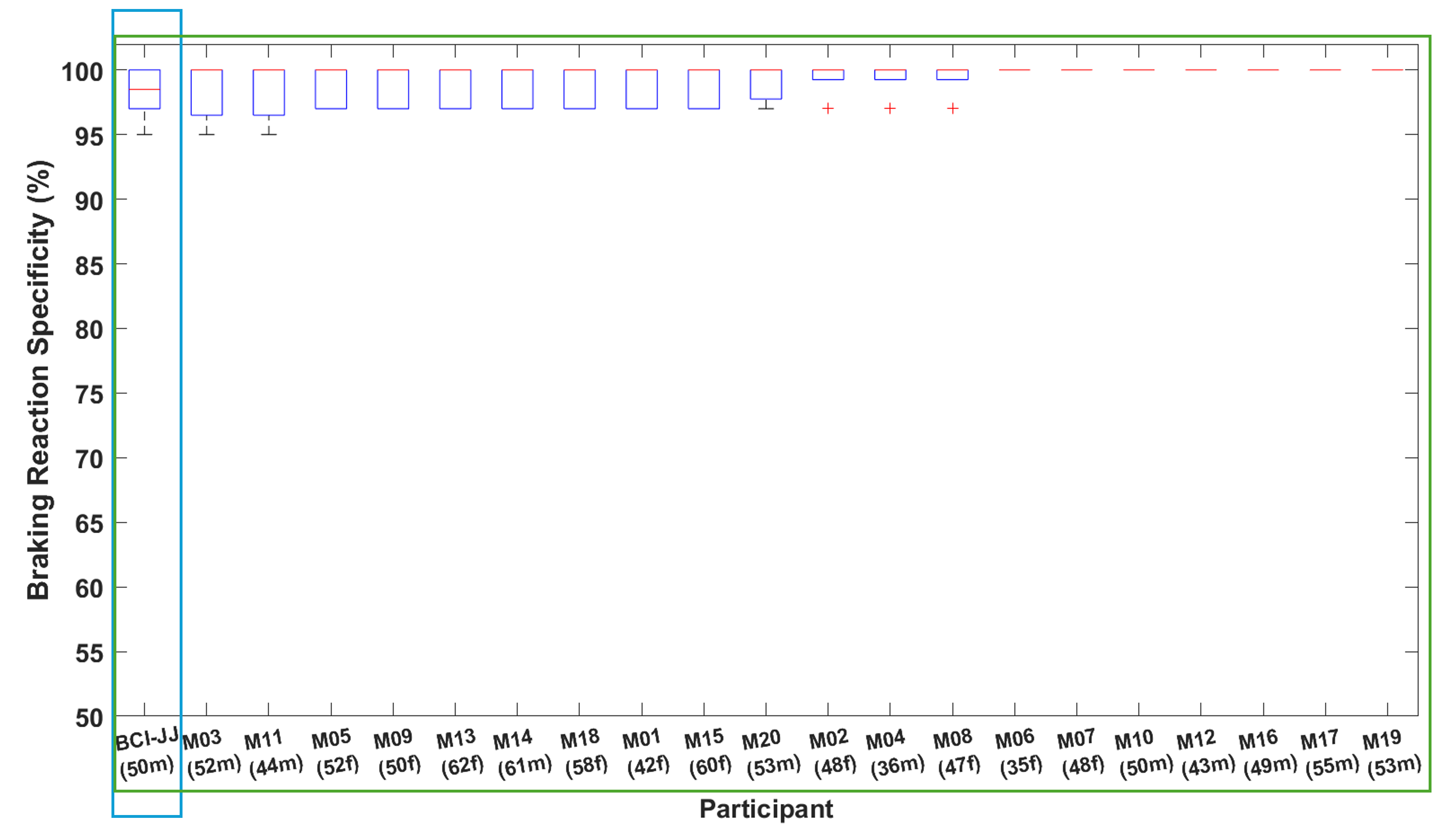}
         \label{Sfig:21brakingspec}
     \end{subfigure}
     \captionsetup{aboveskip=0pt}
        \caption{\textbf{Trial-based reaction specificity for simple and braking reaction time tasks among BCI-JJ and all 20 motor intact participants individually with the same right index finger effector.} In the parenthesis under each participant's ID, the number represents their age, and the letter ``f'' or ``m'' represents their gender (female or male). The box plots are sorted according to the average performance of each participant. We conducted pairwise comparisons using a Bonferroni post hoc test following a one-way ANOVA among 21 participants. For participants whose box plots are outside the green rectangular area, their performance was different from BCI-JJ at the corrected $5\%$ significance level. There are comparisons of reaction specificity among all 21 participants for (\textbf{A}) the simple reaction time task and (\textbf{B}) the braking reaction time task.}
        \label{Sfig:21reactSpec}
\end{figure}

\begin{figure}[ht!]
\centering
\includegraphics[width=0.6\textwidth]{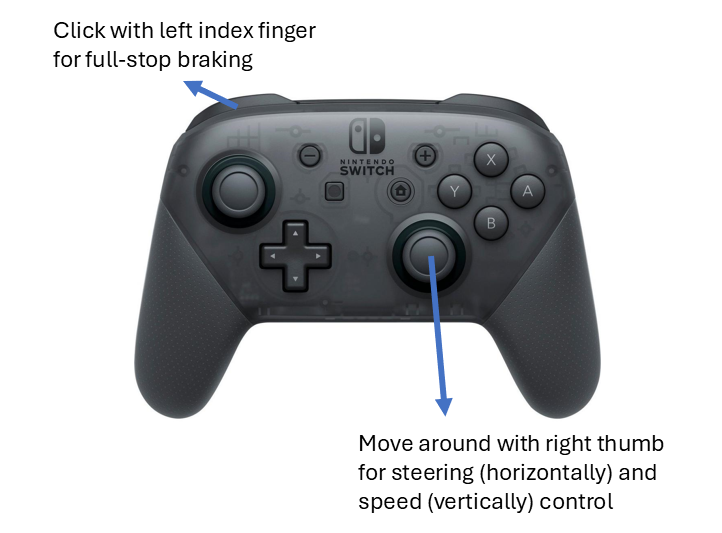}
\caption{\textbf{Joystick programmed for the bimanual cursor-and-click driving control of motor intact participants.} They used the joystick in the simulated town driving task modified from the CARLA Leaderboard 2.0. The top left button was clicked with the left index finger to trigger full-stop braking of the vehicle. The right thumbstick was moved around with the right thumb to trigger the vehicle's steering value changes horizontally and its speed value changes vertically.}
\label{Sfig:joystick}
\end{figure}

\begin{figure}[ht!]
\centering
\includegraphics[width=0.95\textwidth]{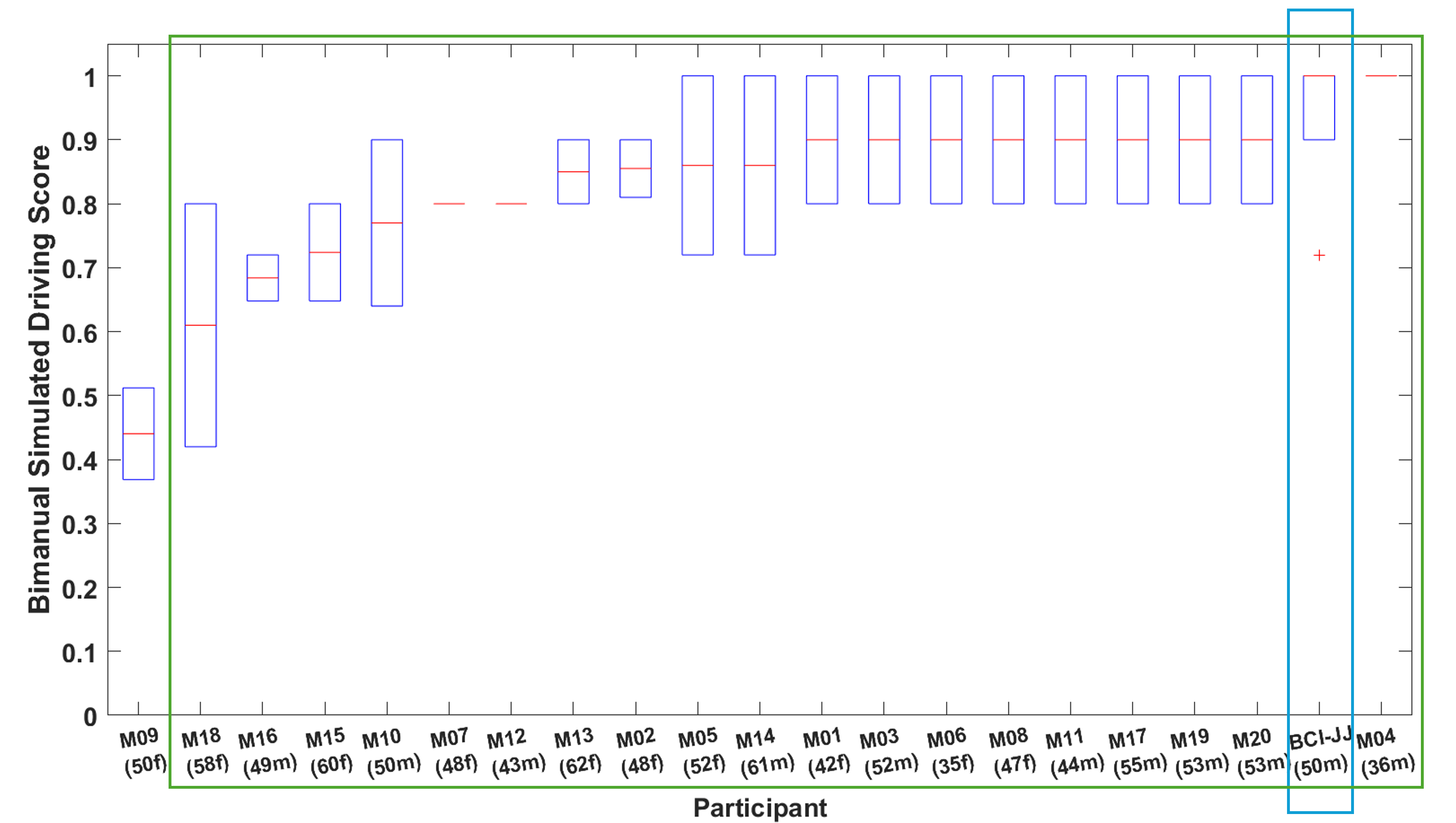}
\caption{\textbf{Trial-based simulated driving scores among BCI-JJ and all 20 motor intact participants individually with the same bimanual cursor-and-click control.} In the parenthesis under each participant's ID, the number represents their age, and the letter ``f'' or ``m'' represents their gender (female or male). The box plots are sorted according to the average performance of each participant. We applied pairwise comparisons with the Bonferroni correction from a multiple comparison test using the information contained from one-way ANOVA among 21 participants. For participants whose box plots are outside the green rectangular area, their performance was different from BCI-JJ at the corrected $5\%$ significance level. }
\label{Sfig:21drivingScore}
\end{figure}

\begin{figure}
     \centering
     \begin{subfigure}[b]{0.48\textwidth}
         \centering
         \caption{}
         \includegraphics[width=\textwidth]{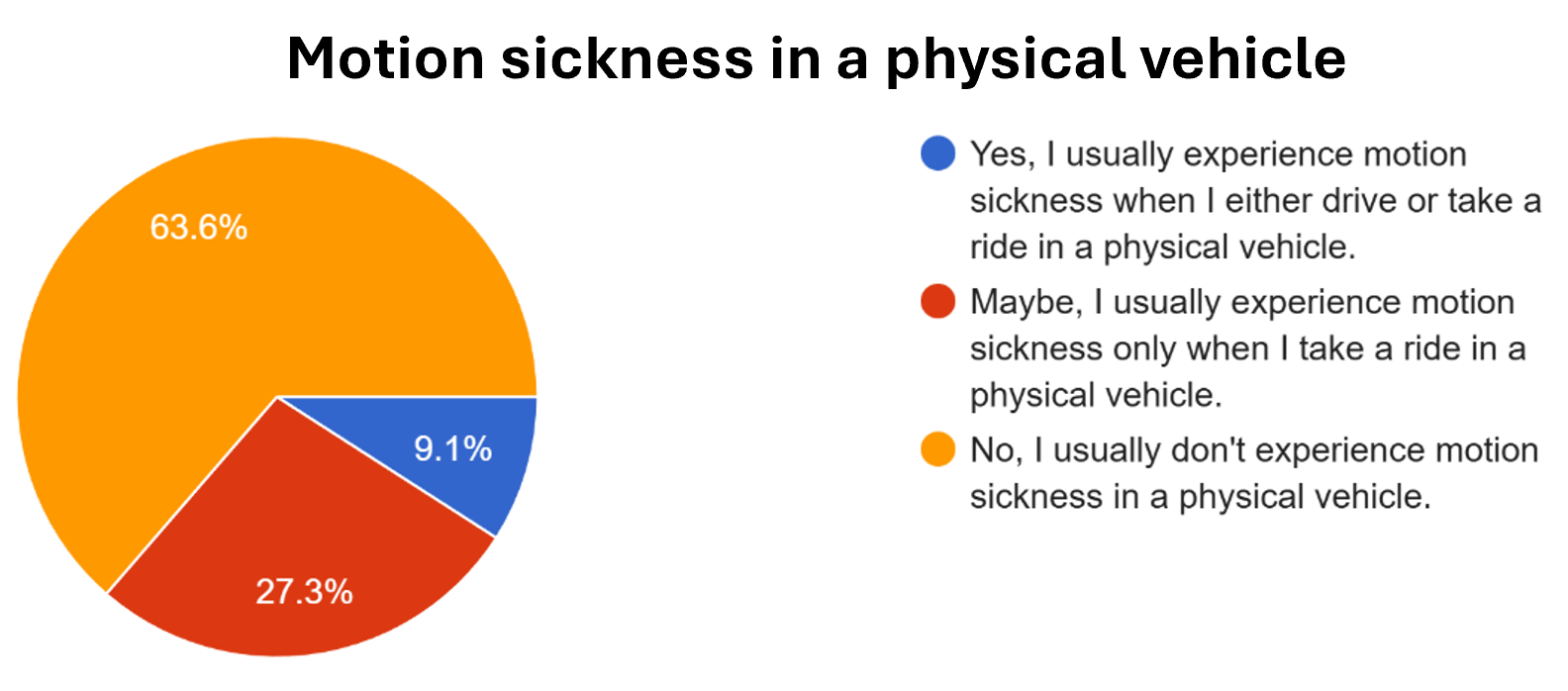}
         \label{Sfig:motionsickness}
     \end{subfigure}
     \hfill
     \begin{subfigure}[b]{0.48\textwidth}
         \centering
         \caption{}
         \includegraphics[width=\textwidth]{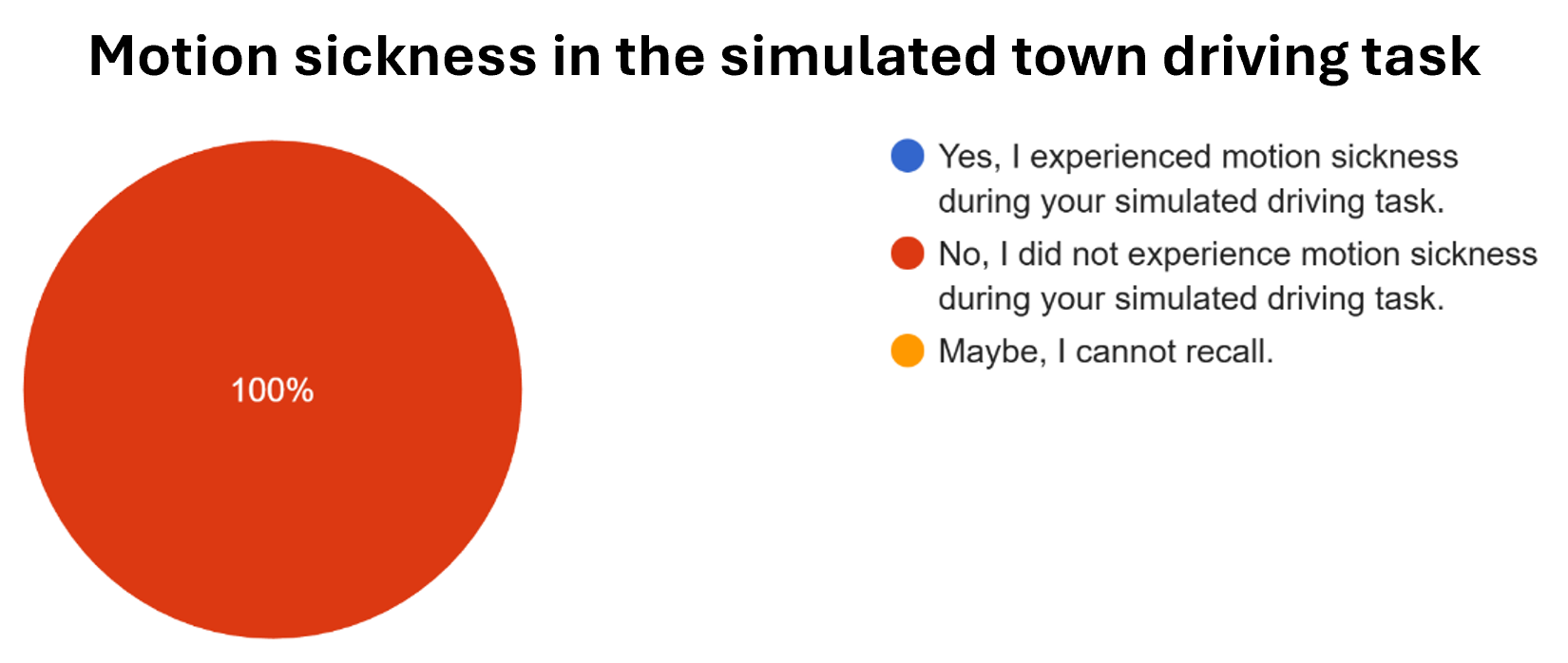}
         \label{Sfig:sicknessInSimulation}
     \end{subfigure}
     
     \begin{subfigure}[b]{0.48\textwidth}
         \centering
         \caption{}
         \includegraphics[width=\textwidth]{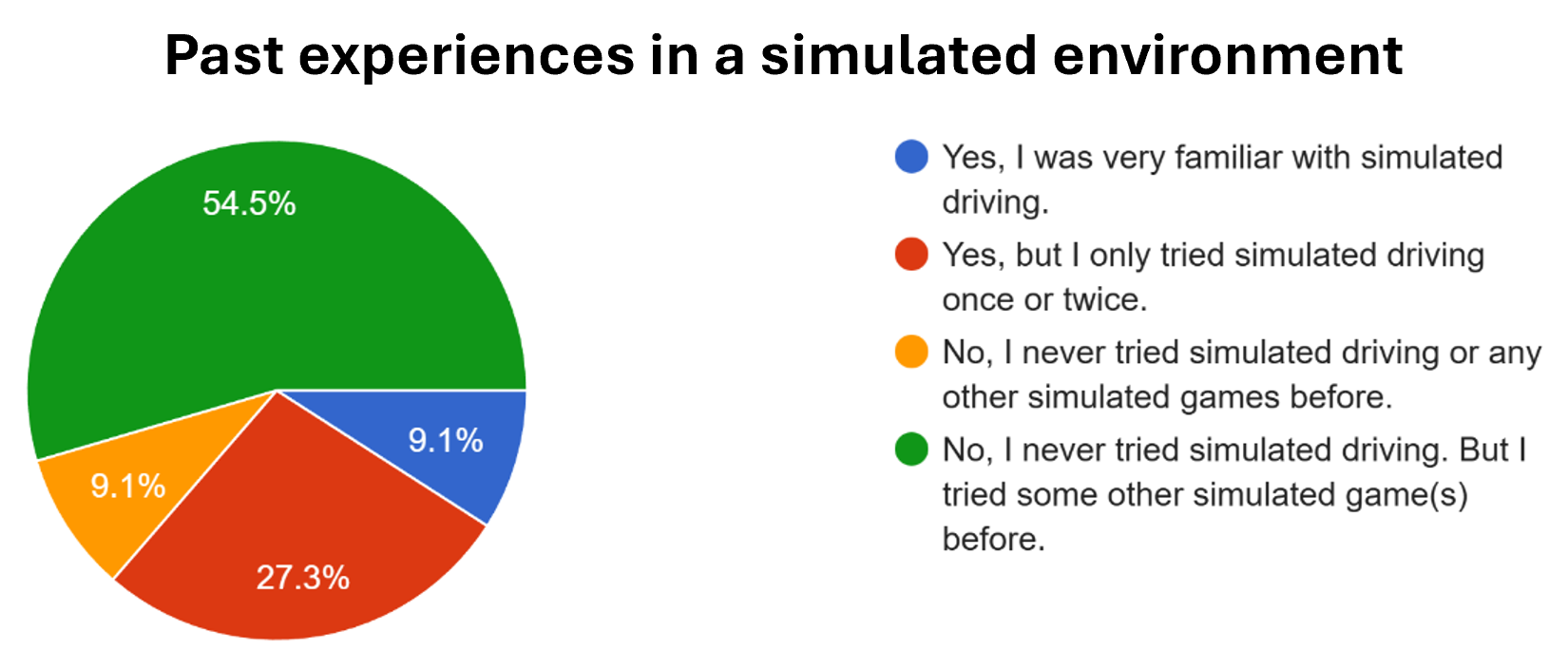}
         \label{Sfig:simulationExperiences}
     \end{subfigure}
     \hfill
     \begin{subfigure}[b]{0.48\textwidth}
         \centering
         \caption{}
         \includegraphics[width=\textwidth]{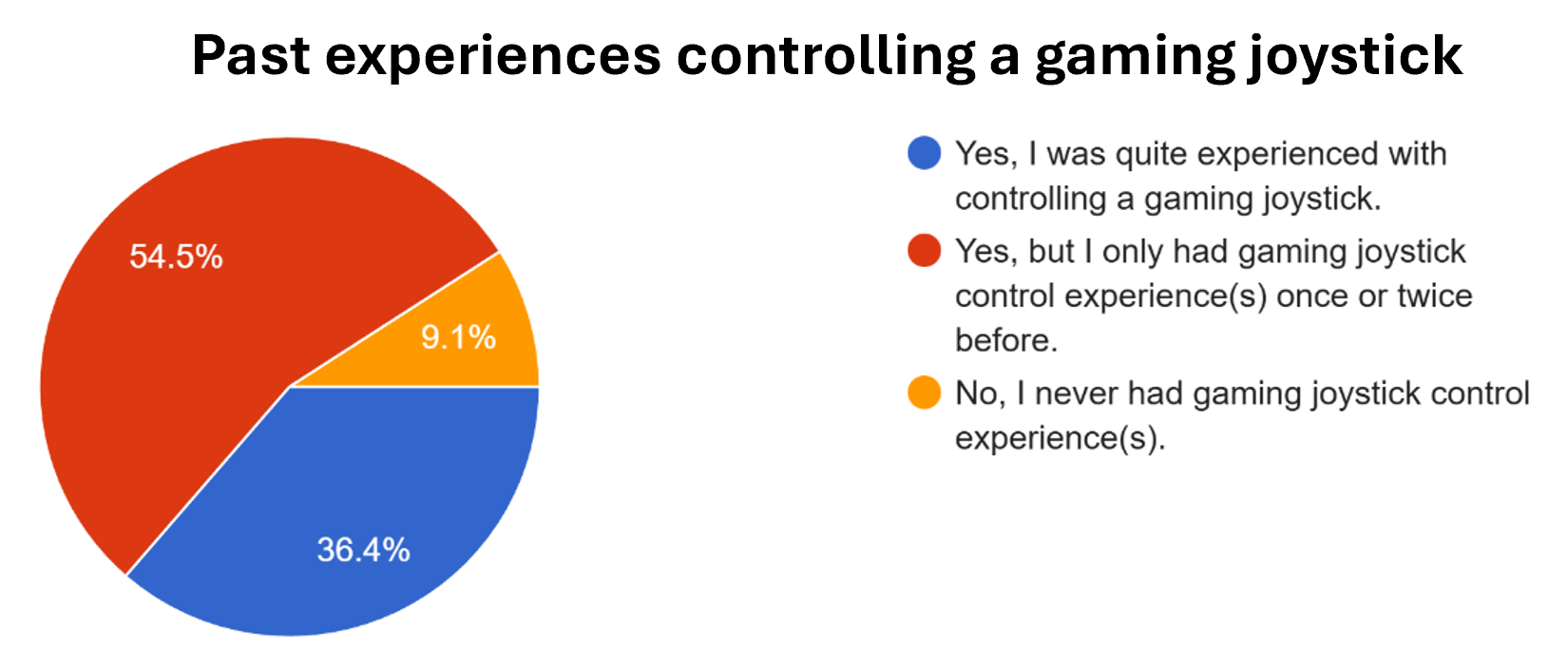}
         \label{Sfig:joystickExperiences}
     \end{subfigure}
     
     \begin{subfigure}[b]{0.48\textwidth}
         \centering
         \caption{}
         \includegraphics[width=\textwidth]{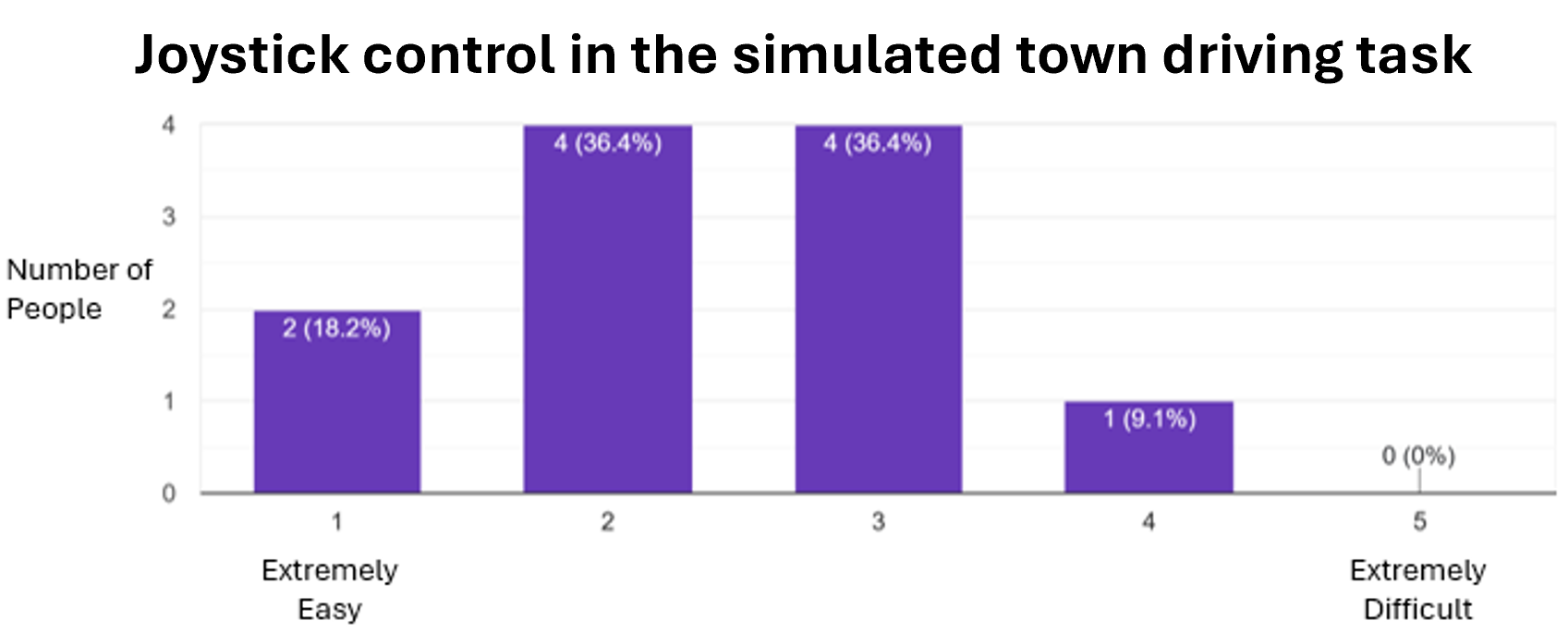}
         \label{Sfig:joystickInSimulation}
     \end{subfigure}
     \hfill
     \begin{subfigure}[b]{0.48\textwidth}
         \centering
         \caption{}
         \includegraphics[width=\textwidth]{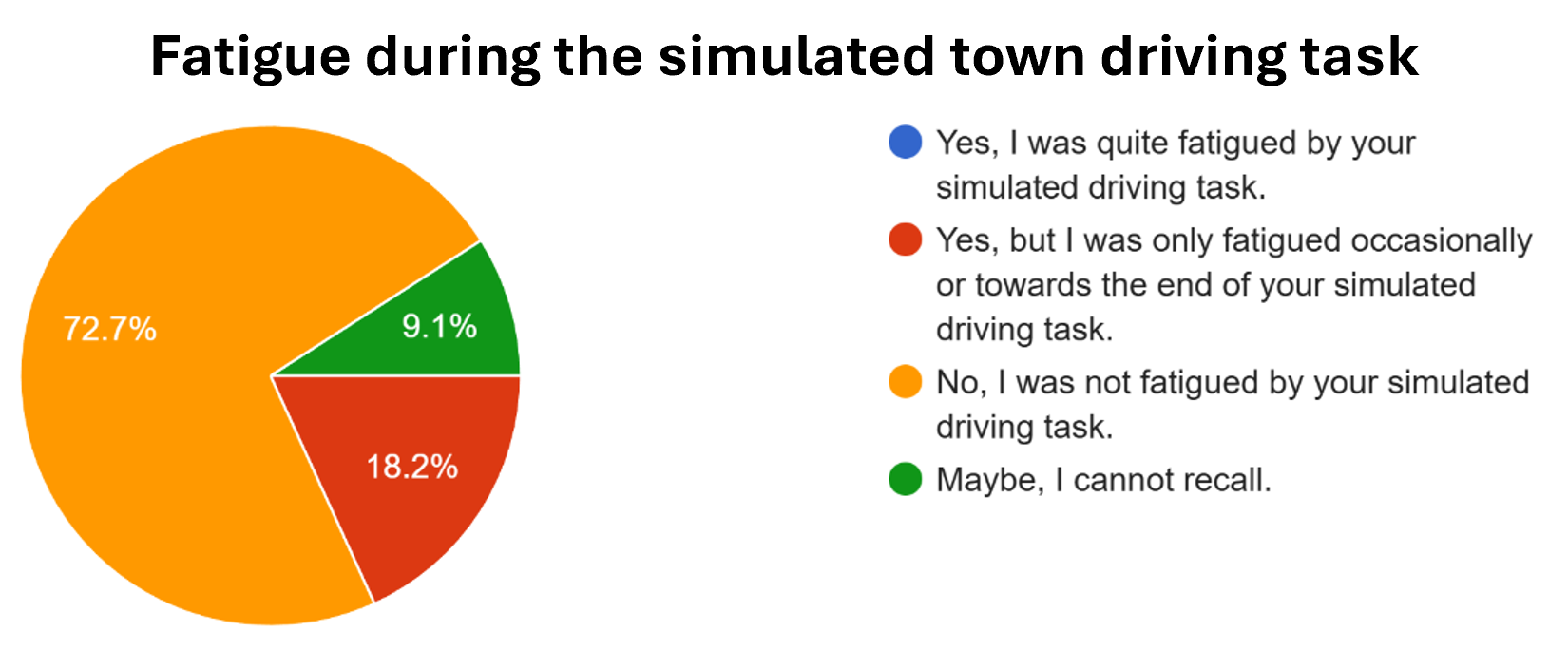}
         \label{Sfig:fatigue}
     \end{subfigure}
        \caption{\textbf{Results from a post-session questionnaire collected from 11 out of 20 motor intact participants after they completed the simulated town driving task.} (\textbf{A}) Whether they usually experience motion sickness in a physical vehicle. (\textbf{B}) Whether they experienced motion sickness during this task. (\textbf{C}) Whether they had any past experience(s) in driving in a simulated environment before participating in this task. (\textbf{D}) Whether they had any past experience(s) in controlling a gaming joystick (similar to the one they used in this task) before participating in this task. (\textbf{E}) How much difficulty they felt to control the gaming joystick during this task. (\textbf{F}) Whether they were fatigued by this task.}
        \label{Sfig:questionnaireResults}
\end{figure}

\begin{figure}
     \centering
     \begin{minipage}[b]{0.54\textwidth}
     \begin{subfigure}[b]{\textwidth}
         \centering
         \caption{}
         \includegraphics[width=\textwidth]{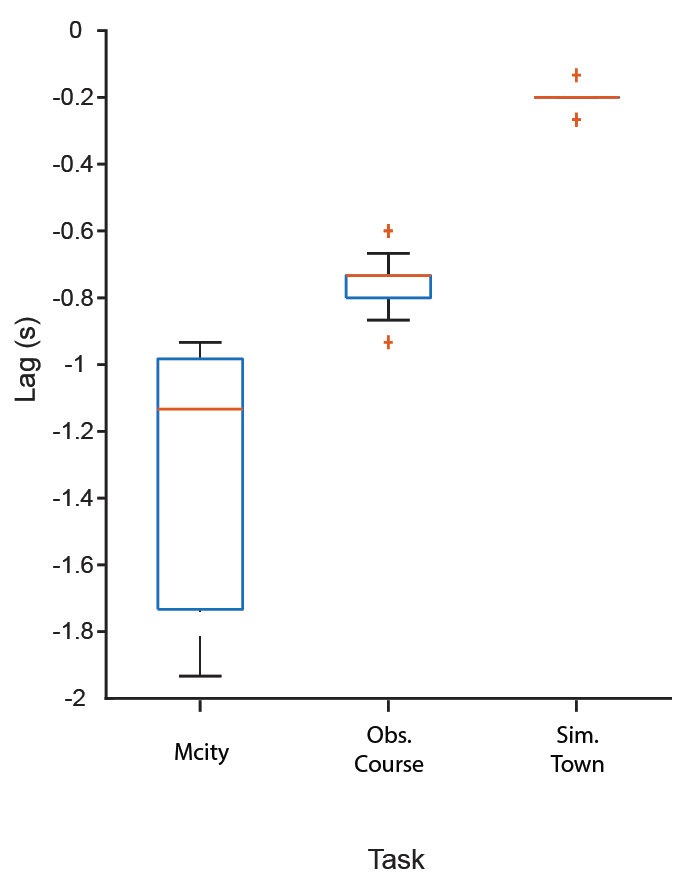}
         \label{Sfig:drivinglags}
     \end{subfigure}
     \end{minipage}
     \hfill
     \begin{minipage}[b]{0.44\textwidth}
     \begin{subfigure}[b]{\textwidth}
         \centering
         \caption{}
         \includegraphics[width=\textwidth]{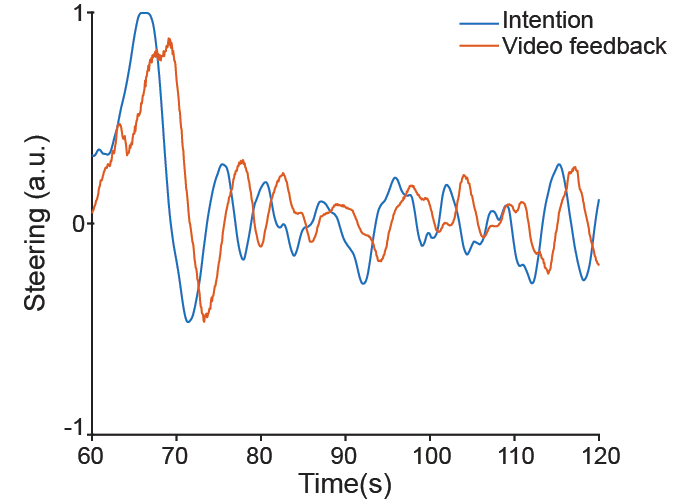}
         \label{Sfig:drivinglags-eg1}
     \end{subfigure}
     
     \begin{subfigure}[b]{\textwidth}
         \centering
         \caption{}
         \includegraphics[width=\textwidth]{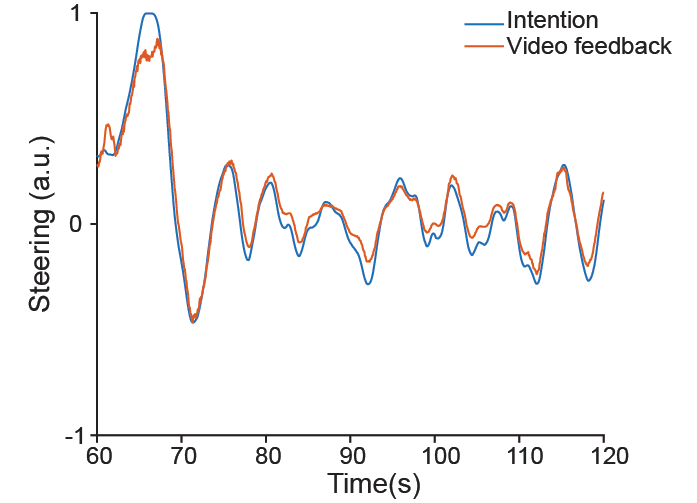}
         \label{Sfig:drivinglags-eg2}
     \end{subfigure}
     \end{minipage}
     \captionsetup{aboveskip=0pt}
        \caption{\textbf{Video lag analysis for three driving tasks.} (\textbf{A}) Boxplot of the lags between intention and video feedback during BCI driving sessions grouped by task, including the Mcity teledriving task, the obstacle-course teledriving task, and the simulated town driving task. Intention was obtained from the overlay state, and execution was calculated from the movement of the vehicle in videos. (\textbf{B}) Segment of the originally extracted steering intention and video feedback values from video recording in the Mcity teledriving task. (\textbf{C}) Aligned steering intention and video feedback values for the segment in (B) with an estimated lag of -193 ms.}
        \label{Sfig:lagResults}
\end{figure}

\clearpage


\begin{table}[ht!]
  \caption{\textbf{BCI trial-based simple reaction time task metrics, mean (SD), among different hand effectors of BCI-JJ.} We collected 10 runs for each of the six effectors (i.e., right and left index fingers, ring fingers, and power grips) consisting of 40 GO trials and 10 randomly interleaved NO-GO catch trials per run. The valid reaction time (RT) ranged from 50 ms to 1000 ms in GO trials.}
  \label{Stab:JJsimple}
  \centering
  \begin{tabular}{l|c|c|c|c}
    \toprule
    & Simple Valid RT (ms) & Accuracy ($\%$) & Sensitivity ($\%$) & Specificity ($\%$)\\
    \midrule
    Right index & 137 (26.8) & 90.8 (4.13) & 98.4 (2.82) & 73.3 (13.8)\\
    \midrule
    Left index & 151 (30.3) & 89.4 (5.42) & 98.9 (1.47) & 68.3 (14.1)\\
    \midrule
    Right Ring & 154 (31.0) & 88.0 (5.42) & 97.7 (3.14) & 66.1 (12.3)\\
    \midrule
    Left Ring & 157 (38.9) & 85.0 (9.35) & 98.1 (1.82) & 61.4 (17.2)\\
    \midrule
    Right Power & 168 (43.3) & 93.6 (2.46) & 96.4 (2.51) & 85.0 (6.19)\\
    \midrule
    Left Power & 167 (34.6) & 89.8 (4.85) & 96.7 (3.47) & 71.6 (9.80)\\
  \bottomrule
\end{tabular}
\end{table}

\begin{table}[ht!]
  \caption{\textbf{Pairwise comparison $p$ values for BCI trial-based simple reaction metrics among six hand effectors of BCI-JJ.} The $p$ values were obtained from a multiple comparison test using the results of one-way ANOVA. We collected 10 runs for each of the six effectors (i.e., right and left index fingers, ring fingers, and power grips) consisting of 40 GO trials and 10 randomly interleaved NO-GO catch trials per run. The valid reaction time (RT) ranged from 50 ms to 1000 ms in GO trials. Bonferroni correction for multiple comparisons was applied by setting the significance level to the corrected $p < 0.05$.}
  \label{Stab:JJsimpleANOVA}
  \centering
  \begin{tabular}{l|l|c|c|c|c}
    \toprule
    & & Simple Valid RT & Accuracy & Sensitivity & Specificity\\
    \midrule
    Right index & Left Index & $1.9\mathrm{e}{-6}$ & 1 & 1 & 1 \\
    Right index & Right Ring & $8.3\mathrm{e}{-10}$ & 1 & 1 & 1 \\
    Right index & Left Ring & $4.8\mathrm{e}{-13}$ & 0.39 & 1 & 0.62 \\
    Right index & Right Power & $1.2\mathrm{e}{-32}$ & 1 & 1 & 0.66 \\
    Right index & Left Power & $2.1\mathrm{e}{-29}$ & 1 & 1 & 1 \\
    \midrule
    Left index & Right Ring & 1 & 1 & 1 & 1 \\
    Left index & Left Ring & 0.26 & 1  & 1 & 1 \\
    Left index & Right Power & $9.5\mathrm{e}{-11}$ & 1  & 0.56 & 0.073 \\
    Left index & Left Power & $5.2\mathrm{e}{-9}$ & 1  & 1 & 1 \\
    \midrule
    Right Ring & Left Ring & 1 & 1  & 1 & 1 \\
    Right Ring & Right Power & $4.8\mathrm{e}{-7}$ & 0.47 & 1 & 0.023 \\
    Right Ring & Left Power & $1.1\mathrm{e}{-5}$ & 1 & 1 & 1 \\
    \midrule
    Left Ring & Right Power & $1.5\mathrm{e}{-4}$ & 0.020 & 1 & 0.0017 \\
    Left Ring & Left Power & 0.0019 & 0.95 & 1 & 1 \\
    \midrule
    Right Power & Left Power & 1 & 1 & 1 & 0.33 \\
  \bottomrule
\end{tabular}
\end{table}

\clearpage
\begin{table}[ht!]
  \caption{\textbf{BCI trial-based braking reaction time task metrics, mean (SD), between two index finger effectors of BCI-JJ.} We collected 10 runs per right or left index finger consisting of 40 trials per run, with one NO-GO phase and one GO phase in each trial. The valid reaction time (RT) ranged from 50 ms to 1000 ms in GO phases. }
  \label{Stab:JJbraking}
  \centering
  \begin{tabular}{l|c|c|c|c}
    \toprule
    & Braking Valid RT (ms) & Accuracy ($\%$) & Sensitivity ($\%$) & Specificity ($\%$)\\
    \midrule
    Right index & 290 (64.9) & 99.2 (0.919) & 99.7 (0.949) & 98.3 (1.89) \\
    \midrule
    Left index & 338 (82.2) & 99.9 (0.316) & 1 (0) & 99.7 (0.949) \\
  \bottomrule
\end{tabular}
\end{table}

\begin{table}[ht!]
  \caption{\textbf{Two-sample $t$-test $p$ values for trial-based braking reaction metrics, when equal variances were not assumed, between the right and left index finger effectors of BCI-JJ.} We collected 10 runs per right or left index finger consisting of 40 trials per run, with one NO-GO phase and one GO phase in each trial. The valid reaction time (RT) ranged from 50 ms to 1000 ms in GO phases. We used a left-tailed $t$-test to test against the alternative hypothesis that the population mean of the valid RT for the right index finger effector of BCI-JJ was less than that for his left index finger effector. We used a both-tailed $t$-test to test against the alternative hypothesis that the population means for each performance measure were not equal between right and left index finger effectors of BCI-JJ. If $p < 0.05$, the metrics had a significant impact on the performance.}
  \label{Stab:JJbrakingTTests}
  \centering
  \begin{tabular}{l|l|c|c|c|c}
    \toprule
    & & Braking Valid RT & Accuracy & Sensitivity & Specificity\\
    \midrule
    Right index & Left Index & $8.2\mathrm{e}{-19}$ & 0.044 & 0.34 & 0.056 \\
  \bottomrule
\end{tabular}
\end{table}

\begin{table}[ht!]
  \caption{\textbf{Trial-based simple reaction metrics, mean (SD), among BCI-JJ and 20 motor intact participants as a control group and as each individual with the same right index finger effector.} We collected 10 runs from BCI-JJ and 4 runs from each motor intact participant (M01\textendash M20). In the parenthesis after each participant's ID, the number represents their age, and the letter ``f'' or ``m'' represents their gender (female or male). The valid reaction time (RT) ranged from 50 ms to 1000 ms in GO trials.}
  \label{Stab:GroupSimple}
  \centering
  {\renewcommand{\arraystretch}{0.7}
  \begin{tabular}{l|c|c|c|c}
    \toprule
    & Simple Valid & Accuracy & Sensitivity & Specificity \\
    & RT (ms) & ($\%$) & ($\%$) & ($\%$)\\
    \midrule
    BCI-JJ (50m) & 137 (26.8) & 90.8 (4.13) & 98.4 (2.82) & 73.3 (13.8) \\
    \midrule
    Motor Intact M01\textendash M20 & 165 (52.3) & 96.2 (3.97) & 99.5 (1.24) & 87.5 (12.5) \\
    \midrule
    \midrule
    M01 (42f) & 182 (73.0) & 97.5 (2.52) & 100 (0) & 89.7 (9.53) \\
    \midrule
    M02 (48f) & 182 (35.3) & 99.0 (1.15) & 99.4 (1.25) & 97.7 (4.55) \\
    \midrule
    M03 (52m) & 148 (54.2) & 91.5 (3.79) & 99.3 (1.32) & 72.6 (10.4) \\
    \midrule
    M04 (36m) & 166 (46.2) & 99.0 (1.15) & 100 (0) & 95.5 (5.25) \\
    \midrule
    M05 (52f) & 182 (38.0) & 100 (0) & 100 (0) & 100 (0) \\
    \midrule
    M06 (35f) & 154 (33.3) & 96.0 (4.32) & 98.0 (3.95) & 89.4 (7.92) \\
    \midrule
    M07 (48f) & 129 (23.0) & 90.0 (1.63) & 100 (0) & 66.8 (3.65) \\
    \midrule
    M08 (47f) & 149 (42.9) & 96.5 (1.91) & 98.7 (1.46) & 89.7 (9.53) \\
    \midrule
    M09 (50f) & 229 (90.8) & 96.5 (3.00) & 98.7 (1.52) & 89.4 (7.92) \\
    \midrule
    M10 (50m) & 133 (23.6) & 92.0 (7.12) & 100 (0) & 75.0 (18.6) \\
    \midrule
    M11 (44m) & 146 (37.3) & 96.5 (1.91) & 100 (0) & 85.5 (6.75) \\
    \midrule
    M12 (43m) & 147 (30.7) & 95.0 (2.58) & 98.7 (1.48) & 84.5 (11.7) \\
    \midrule
    M13 (62f) & 143 (30.0) & 97.0 (2.58) & 100 (0) & 87.8 (9.95) \\
    \midrule
    M14 (61m) & 210 (57.7) & 98.5 (1.00) & 98.8 (1.44) & 97.9 (4.17) \\
    \midrule
    M15 (60f) & 197 (64.0) & 99.0 (1.15) & 100 (0) & 95.5 (5.25) \\
    \midrule
    M16 (49m) & 139 (35.2) & 89.0 (3.46) & 100 (0) & 65.1 (6.73) \\
    \midrule
    M17 (55m) & 161 (31.7) & 98.5 (1.91) & 99.4 (1.28) & 95.5 (5.25) \\
    \midrule
    M18 (58f) & 177 (31.8) & 99.0 (1.15) & 100 (0) & 95.5 (5.25) \\
    \midrule
    M19 (53m) & 162 (44.3) & 96.0 (2.83) & 99.4 (1.25) & 86.4 (12.1) \\
    \midrule
    M20 (53m) & 154 (28.1) & 97.5 (2.52) & 100 (0) & 89.7 (9.53) \\
  \bottomrule
\end{tabular}}
\end{table}

\begin{table}[ht!]
  \caption{\textbf{Two-sample $t$-test $p$ values for trial-based simple reaction metrics, when equal variances were not assumed, between BCI-JJ and the motor intact control group with the same right index finger effector.} We collected 10 runs from BCI-JJ and 4 runs from each of the 20 motor intact participants whose average age was the same as the age of BCI-JJ. There was no gender bias within the motor intact control group. During each run, there were 40 GO trials and 10 randomly interleaved NO-GO catch trials. The valid reaction time (RT) ranged from 50 ms and 1000 ms in GO trials. We used a left-tailed $t$-test to test against the alternative hypothesis that the population mean of the valid RT for BCI-JJ was less than that for the motor intact control group of 20 participants. We used a both-tailed $t$-test to test against the alternative hypothesis that the population means for each performance measure were not equal between BCI-JJ and the motor intact control group. If $p < 0.05$, the metrics had a significant impact on the performance.}
  \label{Stab:GroupSimpleTTest2}
  \centering
  \begin{tabular}{l|l|c|c|c|c}
    \toprule
    & & Simple Valid RT & Accuracy & Sensitivity & Specificity\\
    \midrule
   BCI-JJ & Motor Intact & $3.3\mathrm{e}{-53}$ & $2.3\mathrm{e}{-3}$ & 0.26 & 0.010 \\
  \bottomrule
\end{tabular}
\end{table}

\begin{table}[ht!]
  \caption{\textbf{Pairwise comparison $p$ values for trial-based simple reaction metrics between BCI-JJ and each of the 20 motor intact control participants with the same right index finger effector.} The $p$ values were obtained from a multiple comparison test using the results of one-way ANOVA. We collected 10 runs from BCI-JJ and 4 runs from each motor intact participant (M01\textendash M20). In the parenthesis after each participant's ID, the number represents their age, and the letter ``f'' or ``m'' represents their gender (female or male). The valid reaction time (RT) ranged from 50 ms and 1000 ms in GO trials. Bonferroni correction for multiple comparisons was applied by setting the significance level to the corrected $p < 0.05$.}
  \label{Stab:AGSimpleANOVA}
  \centering
  {\renewcommand{\arraystretch}{0.7}
  \begin{tabular}{l|l|c|c|c|c}
    \toprule
     & & Simple Valid RT & Accuracy & Sensitivity & Specificity\\
    \midrule
    BCI-JJ (50m) & M01 (42f) & $1.1\mathrm{e}{-23}$ & 0.0767 & 1 & 0.97 \\
    \midrule
    BCI-JJ (50m) & M02 (48f) & $1.9\mathrm{e}{-24}$ & 0.0041 & 1 & 0.0091 \\
    \midrule
    BCI-JJ (50m) & M03 (52m) & 1 & 1 & 1 & 1 \\
    \midrule
    BCI-JJ (50m) & M04 (36m) & $7.5\mathrm{e}{-10}$ & 0.0041 & 1 & 0.038 \\
    \midrule
    BCI-JJ (50m) & M05 (52f) & $4.5\mathrm{e}{-24}$ & $5.0\mathrm{e}{-4}$ & 1 & 0.0021 \\
    \midrule
    BCI-JJ (50m) & M06 (35f) & 0.0092 & 1 & 1 & 1 \\
    \midrule
    BCI-JJ (50m) & M07 (48f) & 1 & 1 & 1 & 1 \\
    \midrule
    BCI-JJ (50m) & M08 (47f) & 1 & 0.45 & 1 & 0.97 \\
    \midrule
    BCI-JJ (50m) & M09 (50f) & $1.1\mathrm{e}{-96}$ & 0.45 & 1 & 1 \\
    \midrule
    BCI-JJ (50m) & M10 (50m) & 1 & 1 & 1 & 1 \\
    \midrule
    BCI-JJ (50m) & M11 (44m) & 1 & 0.45 & 1 & 1 \\
    \midrule
    BCI-JJ (50m) & M12 (43m) & 1 & 1 & 1 & 1 \\
    \midrule
    BCI-JJ (50m) & M13 (62f) & 1 & 0.19 & 1 & 1 \\
    \midrule
    BCI-JJ (50m) & M14 (61m) & $1.6\mathrm{e}{-62}$ & 0.011 & 1 & 0.0081 \\
    \midrule
    BCI-JJ (50m) & M15 (60f) & $2.1\mathrm{e}{-43}$ & 0.0041 & 1 & 0.038 \\
    \midrule
    BCI-JJ (50m) & M16 (49m) & 1 & 1 & 1 & 1 \\
    \midrule
    BCI-JJ (50m) & M17 (55m) & $1.3\mathrm{e}{-6}$ & 0.0112 & 1 & 0.038 \\
    \midrule
    BCI-JJ (50m) & M18 (58f) & $2.8\mathrm{e}{-19}$ & 0.0041 & 1 & 0.038 \\
    \midrule
    BCI-JJ (50m) & M19 (53m) & $4.4\mathrm{e}{-7}$ & 1 & 1 & 1 \\
    \midrule
    BCI-JJ (50m) & M20 (53m) & 0.0164 & 0.077 & 1 & 0.97 \\
  \bottomrule
\end{tabular}}
\end{table}

\begin{table}[ht!]
  \caption{\textbf{Trial-based braking reaction metrics, mean (SD), among BCI-JJ and 20 motor intact participants as a control group and as each individual with the same right index finger effector.} We collected 10 runs from BCI-JJ and 5 runs from each motor intact participant (M01\textendash M20). In the parenthesis after each participant's ID, the number represents their age, and the letter ``f'' or ``m'' represents their gender (female or male). The valid reaction time (RT) ranged from 50 ms to 1000 ms in GO phases. }
  \label{Stab:GroupBraking}
  \centering
  {\renewcommand{\arraystretch}{0.7}
  \begin{tabular}{l|c|c|c|c}
    \toprule
    & Braking Valid & Accuracy & Sensitivity & Specificity \\
    & RT (ms) & ($\%$) & ($\%$) & ($\%$)\\
    \midrule
    BCI-JJ (50m) & 290 (64.9) & 99.2 (0.919) & 99.7 (0.949) & 98.3 (1.89) \\
    \midrule
    Motor Intact M01\textendash M20 & 370 (89.5) & 99.6 (0.770) & 99.8 (0.809) & 99.3 (1.35) \\
    \midrule
    \midrule
    M01 (42f) & 391 (115) & 99.7 (0.516) & 100 (0) & 99.0 (1.55) \\
    \midrule
    M02 (48f) & 397 (73.5) & 99.8 (0.447) & 100 (0) & 99.4 (1.34) \\
    \midrule
    M03 (52m) & 411 (122) & 98.8 (1.64) & 99.4 (1.34) & 98.4 (2.30) \\
    \midrule
    M04 (36m) & 336 (74.5) & 99.8 (0.447) & 100 (0) & 99.4 (1.34) \\
    \midrule
    M05 (52f) & 400 (56.8) & 99.6 (0.548) & 100 (0) & 98.8 (1.64) \\
    \midrule
    M06 (35f) & 288 (40.9) & 100 (0) & 100 (0) & 100 (0) \\
    \midrule
    M07 (48f) & 313 (60.3) & 100 (0) & 100 (0) & 100 (0) \\
    \midrule
    M08 (47f) & 348 (68.9) & 99.8 (0.447) & 100 (0) & 99.4 (1.34) \\
    \midrule
    M09 (50f) & 465 (105) & 99.6 (0.548) & 100 (0) & 98.8 (1.64) \\
    \midrule
    M10 (50m) & 382 (57.5) & 100 (0) & 100 (0) & 100 (0) \\
    \midrule
    M11 (44m) & 337 (80.5) & 99.2 (1.30) & 100 (0) & 98.4 (2.30) \\
    \midrule
    M12 (43m) & 329 (67.3) & 100 (0) & 100 (0) & 100 (0) \\
    \midrule
    M13 (62f) & 335 (66.7) & 99.6 (0.548) & 100 (0) & 98.8 (1.64) \\
    \midrule
    M14 (61m) & 436 (87.4) & 97.8 (1.10) & 96.6 (0.894) & 98.8 (1.64) \\
    \midrule
    M15 (60f) & 369 (99.8) & 99.7 (0.516) & 100 (0) & 99.0 (1.55) \\
    \midrule
    M16 (49m) & 385 (79.2) & 100 (0) & 100 (0) & 100 (0) \\
    \midrule
    M17 (55m) & 336 (50.4) & 100 (0) & 100 (0) & 100 (0) \\
    \midrule
    M18 (58f) & 408 (84.1) & 99.6 (0.548) & 100 (0) & 98.8 (1.64) \\
    \midrule
    M19 (53m) & 389 (80.8) & 100 (0) & 100 (0) & 100 (0) \\
    \midrule
    M20 (53m) & 346 (43.6) & 99.6 (0.548) & 100 (0) & 99.0 (1.41) \\
  \bottomrule
\end{tabular}}
\end{table}

\begin{table}[ht!]
  \caption{\textbf{Two-sample $t$-test $p$ values for trial-based braking reaction metrics, when equal variances were not assumed, between BCI-JJ and the motor intact control group with the same right index finger effector.} We collected 10 runs from BCI-JJ and 5 runs from each of the 20 motor intact participants whose average age was the same as the age of BCI-JJ. There was no gender bias within the motor intact control group. There were 40 trials per run, with one NO-GO phase and one GO phase in each trial. The valid reaction time (RT) ranged from 50 ms to 1000 ms in GO phases. We used a left-tailed $t$-test to test against the alternative hypothesis that the population mean of the valid RT for BCI-JJ was less than that for the motor intact control group of 20 participants. We used a both-tailed $t$-test to test against the alternative hypothesis that the population means for each performance measure were not equal between BCI-JJ and the motor intact control group. If $p < 0.05$, the metrics had a significant impact on the performance.}
  \label{Stab:GroupBrakingTTest2}
  \centering
  \begin{tabular}{l|l|c|c|c|c}
    \toprule
    & & Braking Valid RT & Accuracy & Sensitivity & Specificity\\
    \midrule
    BCI-JJ & Motor Intact & $1.4\mathrm{e}{-79}$ & 0.18 & 0.74 & 0.14 \\
  \bottomrule
\end{tabular}
\end{table}

\begin{table}[ht!]
  \caption{\textbf{Pairwise comparison $p$ values for trial-based simple reaction metrics between BCI-JJ and each of the 20 motor intact control participants with the same right index finger effector.} The $p$ values were obtained from a multiple comparison test using the results of one-way ANOVA. We collected 10 runs from BCI-JJ and 5 runs from each motor intact participant (M01\textendash M20). In the parenthesis after each participant's ID, the number represents their age, and the letter ``f'' or ``m'' represents their gender (female or male). The valid reaction time (RT) ranged from 50 ms to 1000 ms in GO trials. Bonferroni correction for multiple comparisons was applied by setting the significance level to the corrected $p < 0.05$.}
  \label{Stab:AGBrakingANOVA}
  \centering
  {\renewcommand{\arraystretch}{0.7}
  \begin{tabular}{l|l|c|c|c|c}
    \toprule
     & & Braking Valid RT & Accuracy & Sensitivity & Specificity\\
    \midrule
    BCI-JJ (50m) & M01 (42f) & $3.5\mathrm{e}{-52}$ & 1 & 1 & 1 \\
    \midrule
    BCI-JJ (50m) & M02 (48f) & $1.2\mathrm{e}{-52}$ & 1 & 1 & 1 \\
    \midrule
    BCI-JJ (50m) & M03 (52m) & $8.8\mathrm{e}{-64}$ & 1 & 1 & 1 \\
    \midrule
    BCI-JJ (50m) & M04 (36m) & $3.3\mathrm{e}{-9}$ & 1 & 1 & 1 \\
    \midrule
    BCI-JJ (50m) & M05 (52f) & $1.1\mathrm{e}{-55}$ & 1 & 1 & 1 \\
    \midrule
    BCI-JJ (50m) & M06 (35f) & 1 & 1 & 1 & 1 \\
    \midrule
    BCI-JJ (50m) & M07 (48f) & 0.13 & 1 & 1 & 1 \\
    \midrule
    BCI-JJ (50m) & M08 (47f) & $2.1\mathrm{e}{-15}$ & 1 & 1 & 1 \\
    \midrule
    BCI-JJ (50m) & M09 (50f) & $1.4\mathrm{e}{-133}$ & 1 & 1 & 1 \\
    \midrule
    BCI-JJ (50m) & M10 (50m) & $2.7\mathrm{e}{-39}$ & 1 & 1 & 1 \\
    \midrule
    BCI-JJ (50m) & M11 (44m) & $1.0\mathrm{e}{-9}$ & 1 & 1 & 1 \\
    \midrule
    BCI-JJ (50m) & M12 (43m) & $2.1\mathrm{e}{-6}$ & 1 & 1 & 1 \\
    \midrule
    BCI-JJ (50m) & M13 (62f) & $6.5\mathrm{e}{-9}$ & 1 & 1 & 1 \\
    \midrule
    BCI-JJ (50m) & M14 (61m) & $5.1\mathrm{e}{-94}$ & 0.054 & $3.0\mathrm{e}{-19}$ & 1 \\
    \midrule
    BCI-JJ (50m) & M15 (60f) & $2.5\mathrm{e}{-29}$ & 1 & 1 & 1 \\
    \midrule
    BCI-JJ (50m) & M16 (49m) & $1.7\mathrm{e}{-41}$ & 1 & 1 & 1 \\
    \midrule
    BCI-JJ (50m) & M17 (55m) & $1.4\mathrm{e}{-9}$ & 1 & 1 & 1 \\
    \midrule
    BCI-JJ (50m) & M18 (58f) & $4.3\mathrm{e}{-64}$ & 1 & 1 & 1 \\
    \midrule
    BCI-JJ (50m) & M19 (53m) & $1.6\mathrm{e}{-45}$ & 1 & 1 & 1 \\
    \midrule
    BCI-JJ (50m) & M20 (53m) & $1.2\mathrm{e}{-14}$ & 1 & 1 & 1 \\
  \bottomrule
\end{tabular}}
\end{table}

\begin{table}[ht!]
  \caption{\textbf{Simulated driving scores and infractions, mean (SD), and two-sample $t$-test $p$ values between BCI-JJ and the motor intact control group using the same bimanual control.} We collected 10 runs from BCI-JJ and 2 runs from each of the 20 motor intact participants whose average age was the same as the age of BCI-JJ. There was no gender bias within the motor intact control group. During each run of navigating a virtual vehicle in a simulated town environment with heavy traffic and a route modified from the CARLA Leaderboard 2.0, a participant used the right thumb for cursor movement to control its steering and speed and the left index finger for clicks to control its full-stop braking. The simulated driving scores ranged from 0 to 1.0. The proportion of the completed route distance (i.e., $C$ for route completion) could be a decimal between 0 and 1.0 with 1.0 as full completion of the route. The rest infraction components, which described the occurrences of collisions, running red lights, and lane deviations (i.e., $N_c$, $N_l$ and $N_s$ respectively) per run, were all nonnegative numbers. We used a right-tailed $t$-test to test against the alternative hypothesis that the population mean of the simulated driving score for BCI-JJ was greater than that for the motor intact control group of 20 participants. We used a both-tailed $t$-test to test against the alternative hypothesis that the population means for each infraction component were not equal between BCI-JJ and the motor intact control group. If $p < 0.05$, the metrics had a significant impact on the performance.}
  \label{Stab:GroupLeaderboard}
  \centering
  \begin{tabular}{l|c|c|c}
    \toprule
    & BCI-JJ & Motor Intact & ($t$-test $p$-value)\\
    \midrule
    Simulated Driving Score & 0.924 (0.115) & 0.823 (0.160) & 0.017 \\
    \midrule
    Route Completion ($C$) & 1 (0) & 1 (0) & NaN \\
    \midrule
    Number of Collisions ($N_c$) & 0.200 (0.422) & 0.775 (0.862) & $5.2\mathrm{e}{-3}$ \\
    \midrule
    Number of Lane Deviations ($N_l$) & 0.100 (0.316) & 0.075 (0.267) & 0.82 \\
    \midrule
    Number of Running Red Lights ($N_s$) & 0.300 (0.483) & 0.350 (0.700) & 0.79 \\
  \bottomrule
\end{tabular}
\end{table}

\begin{table}[ht!]
  \caption{\textbf{Route lengths for three driving tasks.} The route lengths were estimated from route maps for the Mcity teledriving task, the obstacle-course teledriving task, and the simulated town driving task. CW: clockwise. CCW: counterclockwise.}
  \label{Stab:route-length}
  \centering
  \begin{tabular}{l|l|c}
    \toprule
    \bf{Task Name} & \bf{Route Option} & \bf{Route Length (mi)} \\
    \midrule
    \multirow{4}{*}{Mcity Teledriving} & Mcity Route 1 & 0.142 \\ \cline{2-3}
     & Mcity Route 2 & 0.440 \\ \cline{2-3}
     & Mcity Route 3 & 0.197 \\ \cline{2-3}
     & Mcity Route 4 & 0.165 \\ 
    \midrule
    \multirow{2}{*}{Obstacle-course Teledriving} & \makecell[l]{A course loop \\ with a lane switch \\ (either CW or CCW)} & 0.116 \\ \cline{2-3}
     & \makecell[l]{A course loop \\ without any lane switch \\ (either CW or CCW)} & 0.115 \\ 
    \midrule
    Simulated Town Driving & A virtual town route & 0.967 \\
  \bottomrule
\end{tabular}
\end{table}

\begin{table}[ht!]
  \caption{\textbf{Safety criteria for driving tasks.} We set up the safety criteria for all three driving tasks to include some basic components of a standard driving test, in accordance with part of the evaluation and metrics from CARLA Autonomous Driving Leaderboard 2.0 and part of the checklist in the ``Driving Performance Evaluation Score Sheet'' from the California Department of Motor Vehicles (DMV) \cite{department_of_motor_vehicles_california_2025}.}
  \label{Stab:safety-components}
  \centering
  \begin{tabular}{l|l|l|l}
    \toprule
    \bf{Infraction} & \bf{Infraction} & \bf{Corresponding} & \bf{California DMV Related} \\
    \bf{Components} & \bf{Descriptions} & \bf{Safety Criteria} & \bf{Performance Checklist} \\
    \midrule
    Collisions & \makecell[l]{Collisions with \\ other vehicles \\ (simulation), or \\
static elements \\ (e.g., cones, \\ bumps)} & \makecell[l]{No collisions with any \\ vehicle or static elements \\ (e.g., bumps, buildings) in \\ the simulator. No actual or \\ potential collisions with \\ any manual vehicle reverse \\ during teledriving.} & \makecell[l]{\textbullet~``Keep an adequate space \\ cushion between vehicles'' \\ \textbullet~``Yield to oncoming traffic \\ when appropriate''} \\
    \midrule
    \makecell[l]{Lane \\ Deviations} & \makecell[l]{Off-lane driving \\ from a straight \\ or curved lane \\ (e.g., a turn, a \\ roundabout)} & \makecell[l]{In-lane driving on a straight \\ or curved lane (e.g., a turn, \\ a roundabout) for all four \\ wheels.} & \makecell[l]{\textbullet~``Begin and end turns in \\ correct lane'' \\ \textbullet~``Do not cut turns too short'' \\ \textbullet~``Do not make turns too \\ wide'' \\ \textbullet~``Keep in center of lane'' \\ \textbullet~``Do not under-steer'' \\ \textbullet~``Do not over-steer''} \\
    \midrule
    \makecell[l]{Running Red \\ Traffic Lights \\ or Stop Signs} & \makecell[l]{Running a red \\light or a stop \\ sign without a \\ full stop behind \\ the limit line} & \makecell[l]{An in-time full stop right \\ behind the limit line at a \\ stop sign or while waiting \\ at the red traffic light.} & \makecell[l]{\textbullet~``Make full stops behind \\ limit lines'' \\ \textbullet~``Do not make unnecessary \\ stops'' \\ \textbullet~``Make smooth safe stops''} \\
  \bottomrule
\end{tabular}
\end{table}


\clearpage 

\makeatletter
\renewcommand{\fnum@figure}{\textbf{Caption for Movie \thefigure}}
\makeatother

\renewcommand{\thefigure}{S\arabic{figure}}
\setcounter{figure}{0}

\begin{figure}[ht!]
\centering
\includegraphics[width=0.29\textwidth]{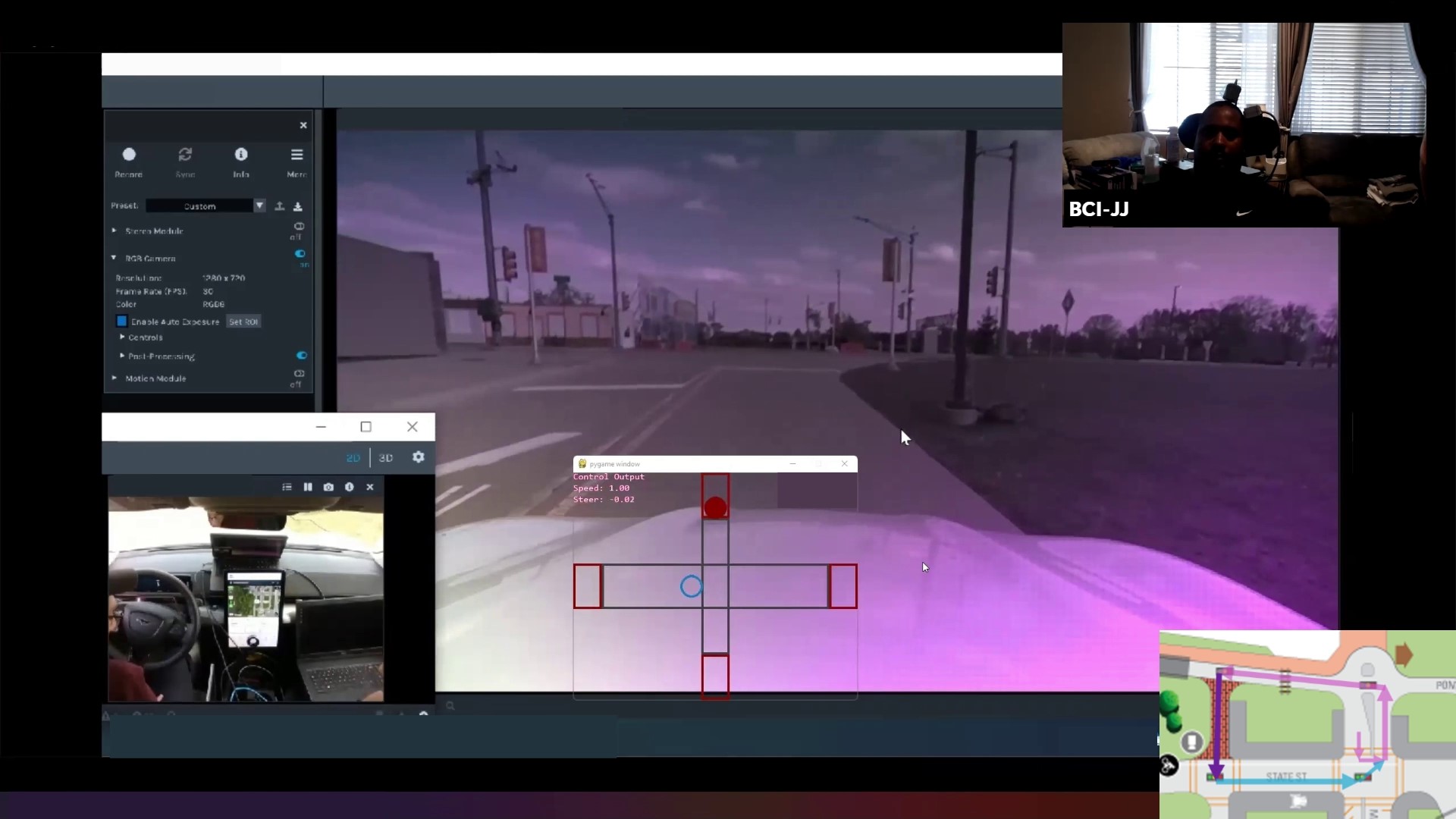}
\caption{\textbf{BCI teledriving of a commercially available Ford Mustang Mach-E via Route 1 in Mcity by BCI-JJ using the single-effector cursor control.} We used a camera mounted outside the front windshield of the vehicle in Michigan for continuous recording of the view ahead and real-time video feedback to BCI-JJ (top-right) in California. As shown on the overlay (bottom-middle), the cursor movement was decoded with BCI-JJ's right thumb for steering and speed control of the vehicle. If the vehicle was about to collide with some dangerous obstacle and/or experienced more than two seconds of aggregated lags, the safety driver inside the vehicle (bottom-left) would take over the vehicle control for a few seconds to move it back to the right track before the BCI teledriving could resume. Route 1 map (bottom-right) has the current segment labeled in dark purple. The video is played back at 4 times the speed of the original recording.}
\label{Smovie:mcity-route1}
\end{figure}

\begin{figure}[ht!]
\centering
\includegraphics[width=0.29\textwidth]{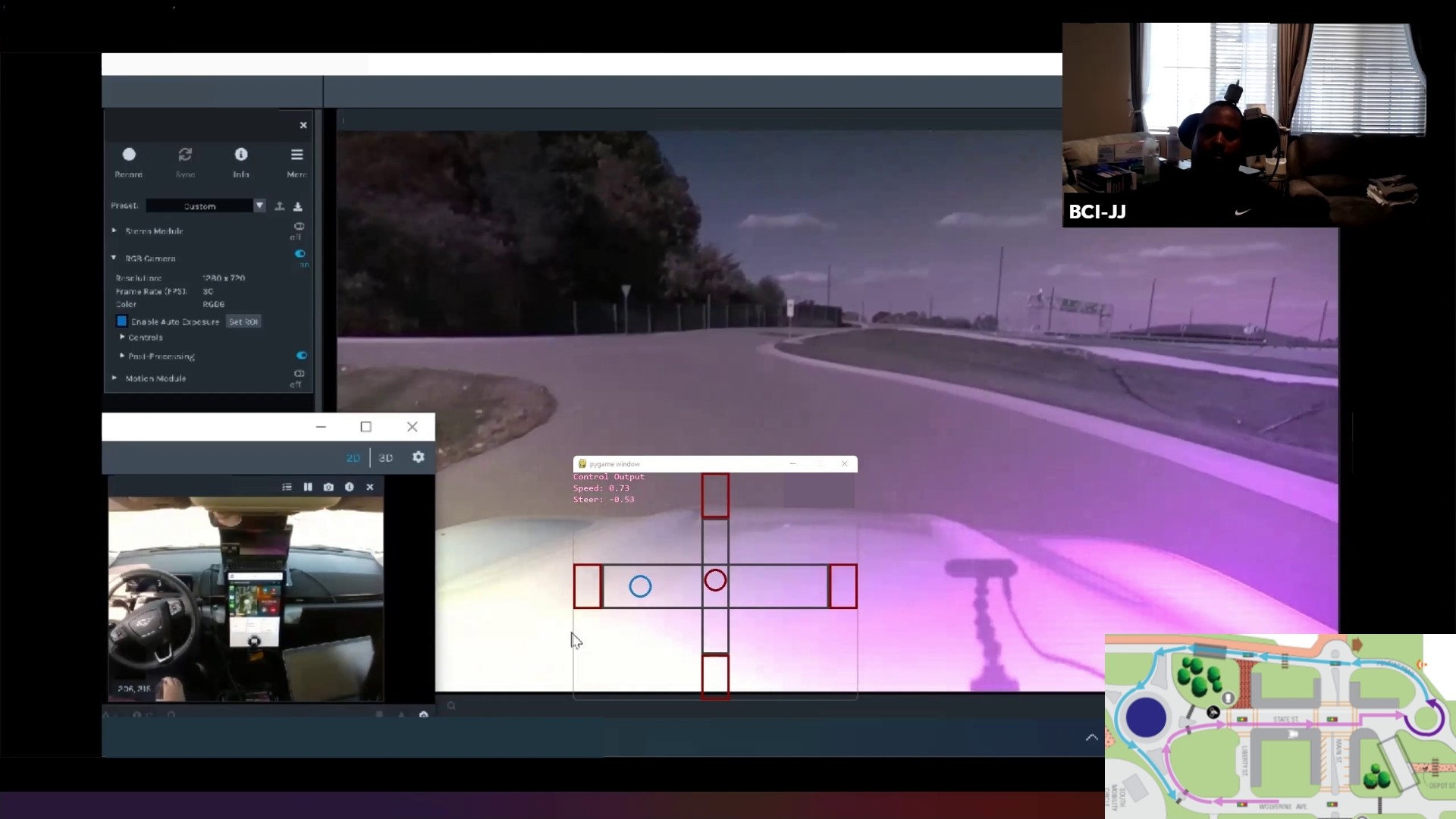}
\caption{\textbf{BCI teledriving of a commercially available Ford Mustang Mach-E via Route 2 in Mcity by BCI-JJ using the single-effector cursor control.} We used a camera mounted outside the front windshield of the vehicle in Michigan for continuous recording of the view ahead and real-time video feedback to BCI-JJ (top-right) in California. As shown on the overlay (bottom-middle), the cursor movement was decoded with BCI-JJ's right thumb for steering and speed control of the vehicle. If the vehicle was about to collide with some dangerous obstacle and/or experienced more than two seconds of aggregated lags, the safety driver inside the vehicle (bottom-left) would take over the vehicle control for a few seconds to move it back to the right track before the BCI teledriving could resume. Route 2 map (bottom-right) has the current segment labeled in dark purple. The video is played back at 4 times the speed of the original recording.}
\label{Smovie:mcity-route2}
\end{figure}

\clearpage
\begin{figure}[ht!]
\centering
\includegraphics[width=0.29\textwidth]{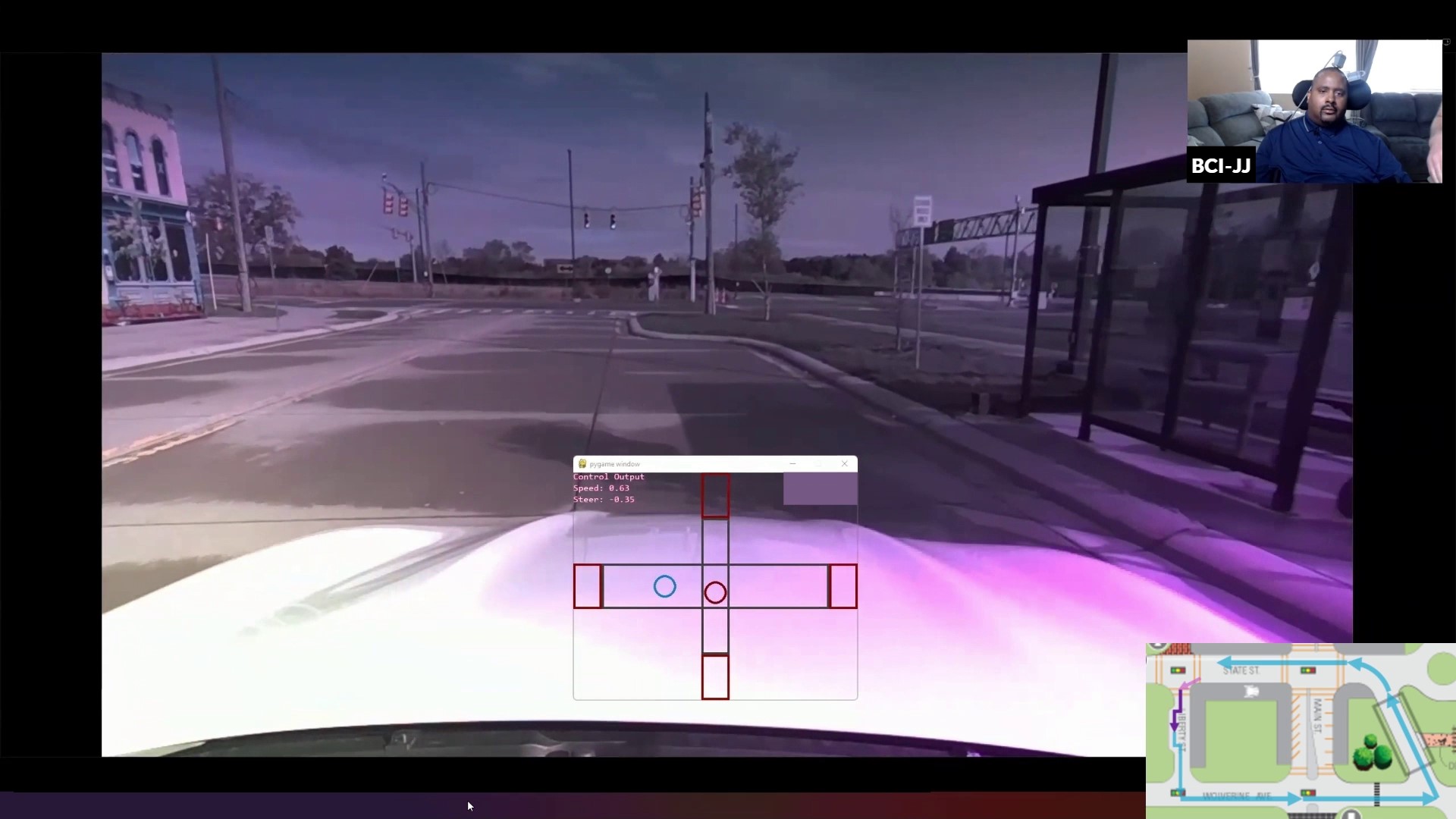}
\caption{\textbf{BCI teledriving of a commercially available Ford Mustang Mach-E via Route 3 in Mcity by BCI-JJ using the single-effector cursor control.} We used a camera mounted outside the front windshield of the vehicle in Michigan for continuous recording of the view ahead and real-time video feedback to BCI-JJ (top-right) in California. As shown on the overlay (bottom-middle), the cursor movement was decoded with BCI-JJ's right thumb for steering and speed control of the vehicle. If the vehicle was about to collide with some dangerous obstacle and/or experienced more than two seconds of aggregated lags, the safety driver inside the vehicle would take over the vehicle control for a few seconds to move it back to the right track before the BCI teledriving could resume. Route 3 map (bottom-right) has the current segment labeled in dark purple. The video is played back at 4 times the speed of the original recording.}
\label{Smovie:mcity-route3}
\end{figure}

\begin{figure}[ht!]
\centering
\includegraphics[width=0.29\textwidth]{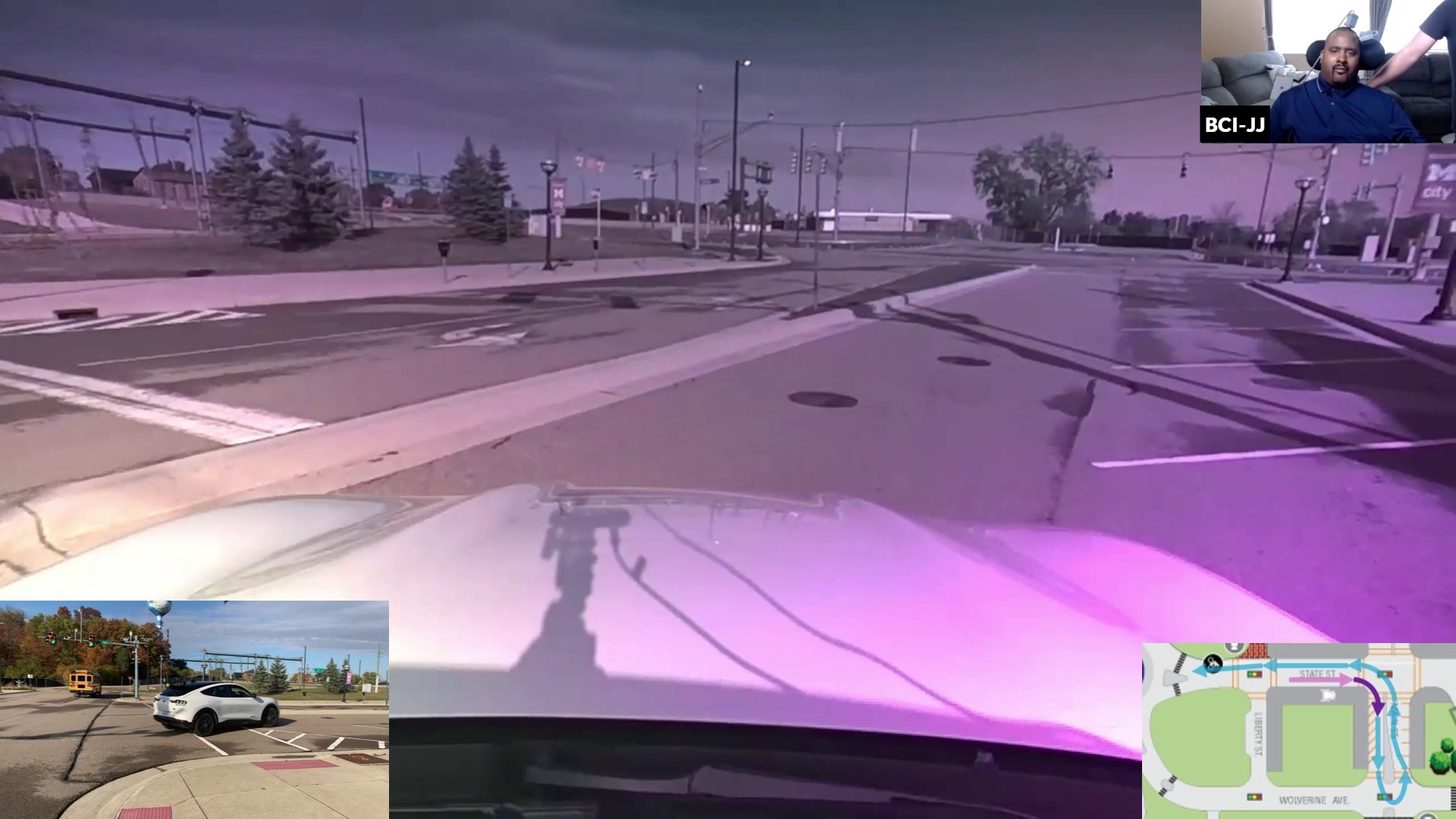}
\caption{\textbf{BCI teledriving of a commercially available Ford Mustang Mach-E via Route 4 in Mcity by BCI-JJ using the single-effector cursor control.} We used a camera mounted outside the front windshield of the vehicle in Michigan for continuous recording of the view ahead and real-time video feedback to BCI-JJ (top-right) in California. The cursor movement was decoded with BCI-JJ's right thumb for steering and speed control of the vehicle. If the vehicle was about to collide with some dangerous obstacle and/or experienced more than two seconds of aggregated lags, the safety driver inside the vehicle (bottom-left) would take over the vehicle control for a few seconds to move it back to the right track before the BCI teledriving could resume. Route 4 map (bottom-right) has the current segment labeled in dark purple. The video is played back at 4 times the speed of the original recording.}
\label{Smovie:mcity-route4}
\end{figure}

\begin{figure}[ht!]
\centering
\includegraphics[width=0.35\textwidth]{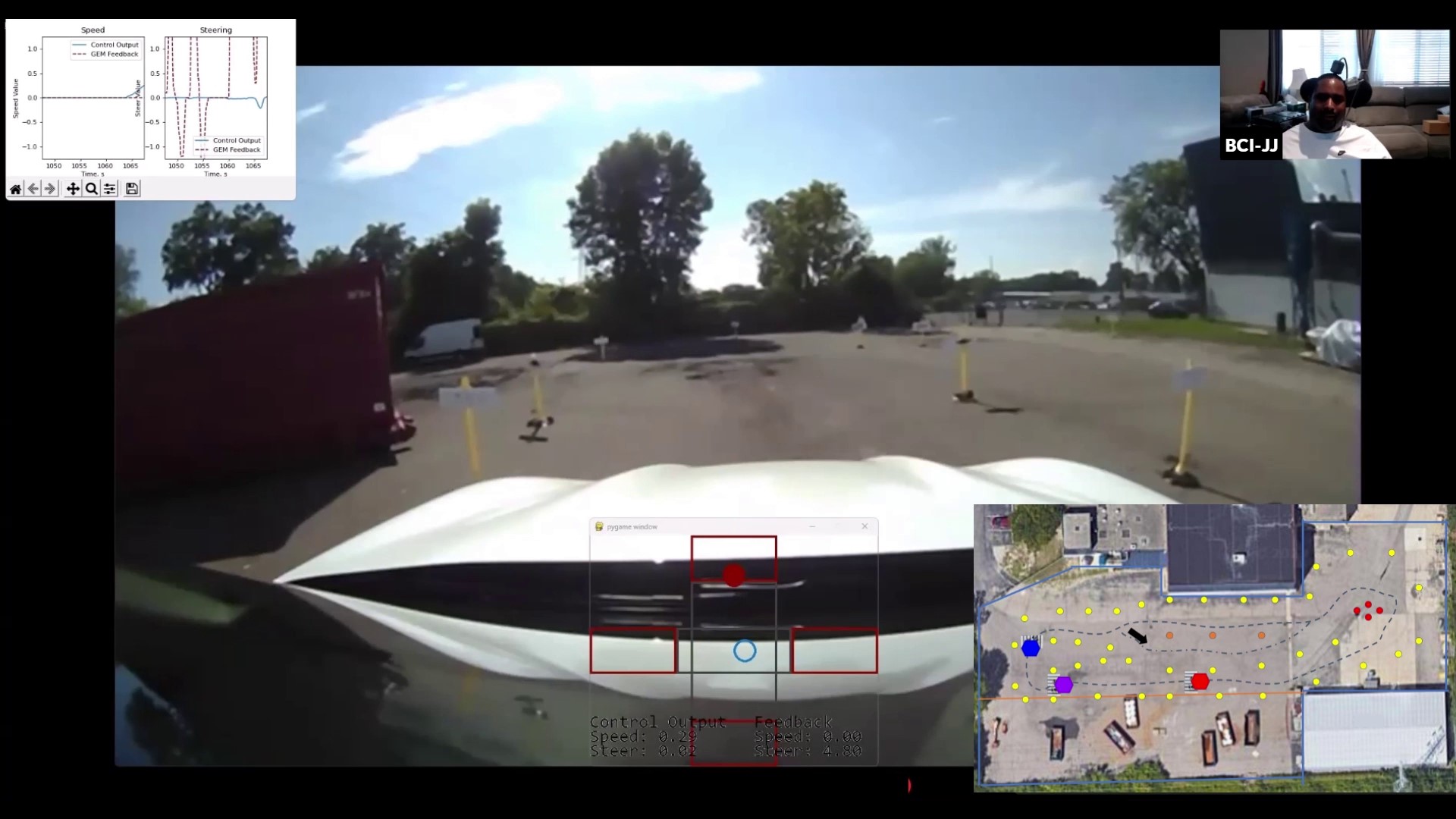}
\caption{\textbf{BCI teledriving of a commercially available Ford Mustang Mach-E through the obstacle course clockwise with a lane-switch segment by BCI-JJ using the single-effector cursor control.} The example run had its teledriving score closest to the average scores for this route option. We used the camera mounted within the vehicle in Michigan for continuous recording of the view ahead and real-time video feedback to BCI-JJ (top-right) in California. As shown on the overlay (bottom-middle), the cursor movement was decoded with BCI-JJ's right thumb for steering and speed control of the vehicle, whereas the click information was decoded with his left index finger for its full-stop braking. The route map (bottom-right) has the chosen direction of the loop and the chosen lane switch option labeled with the black arrow. The video is played back at twice the speed of the original recording.}
\label{Smovie:obstacle-course-cw-ls}
\end{figure}

\begin{figure}[ht!]
\centering
\includegraphics[width=0.35\textwidth]{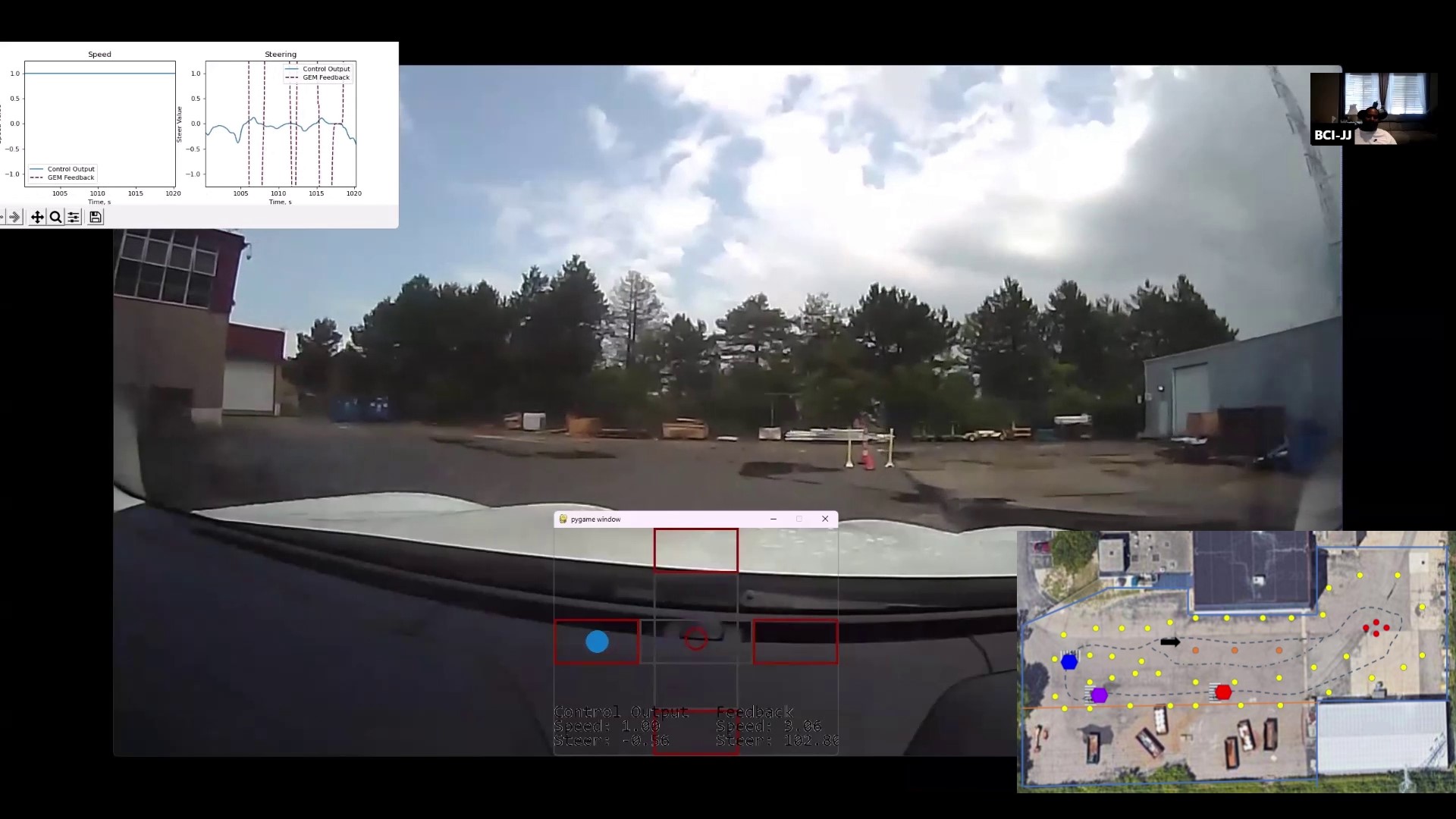}
\caption{\textbf{BCI teledriving of a commercially available Ford Mustang Mach-E through the obstacle course clockwise without any lane switch by BCI-JJ using the single-effector cursor control.} The example run had its teledriving score closest to the average scores for this route option. We used the camera mounted within the vehicle in Michigan for continuous recording of the view ahead and real-time video feedback to BCI-JJ (top-right) in California. As shown on the overlay (bottom-middle), the cursor movement was decoded with BCI-JJ's right thumb for steering and speed control of the vehicle, whereas the click information was decoded with his left index finger for its full-stop braking. The route map (bottom-right) has the chosen direction of the loop and the chosen lane switch option labeled with the black arrow. The video is played back at twice the speed of the original recording.}
\label{Smovie:obstacle-course-cw-nols}
\end{figure}

\begin{figure}[ht!]
\centering
\includegraphics[width=0.35\textwidth]{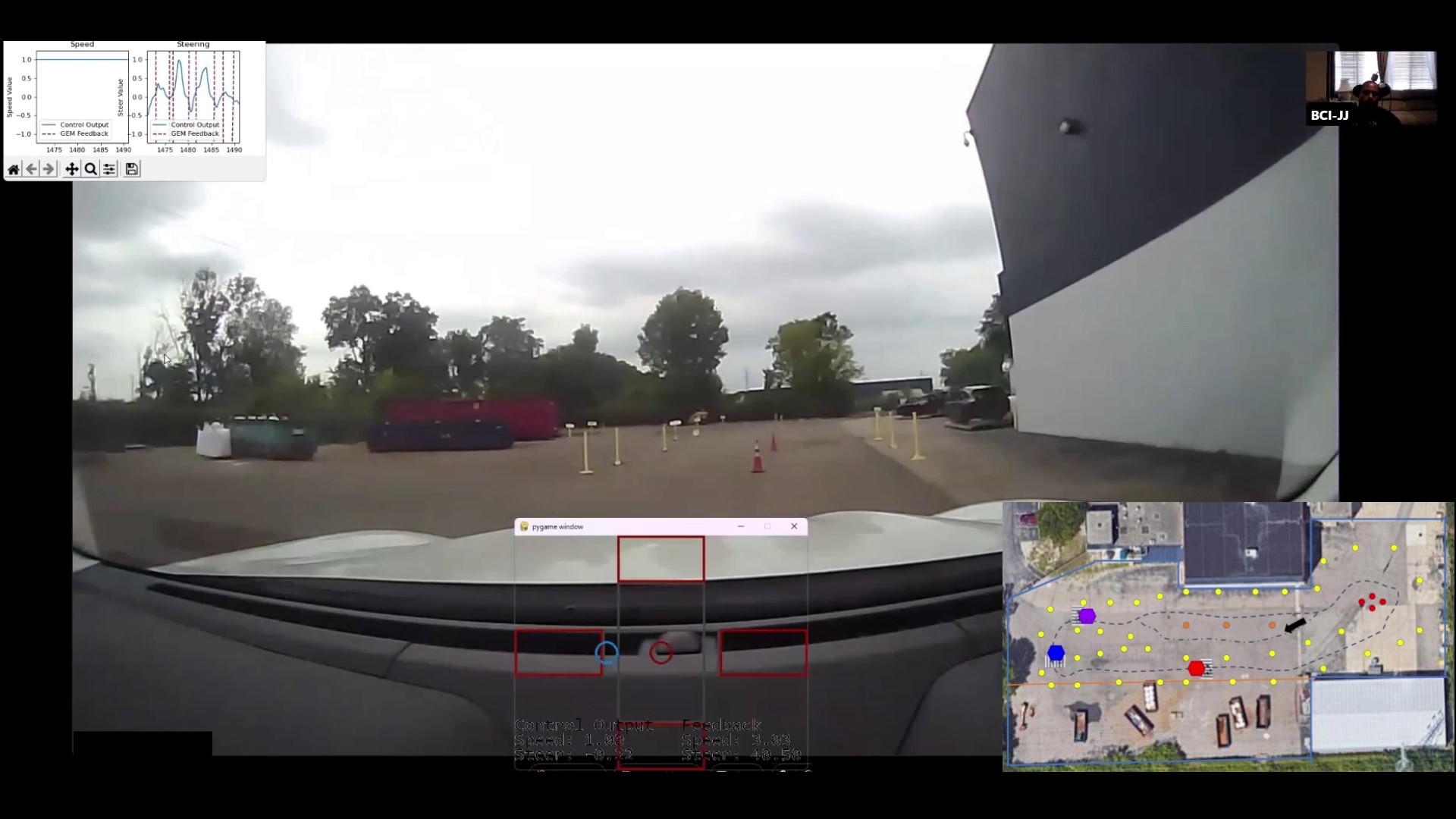}
\caption{\textbf{BCI teledriving of a commercially available Ford Mustang Mach-E through the obstacle course counterclockwise with a lane-switch segment by BCI-JJ using the single-effector cursor control.} The example run had its teledriving score closest to the average scores for this route option. We used the camera mounted within the vehicle in Michigan for continuous recording of the view ahead and real-time video feedback to BCI-JJ (top-right) in California. As shown on the overlay (bottom-middle), the cursor movement was decoded with BCI-JJ's right thumb for steering and speed control of the vehicle, whereas the click information was decoded with his left index finger for its full-stop braking. The route map (bottom-right) has the chosen direction of the loop and the chosen lane switch option labeled with the black arrow. The video is played back at twice the speed of the original recording.}
\label{Smovie:obstacle-course-ccw-ls}
\end{figure}

\begin{figure}[ht!]
\centering
\includegraphics[width=0.35\textwidth]{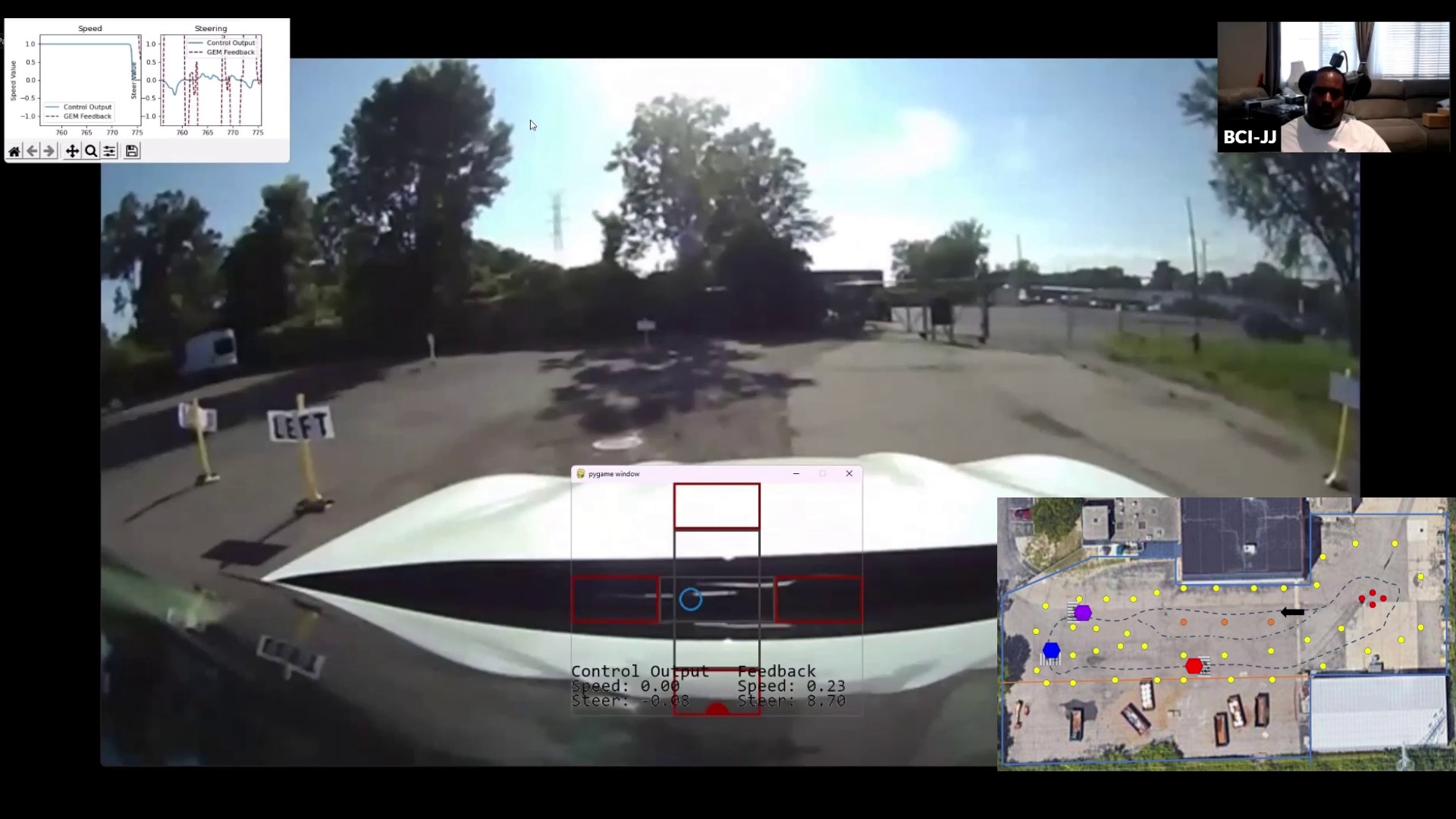}
\caption{\textbf{BCI teledriving of a commercially available Ford Mustang Mach-E through the obstacle course counterclockwise without any lane switch by BCI-JJ using the single-effector cursor control.} The example run had its teledriving score closest to the average scores for this route option. We used the camera mounted within the vehicle in Michigan for continuous recording of the view ahead and real-time video feedback to BCI-JJ (top-right) in California. As shown on the overlay (bottom-middle), the cursor movement was decoded with BCI-JJ's right thumb for steering and speed control of the vehicle, whereas the click information was decoded with his left index finger for its full-stop braking. The route map (bottom-right) has the chosen direction of the loop and the chosen lane switch option labeled with the black arrow. The video is played back at twice the speed of the original recording.}
\label{Smovie:obstacle-course-ccw-nols}
\end{figure}

\begin{figure}[ht!]
\centering
\includegraphics[width=0.5\textwidth]{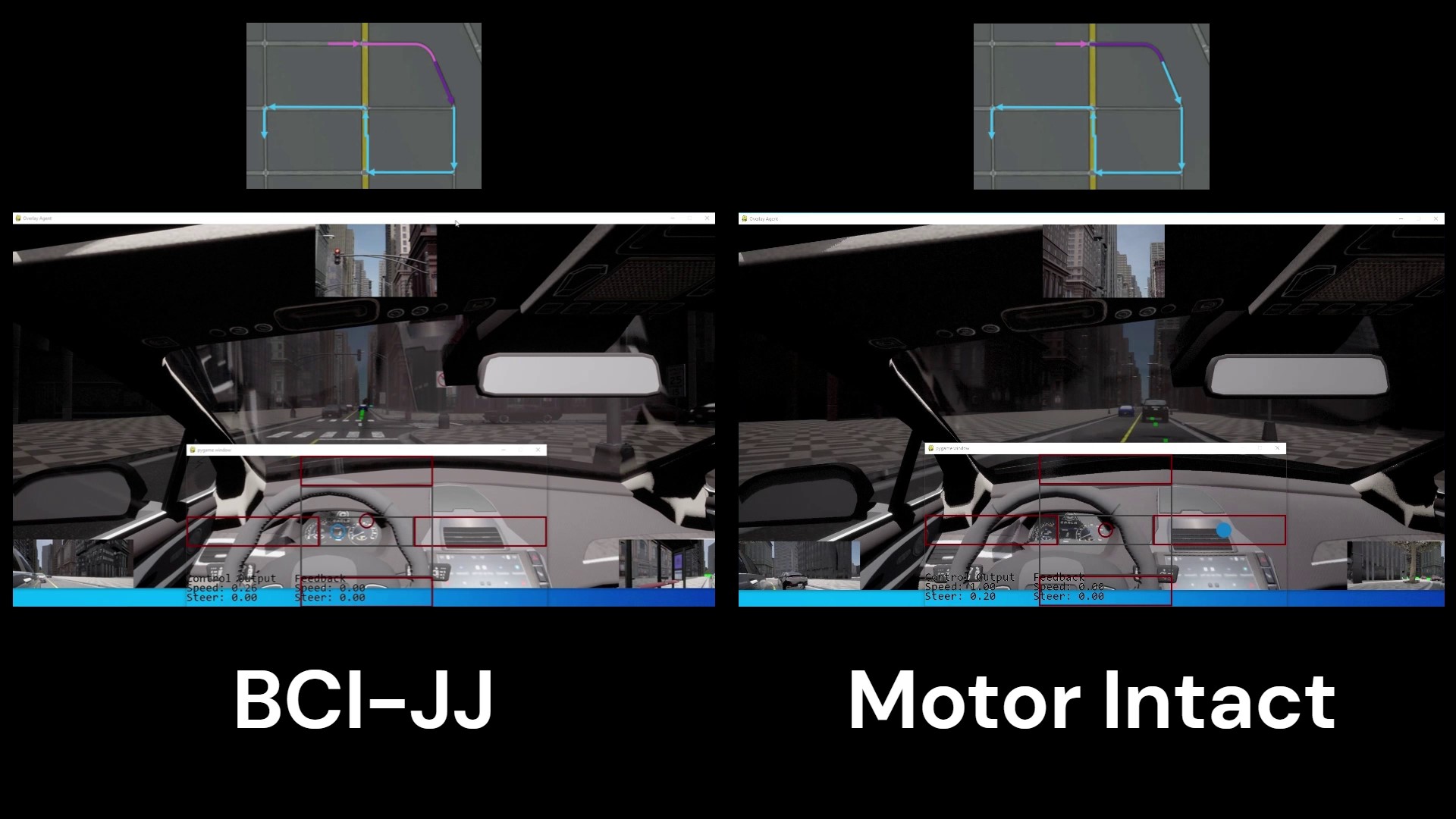}
\caption{\textbf{Side-by-side comparisons of the average simulated town driving performance between BCI-JJ and a motor intact participant using the bimanual cursor-and-click control.} We showed two example runs from BCI-JJ and one motor intact participant on the town route modified from the CARLA Leaderboard 2.0. Among all the collected runs, these two example runs had their simulated driving scores closest to the average scores for BCI-JJ and the motor intact control group. We retrieved the cursor movement and click information from either the BCI decoder for BCI-JJ or the joystick for motor intact participants. With the overlay instance floating above the vehicle simulation window, each participant got real-time visual feedback of their cursor and click control, the view ahead of the vehicle, and the side views. The cursor movement with the right thumb controlled steering and speed of the virtual vehicle, whereas the clicks with the left index finger controlled its full-stop braking. Each route map (top) corresponding to each participant's driving recording has the current segment labeled in dark purple. The video is played back at 16 times the speed of the original recording.}
\label{Smovie:simulated-driving}
\end{figure}

\end{document}